%% file: main.tex
\documentclass[epjST]{svjour}
\usepackage{graphics}
\usepackage[sort&compress,numbers]{natbib}
\usepackage{graphicx,amsmath,amssymb,soul,bbm,slashed}
\usepackage{bbm}
\usepackage{amsfonts}
\usepackage{mathtools}
\usepackage{xspace}
\usepackage[utf8]{inputenc}
\usepackage{dcolumn,booktabs,multirow,array,float}
\usepackage{multicol}
\usepackage{mfirstuc}

\usepackage[usenames,dvipsnames,svgnames,table]{xcolor} 

\usepackage[pdftex,
            pdftitle={Review sigma and kappa.},
            pdfauthor={J.R. Pelaez, A. Rodas, J. Ruiz de Elvira},
            bookmarks=false,
            colorlinks,
            linkcolor=SeaGreen,
            citecolor=SeaGreen,
            menucolor=black,
            urlcolor=SeaGreen,
            plainpages=false,
            pdfpagelabels,
            hypertexnames=false]{hyperref}

\graphicspath{ {figures/} }
\newcommand{\degree}{{\rm o}}
\newcommand{\im}{\mbox{Im}}
\newcommand{\re}{\mbox{Re}}

\newcommand{\mev}{{\mbox{MeV}\,}}

\newcommand{\addReviewer}[2]{
  \expandafter\newcommand\csname #1\endcsname[1]{{\bf \color{#2} \capitalisewords{#1}:\,##1}}
  \expandafter\newcommand\csname #1cor\endcsname[2]{{\color{#2} \capitalisewords{#1}:\,\st{##1}{\bf ##2}}}
  \expandafter\newcommand\csname #1color\endcsname{#2}
}

\definecolor{chromeyellow}{rgb}{1.0, 0.65, 0.0}
\definecolor{DodgeBlue}{rgb}{0.118, 0.565,1.000}
\definecolor{asparagus}{rgb}{0.53, 0.66, 0.42}
\definecolor{cadmiumgreen}{rgb}{0.0, 0.42, 0.24}

\addReviewer{CP}{orange}
\addReviewer{AR}{asparagus}
\addReviewer{JR}{red}
\addReviewer{JRE}{blue}

\newcommand{\sig}
{\ensuremath{\sigma/f_0(500)}\xspace}
\newcommand{\kap}
{\ensuremath{\kappa/K_0^*(700)}\xspace}
\newcommand{\pipi}
{\ensuremath{\pi \pi \rightarrow \pi \pi}\xspace}
\newcommand{\pik}
{\ensuremath{\pi K \rightarrow \pi K}\xspace}

\begin{document}
\title{Precision dispersive approaches versus unitarized Chiral Perturbation Theory  for the lightest scalar resonances
$\sig$  and $\kap$. 
}
\headnote{\hfill{\footnotesize{JLAB-THY-21-3306}}}
\author{Jos\'e R. Pel\'aez\inst{1}\fnmsep\thanks{\email{jrpelaez@fis.ucm.es}} 
\and Arkaitz Rodas\inst{2,3}\fnmsep\thanks{\email{arodas@wm.edu}}
\and Jacobo Ruiz de Elvira\inst{4}\fnmsep\thanks{\email{elvira@itp.unibe.ch}}
}
\institute{Departamento de F\'{\i}sica Te\'orica. Universidad Complutense and IPARCOS. 28040 Madrid. SPAIN \and
Department  of  Physics, College  of  William  and  Mary,  Williamsburg,  VA  23187,  USA
\and Thomas Jefferson National Accelerator Facility, 12000 Jefferson Avenue, Newport News, VA 23606, USA
\and Albert Einstein Center for Fundamental Physics, Institute for Theoretical Physics, University of Bern, Sidlerstrasse 5, 3012 Bern, Switzerland}
%
%
\abstract{ For several decades, the \sig and \kap resonances have been subject to long-standing debate. Both their existence and properties were controversial until very recently.
In this 
tutorial review 
we compare model-independent dispersive and analytic techniques  versus unitarized Chiral Perturbation Theory, when applied to the lightest scalar mesons $\sig$ and $\kap$. Generically, the former have settled the long-standing controversy about the existence of these states, providing a precise determination of their parameters, whereas unitarization of chiral effective theories allows us to understand their nature, spectroscopic classification and dependence on QCD parameters. Here we review in a pedagogical way their uses, advantages and caveats.
} 
\maketitle

\section{Introduction}
\label{sec:intro}
\input{sections/introduction.tex}


\section{ The $\sigma$ and $\kappa$ controversy: the data and model problems}
\label{sec:problems}
\input{sections/controversy.tex}




\section{Analytic structure of amplitudes and Dispersion relations}
\label{sec:AnsDR}
\input{sections/dispre.tex}


\section{Precision dispersive analyses}
\label{sec:precisionDR}
\input{sections/dispanalysis.tex}


\section{Unitarized Chiral Perturbation Theory}
\label{sec:UChPT}
\input{sections/uchpt.tex}


\section{Summary}

After reviewing the phenomenological interest on the lightest scalar mesons $\sig$ and $\kap$ for hadron spectroscopy, 
we have provided a tutorial review of the advantages and caveats of the fully dispersive methods versus unitarized Chiral Perturbation Theory (UChPT) to tackle their existence and properties, always in the isospin-conserving approximation and neglecting electroweak effects.

Very briefly, the fully dispersive methods have provided the rigorous mathematical treatment to settle, once for all, their existence, which had nevertheless been strongly suggested before by many phenomenological approaches, including UChPT. In general, dispersion relations aim at precision and model independence, respecting crossing symmetry. They also provide precise values for threshold parameters of relevance for ChPT. The price to pay is little insight into the underlying dynamics, complicated integral equations and the need for input from data, together with reliable estimates of their uncertainties, in many regions. 

We have seen that the analyticity properties of unitarization methods can be justified from dispersion relations. Elastic or two-body coupled channel unitarity is exact, fixing the imaginary part of the inverse partial wave (or matrix of partial waves when considering coupled channels). Different unitarization methods correspond to different approximations of the real part of the inverse partial-wave (or matrix of partial waves). Crossing symmetry and the unphysical cuts are just an approximation. Therefore precision and full model independence is not the aim, although with the best approximations to the chiral series, the $\sig$ and $\kap$ poles are very close to their values determined from dispersion relations.
Dynamical information is encoded in the ChPT parameters, whose expansion they match at low energies. 
This provides a connection with Quantum Chromodynamics and allows for the study of the $\sig$ and $\kap$ dependence on  quark-masses and the number of colors. The emerging picture is that both the $\sig$ and $\kap$ are not ordinary quark-antiquark states, but that their formation dynamics is dominated by meson-meson interactions at the scale of spontaneous chiral symmetry breaking.

Both $\pi\pi$ and $\pi K$ scattering are very relevant for many other processes in Hadron Physics, and the recent dispersively constrained amplitudes do provide model independent precise input for any other hadronic process involving at least a pair of these mesons in the final state. The confirmation of the existence of a $\sigma$ and a $\kappa$ should not be ignored also when two pions or a pion and a kaon are exchanged in a given process. For instance, in the NN interaction the role of the $\sigma$ has been revisited with unitarized chiral interactions \cite{Oset:2000gn}, confirming that it provides a moderate attraction although not as much as in conventional sigma-exchange models.

The dispersive treatment of these scattering reactions also provides a template and a demonstration of the power and usefulness of dispersive treatments, whose use is also spreading to other phenomenological analysis in Hadron Physics. We have also argued why it is interesting for future analysis of lattice data.

Chiral unitarization methods, particularly the simplest ones, have also become very popular in other areas of Hadronic Physics, reaching the realm of baryons and heavy mesons, for which effective theories do not converge so well or are less developed than meson ChPT. Many of the predictions and approximations of these simple methods in such contexts may be thus justified, inspired and supported by the robustness of the unitarization approach for $\pi\pi$ and $\pi K$ interactions, where the most elaborated and rigorous derivations exist and we have briefly reviewed here.

\begin{acknowledgement}
 {\bf Acknowledgements} We acknowledge J.A. Oller for encouraging us to write a tutorial review on this topic. We thank I. Danilkin for his corrections on the manuscript. This project has received funding from the Spanish Ministerio de Ciencia e Innovación grant PID2019-106080GB-C21 and the European Union’s Horizon 2020 research and innovation program under grant agreement No 824093 (STRONG2020). AR acknowledges the financial support of the U.S. Department of Energy contract DE-SC0018416 at William \& Mary, and contract DE-AC05-06OR23177, under which Jefferson Science Associates, LLC, manages and operates Jefferson Lab. JRE acknowledges financial support from the Swiss National Science Foundation under Project No. PZ00P2 174228.
\end{acknowledgement}

\footnotesize
\bibliography{largebiblio.bib}

\end{document}

%% file: sections/introduction.tex
The lightest scalar mesons have been a matter of debate since they were proposed around six decades ago. A fairly light neutral scalar field 
was introduced in 1955 by Johnson and Teller \cite{Johnson:1955zz} to explain the attraction between two nuclei. Schwinger \cite{Schwinger:1957em} soon considered it as an isospin singlet and named it $\sigma$, remarking that it would couple strongly to pions and be very unstable and difficult to observe. In the early sixties Gell-Mann \cite{GellMann:1960np} considered it the fourth member of a multiplet together with the three pions to build his famous ``Linear Sigma Model" (L$\sigma$M), describing spontaneous  symmetry breaking and the lightness of pions, identified with massless Nambu-Goldstone Bosons (NGB).
Actually, as pseudo-NGB, since they have a small mass. 
Similarly, a relatively light scalar–isoscalar
resonance, very wide due to its  strong coupling to two pions,  was also generated in  Nambu–Jona–Lasinio (NJL) models \cite{Nambu:1961tp,Nambu:1961fr,Hatsuda:1994pi}, where the $\sigma$ mass is
generically around twice the constituent quark mass, $m_\sigma\sim 2\times300\mev$. With the advent of QCD in the early seventies
and its rigorous low-energy effective theory \cite{Weinberg:1978kz,Gasser:1983yg,Gasser:1984gg}, known as Chiral Perturbation Theory (ChPT),  as well as with better measurements at low-energies \cite{Batley:2010zza}, we understand the L$\sigma$M and NJL just as toy models, which, very roughly, capture the leading order behavior of ChPT, but have further additions that do not fully agree with QCD or experiment. 

Experimental claims for relatively narrow scalar-isoscalar states, not quite as the expected wide $\sigma$, were made as son as 1962, but not confirmed. A wide resonance in the 550-800 MeV region was present in the Review of Particle Properties (RPP)~\cite{pdg} since the late sixties\footnote{ Although we have quoted the last RPP edition, all previous ones can be found in the Particle Data Group web-page at https://pdg.lbl.gov/rpp-archive/.}, but not firmly established. The appearance of high-statistics $\pi\pi$-scattering phase shifts \cite{Protopopescu:1973sh,Hyams:1973zf,Grayer:1974cr} showed no indication for Breit--Wigner-like peaks, as we will see below, and thus  these states were removed from the 1976 RPP edition and the lightest scalar isoscalar state was listed around 1 GeV.  However,  many models showed later that a $\sigma$ resonance as wide as 500 MeV was needed for better data description. In particular, unitarization of meson-meson interactions had also been shown earlier \cite{vanBeveren:1986ea} to be needed to match quark-model predictions to the scattering information (see also the recent \cite{Lukashov:2019dir}), and coupled channel analyses also suggested its presence \cite{Au:1986vs,Zou:1993az,Zou:1994ea,Janssen:1994wn}.
For such wide resonances the rigorous description is made in terms of the associated pole in the second Riemann sheet of the complex plane of the Mandelstamm $s$ variable, identifying its mass $M$ and width $\Gamma$ as $\sqrt{s_{pole}}=M-i\Gamma/2$.
Hence, a light $\sigma$ was resurrected  in the RPP 20 years later, although with an extremely conservative  name $\sigma(400-1200)$ and similarly large uncertainty on the width.  Around the same time, it became clear that chiral constraints together with unitarity and analyticity yield a light and very broad sigma meson \cite{Dobado:1996ps,Oller:1997ti}. Breit-Wigner approximations, devised for narrow resonances, could not describe such a pole and simultaneously the ChPT constraints.  
With the turn of the millennium, further experimental support, complementary to scattering data, appeared from heavy meson decays \cite{Asner:1999kj,Aitala:2000xu,Aitala:2002kr,Ablikim:2004qna,Ablikim:2006bz,Bonvicini:2007tc}, which helped changing the name in the 2002 RPP to $\sigma(600)$. Later on, from the theoretical side, a very precise  $\sigma$ meson was provided in \cite{Caprini:2005zr}, building  on rigorous dispersive analyses
\cite{Ananthanarayan:2000ht,Colangelo:2001df} with Roy equations \cite{Roy:1990hw}, ChPT and data.
These equations implemented crossing and analyticity to determine the $\sigma$ meson pole position, which was shown to lie  within their applicability region \cite{Caprini:2005zr}.
Note that these works do not use data for scalar or vector waves below 800 MeV, and in particular in the $\sig$ region, which is why their $\sigma$ pole can be considered a prediction.
From the experimental side, the accurate methods devised \cite{Colangelo:2008sm} for extracting very reliable low-energy $\pi\pi$ data from $K_{\ell 4}$ decays measured at NA48/2 at CERN \cite{Batley:2010zza}, provided the needed precision for a competitive dispersive determination of the $\sig$ from data \cite{GarciaMartin:2011cn,GarciaMartin:2011jx}. Furthermore those data also excluded many existing models. Consequently, in 2012 the $\sigma$ was finally considered well established in the RPP and called $f_0(500)$, reducing dramatically its estimated uncertainties. A much more detailed account of the $\sigma$ meson history can be found in the review \cite{Pelaez:2015qba}, together with the estimate of its pole taking into account rigorous dispersive analyses only: $(449^{+22}_{-16})-i(275\pm12)\,$MeV. The RPP, however, takes into its estimate less rigorous determinations and provides larger uncertainties for the $t$-matrix pole. In addition, it provides a Breit-Wigner approximation, which, as we will show below, is definitely inappropriate for the $\sig$ but, unfortunately, still popular even in modern analyses.

The first prediction  for a $\kappa$ meson followed relatively soon after that of the $\sigma$, as a result of a quark model with a simple potential proposed by Dalitz \cite{Dalitz:1966fd} in 1965. Following a $q \bar q$ assignment it was crudely expected around 1.1 GeV and forming a nonet with the $\sigma$ meson. In Dalitz's own words: ``Quite apart
from the model discussed here, such $K^*$ states are expected to exist simply on the basis of $SU(3)$ symmetry". 
Around the mid 60's there were several claims and refutations of an scalar isospin-1/2 $\kappa$ state in $\pi K$ scattering around 725 MeV, but with a very tiny width of 20 MeV or less. This extremely narrow state was omitted from the main tables and considered discredited\footnote{Literally, the authors of that compilation wrote: {\it ``We are beginning to think that $\kappa$ should
be classified along with flying saucers, the
Loch Ness Monster, and the Abominable Snowman"}.} in the 1967 compilation of ``Data on Particles and Resonant States" \cite{ROSENFELD:1967zz}, the precursor of our modern Review of Particle Physics. But the $\kappa$ name stuck. The need for a near-threshold  $\kappa$ state besides the $K^*$ observed above 1.3 GeV was also required from attempts to saturate the Adler-Weisberger Axial-charge sum rule for $\pi K$ scattering, although the mass predictions were very crude, ranging from 0.85-1.2 MeV \cite{Mathur:1966zza} or 500-740 MeV \cite{Matsuda}. This predicted state was much wider than 30 MeV, even reaching a 450 MeV width \footnote{See reference 15 in \cite{Trippe:1969zs}.}. As soon as 1967 there were also experimental claims  of a broad scalar $\pi K$ resonance near 1.1 GeV \cite{Trippe:1969zs}, although the authors explicitly state that ``The assumption of a pure Breit-Wigner form for the S-wave is probably a serious oversimplification".
As it  happened with the $\sigma$, the $\kappa$ was also removed from the RPP in 1976. 
At the end of that decade the first high-statistics $\pi K$ scattering phase-shift analysis was obtained at SLAC \cite{Estabrooks:1977xe}, showing 
a strong increase in the $\kappa$ channel near threshold but no evident resonance shape, a situation that was later confirmed with the 1988 high-statistics experiment at the Large  Acceptance  Superconducting  Solenoid  (LASS) Spectrometer, also  at SLAC. Nevertheless, several models and reanalysis of these data still found a wide $\kappa$
\cite{vanBeveren:1986ea,Aitala:2002kr,Bugg:2003kj}, including those using unitarized ChPT \cite{Oller:1998hw,Oller:1998zr,Pelaez:2004xp}. Although some theoretical works \cite{Ishida:1997wn}, using Breit-Wigners, suggested that the $\kappa$ could be as massive as 900 MeV, this possibility was soon discarded in favor of a lighter state \cite{Cherry:2000ut}.
However, for the $\kappa$ it took two more years than for the $\sigma$ to return to the RPP, which happened in 2004, under the name of $K^*_0(800)$. It was nevertheless omitted from the summary table and carried the warning that ``The existence of this state is controversial".  With further and strong experimental evidence for a $\kappa$ 
from heavy-meson decays \cite{Ablikim:2005ni}, such a warning changed to ``Needs confirmation" in 2006 and has been kept as such until this year's 2020 edition. It should be noted that in 2003 Roy-Steiner fixed-t partial-wave dispersion relations were rigorously {\it solved} for the first time in $\pi K$ scattering \cite{Buettiker:2003pp}, using input only above roughly the elastic region. A later analysis by the same group, using hyperbolic dispersion relations \cite{DescotesGenon:2006uk} showed the existence of the $\kappa$ pole within their region of applicability. Note that this approach did not use data on the $\kappa$ region. The predicted pole-mass lied much lower than other analysis, $(658\pm13)-i(279\pm12)\,$MeV. With the aim of providing the required confirmation using data in the $\kappa$ region, two of us started a program to constrain the data analysis with several kinds of dispersion relations \cite{Pelaez:2016tgi,Pelaez:2018qny,Pelaez:2020uiw,Pelaez:2020gnd}. Our use of analytic techniques  to extract poles from data fits constrained with forward dispersion relations \cite{Pelaez:2016tgi} confirmed the pole around 670 MeV \cite{Pelaez:2016klv}, definitely very far from 800 MeV. This growing evidence for an even lighter $\kappa$ pole lead to the present name $K^*_0(700)$, given in the 2018 RPP edition and its inclusion in the summary tables in 2020, although still under the ``Needs confirmation" label.  We recently finished our ``data driven" dispersive program \cite{Pelaez:2020uiw,Pelaez:2020gnd}, confirming the light pole at 
$(648\pm7)-i(280\pm16)\,$MeV. The RPP ``Needs Confirmation" warning for the $\kap$ will be removed in the following update\footnote{C. Hanhart. Private communication.}.

\begin{figure}
\begin{center}
\resizebox{\columnwidth}{!}{%
  \includegraphics{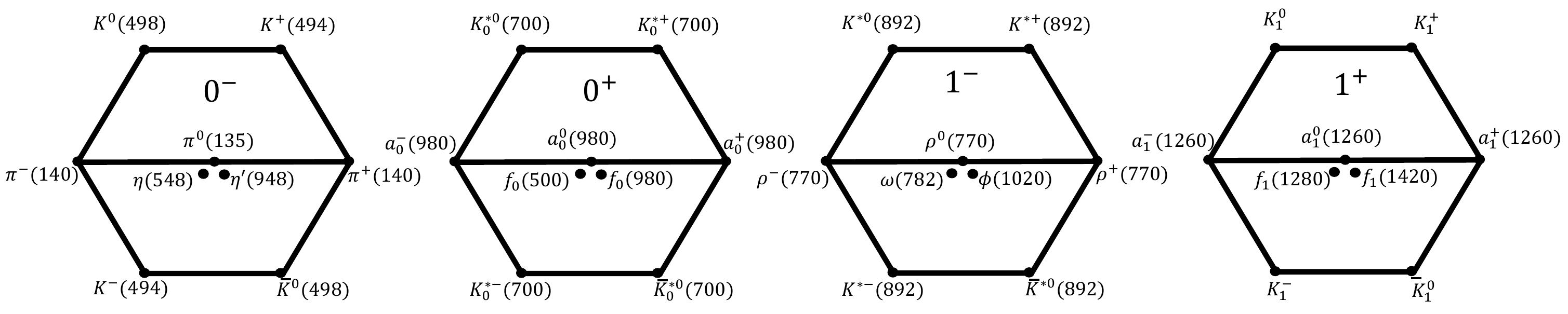} }
\caption{Lightest meson nonets with different $J^P$ numbers. Note that, contrary to the other nonets
and naive expectations of ordinary $q\bar q$ mesons, the $\kap$ is lighter than the isotriplet $a_0(980)$. This ``inverted" hierarchy is closer to states made of two quark-antiquark pairs, generically known as tetraquarks, which could be in the form of ``genuine" (or ``elementary") tetraquarks or meson-meson states. Note also that mesons with the same $J$ but opposite parity have mass differences of several hundred MeV. We refrain from identifying the $K_1$ mass because in the RPP there are two nearby states, $K_1(1270)$ and $K_1(1400)$, which can mix and the former may possibly have two poles \cite{Roca:2005nm}, leading to different interpretations.}
\label{fig:multiplet}       
\end{center}
\end{figure}

Thus far we have commented on the relevant role that light scalars play in nucleon-nucleon attraction and in the spontaneous chiral symmetry breaking of QCD. However, they are also of interest for spectroscopy for several reasons: First, since pions, kaons and etas are so light due to their pseudo-NGB nature and the existence of a mass gap, light scalars become the first ``non-NGB" states after that mass gap. In particular, the pseudo-NGB mass is proportional to the quark mass, and becomes zero in the chiral limit. In contrast, light scalars are the first QCD states whose mass is not protected by the chiral symmetry breaking mechanism, but dominated by the QCD dynamics beyond the symmetry breaking pattern and should not vanish even in the chiral limit. Second, given its non-abelian nature, one of the most salient features of QCD, or more precisely Yang-Mills theories, is the existence of self-interactions between gluons. This suggests the existence of bosonic glueball states, which look like flavorless mesons without isospin. However, in QCD we do not expect pure glue states, but mixtures with other mesonic configurations made of quarks and antiquarks as long as they have the same quantum numbers~\cite{Cohen:2014vta}. Nevertheless, the existence of glueball configurations will lead to an excess of states with respect to the flavor $SU(3)$ multiplets formed just with quarks and antiquarks. The lightest glueball is expected to have zero angular momentum or intrinsic spin, which are the same quantum numbers of the $\sig$ meson. It is, therefore, very relevant to be able to classify  all $SU(3)$ nonets, and see if there are indeed more $f_0$ states than needed to complete them. Of course, 
the number of required nonets can be determined easily by counting the states with strangeness, since they cannot mix with glueballs. This is where the $\kap$  plays a very significant role, because its presence implies the existence of a light nonet, with at least some lighter $f_0$ state, i.e., the $\sig$, which, therefore, cannot be identified with a glueball (apart from the fact that Lattice-QCD calculations predict the predominantly glueball state to lie around 1.5-1.8 GeV \cite{Berg:1982kp,Michael:1988jr,Bali:1993fb,Sexton:1995kd,Morningstar:1999rf,Hart:2001fp,Chen:2005mg,Richards:2010ck,Gregory:2012hu}). Given that the $\kap$ is already confirmed, and is the lightest strange resonance, together with the other lightest meson scalar resonances $\sig$, $f_0(980)$ and  $a_0(980)$, they form the nonet depicted in Fig.~\ref{fig:multiplet}.

After settling which are the members of the lightest scalar nonet, one immediately observes that its mass hierarchy does not match that of ordinary 
$q\bar q$-valence  mesons, since for them one would expect the strange mesons to have one strange quark or antiquark, whereas the isotriplet would contain no strange quarks at all, and should therefore be roughly 200 MeV lighter. However, the opposite is found, since the $\kap$ meson is more than 300 MeV lighter than the $a_0(980)$ (remember the $\kap$ pole is actually closer to 650 than 700 MeV). This possibility had already been contemplated by Jaffe in 1976 \cite{Jaffe:1976ig}, who proposed that these states could correspond to some ``tetraquark"-valence configuration. 
In this case, the isotriplet would be heavier because it would contain one strange quark-antiquark pair, and thus no net strangeness, but the 
$\kappa$ would still have just one strange quark or antiquark and therefore be lighter. Recently, Jaffe has argued that with respect to this argument ``there is no clear distinction between a meson–meson molecule
and a $\bar q\bar q qq$" \cite{Jaffe:2007id}. This interpretation seems favoured when analysing data on scattering and decays with chiral models \cite{Achasov:1994iu,Achasov:2005hm,Giacosa:2006rg,Parganlija:2010fz}. However, different kinds of mesons could be distinguished also from their  different dependence on the number of colors, $N_c$, of QCD \cite{tHooft:1973alw,Witten:1980sp}.
Weinberg \cite{Weinberg:2013cfa} recently showed that 
 ``elementary" tetraquark  and $q\bar q$ mesons have the same behavior.  Such behavior is at odds with the $\sig$ and $\kap$ $1/N_c$-leading dependence obtained from unitarized ChPT \cite{Pelaez:2003dy,Pelaez:2004xp,Pelaez:2006nj,RuizdeElvira:2010cs,Guo:2011pa,Guo:2012ym,Guo:2012yt} which suggests that the 
 predominant dynamics in their formation occurs at the meson scale both for the $\sig$ and the $\kap$ arising from the chiral meson loops responsible for the unitarity cut. Using just their pole position and residues obtained dispersively, the ordinary $q\bar q$ meson and even more the glueball components should be subdominant \cite{Nebreda:2011cp}. 
 See \cite{Pelaez:2015qba} for a review. 
 Within the context of quark models, some form of unitarization of meson-meson interactions 
 seems also essential to reproduce the scattering data and the lightest resonances   \cite{vanBeveren:1986ea,Black:1998wt,Black:1998zc}. See also the recent review in \cite{vanBeveren:2020eis}.

Related to the nature of the scalar mesons, they are also a more technical source of interest, since the low-energy constants (LECs) that appear at each order of the ChPT expansion and encode the information on the underlying theory or smaller scales, are generally understood as the remnants of the exchange of the other resonances that are not explicitly included in the effective theory \cite{Gasser:1983yg,Ecker:1988te}. 
In principle, the contribution to the LECs should be dominated by the lightest resonances that are integrated out.
However, it is known that the LECs are saturated by vector resonances \cite{Gasser:1983yg,Ecker:1988te}, even though the $\sig$ and $\kap$ are lighter and wider than their respective vector counterparts, i.e., the  $\rho(770)$ and $K^*(892)$. For some time this suggested that the lightest scalars should be heavier than the vectors. However, this is also an indication that the dynamics that generates these light scalars occurs at meson-meson interaction scales, i.e., by the unitarization of the LO ChPT, and that is why they do not contribute to the LECs, whereas the vectors are ``genuine" QCD scales.

Finally, light scalars are very relevant not only by themselves, but also because pions and kaons, being the lightest mesons, appear as products of almost all hadronic reactions and, when there are at least two of them, their final state interactions may reshape the whole process. Actually, by Watson's Theorem, the phase of the whole process is given by the phase of the two mesons if they are the only particles that interact strongly.
There is a well-known vector meson dominance, but scalar exchanges also give important contributions, 
and the $\sig$ and $\kap$ dominate near the two particle threshold. Therefore, having a precise description of 
meson-meson scattering partial-wave amplitudes is relevant to describe many other hadronic processes of interest, and even more so now that the present experimental facilities are providing unprecedented statistics.

In conclusion, we have seen the relevant role of the $\sig$ and $\kap$ in Hadron Physics.
Of particular interest for the next sections is the crucial role played by model-independent dispersive determinations of meson-meson scattering and resonance poles to close the debate about the existence of both the $\sig$ and $\kap$ resonances, but also to provide a precise and rigorous determination of their parameters. These methods do not make any assumption about the underlying dynamics and no attempt to model it. Beyond scalar waves, they also provide relatively simple, consistent and {\it precise} parameterizations of meson-meson partial waves that can reach energies between 1.5 and 2 GeV, which are of interest for studying other resonances and to parameterize final-state interactions for further phenomenological and experimental studies.
In contrast, different ChPT unitarization methods 
also yield strong support in these directions, but they contain further simplifying approximations that make them less suited for precision studies. However, they provide connections to fundamental QCD parameters, like quark masses or $N_c$, that allow us to understand the relation between these resonances as well as their nature and spectroscopic classification, which cannot be achieved with the other purely dispersive methods. In this sense both approaches are complementary. 

The aim of this review is to provide a brief pedagogical introduction to these methods, discuss  when and why one method is more appropriate than the other and what have we learned from them about the $\sig$ and $\kap$. But before that, let us detail a little more why these two particular states have been so controversial.

%% file: sections/controversy.tex
On a first approximation, all the reasons behind the longstanding debate about the existence and properties of these two states can be reduced to two main problems:
The difficulty in getting good meson-meson scattering data 
with reliable uncertainties, and the use of too simple models to analyze data either from scattering or decays. The use of these too simple models also hinders the discussion about the classification, interpretation and nature of these states.

\subsection{The data problem}

\begin{figure}
\begin{center}
\resizebox{\columnwidth}{!}{%
  \includegraphics{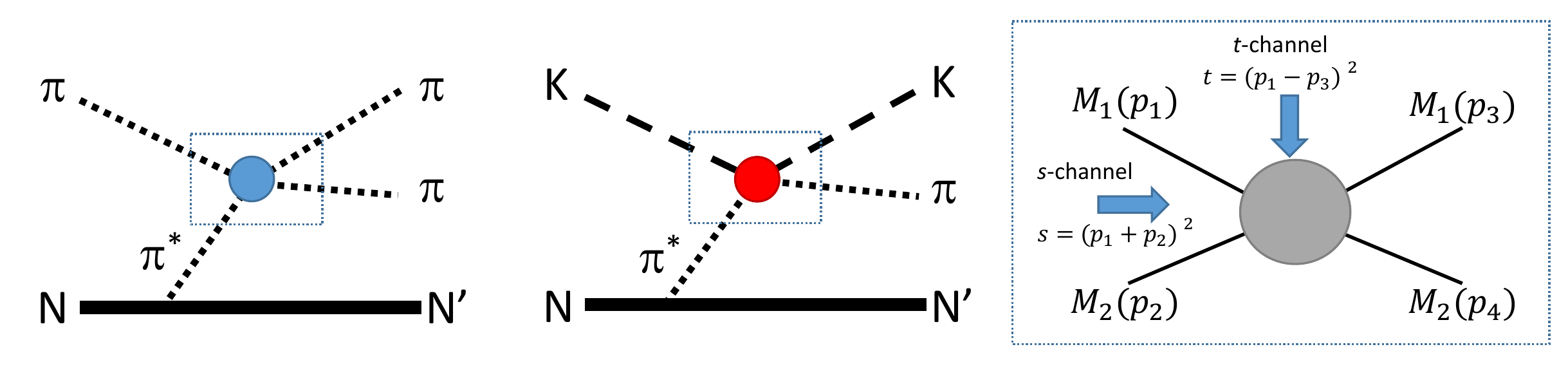} }
\caption{One-pion exchange diagrams for the processes used to obtain data on $\pi\pi$ scattering (Left) and $\pi K$ scattering (Center), assuming that the scattering subprocess (Right, inside a dotted square) factorizes from the full process. In principle these diagrams should dominate the full process when the kinematics are such that the virtual exchanged pion is almost real, i.e., close to the exchanged pion pole. On the right panel we show the generic meson-meson scattering subprocess and the definition of $s$ and $t$ variables and channels.}
\label{fig:mNmmN}       
\end{center}
\end{figure}

Since both kaons and pions are unstable, it is hard to make very luminous beams with them. Thus, lacking direct collisions, the available data is extracted indirectly from $P\, N \rightarrow P\, \pi\, N'$ processes, where $P=\pi,K$ and $N, N'$ are different kinds of nucleons. This was done by looking at the kinematic region where  the one-pion-exchange mechanism \cite{Goebel:1958zz,Chew:1958wd,Estabrooks:1975zw,Estabrooks:1976hc,Estabrooks:1976sp} dominates the whole process and assuming the meson-meson scattering sub-process can be factorized, as illustrated in Fig.~\ref{fig:mNmmN}. 

\begin{figure}
\begin{center}
\resizebox{.325\columnwidth}{!}{%
  \includegraphics{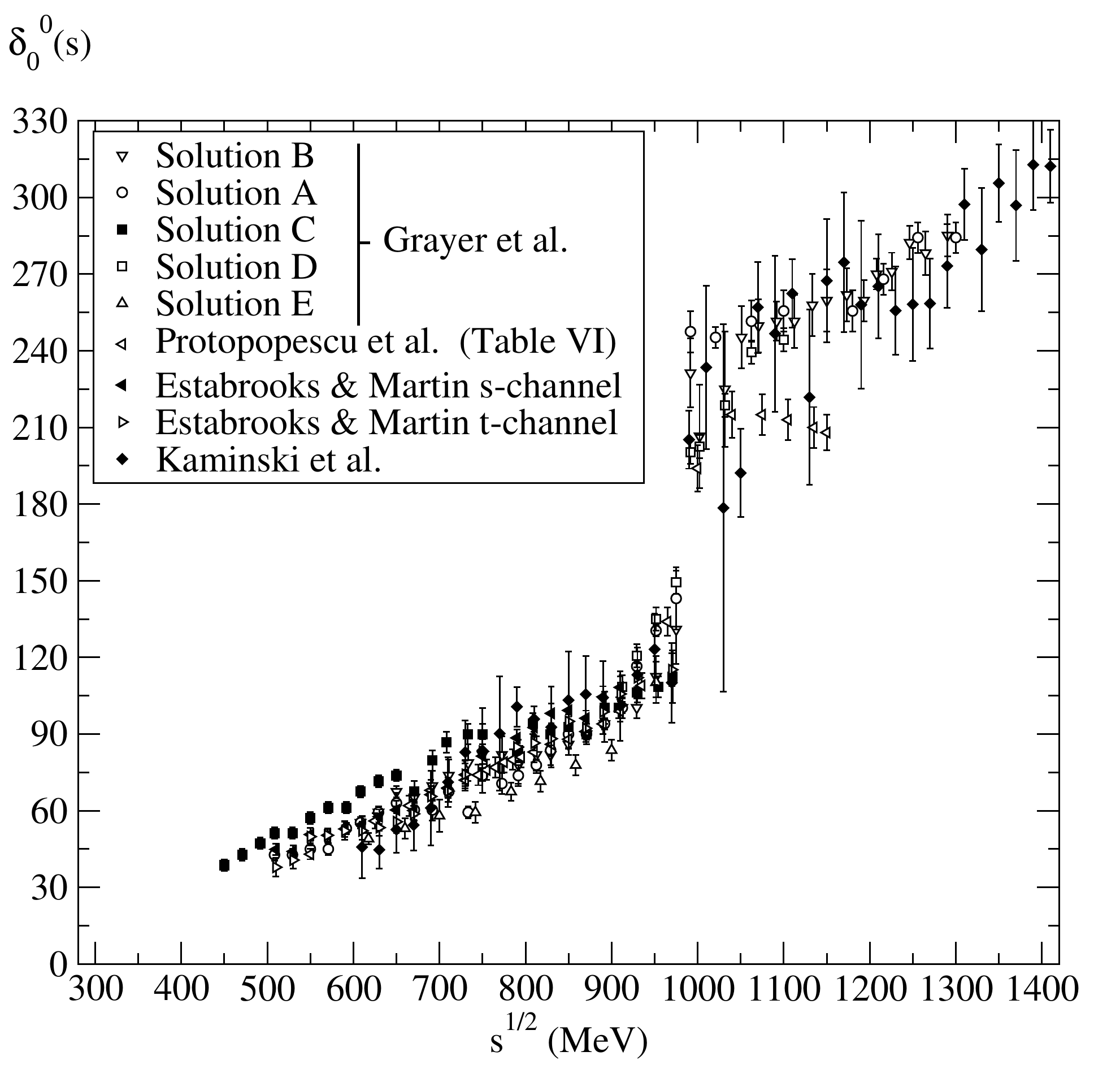} }
  \hspace{-.42cm}
\resizebox{.355\columnwidth}{!}{%
  \includegraphics{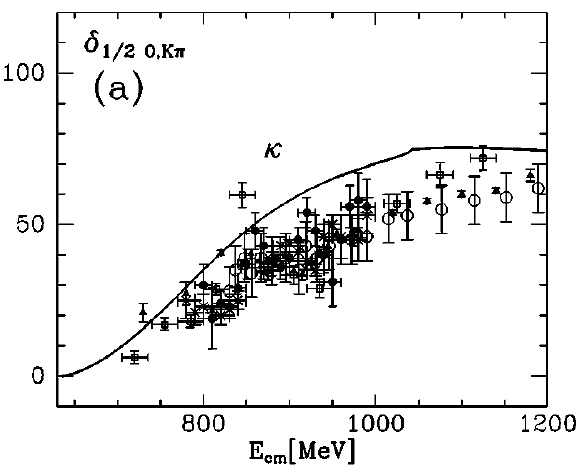} }
  \hspace{-.31cm}
\resizebox{.315\columnwidth}{!}{%
  \includegraphics{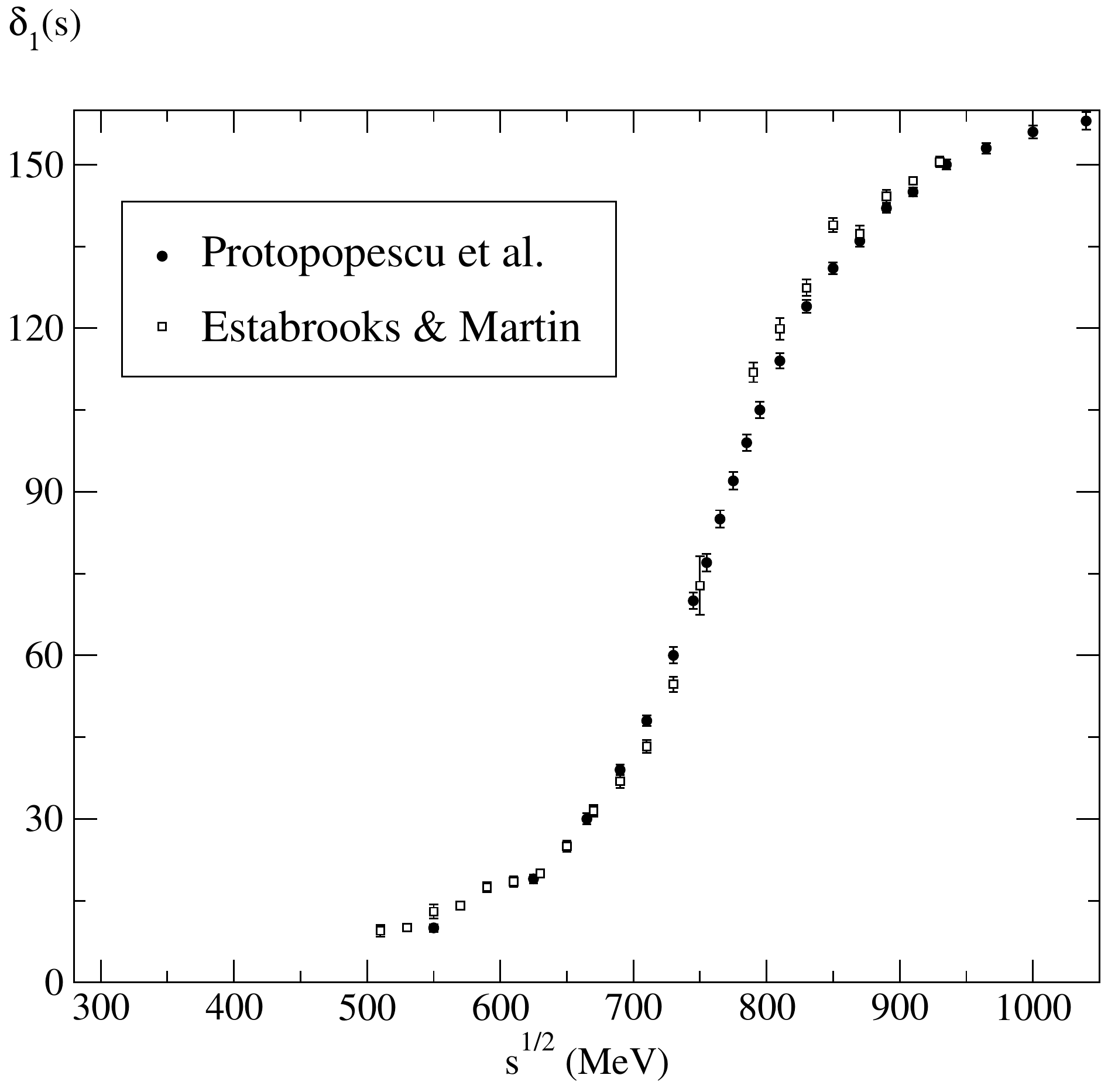} }
\caption{Example of meson-meson scattering phase shifts obtained from $P\,N\rightarrow P\,\pi\, N'$. {\bf Left:} $\pi\pi$ scalar isoscalar phase shift. The data come from Grayer et al. \cite{Grayer:1974cr}, Protopopescu \cite{Protopopescu:1973sh}, Estabrooks and Martin \cite{Estabrooks:1974vu} and Kaminski et al. \cite{Kaminski:1996da}. Note that each experiment may provide different and incompatible solutions. {\bf Center:} $\pi K$ isospin 1/2 scalar channel. Note the evident incompatibilities between data points. Data sources: solid circles \cite{Mercer:1971kn}, crosses \cite{Bingham:1972vy}, open squares \cite{Baker:1974kr}, solid
triangles \cite{Estabrooks:1977xe}, open circles \cite{Aston:1987ir}. {\bf Right:} Isospin-1 vector channel for $\pi\pi$ scattering. Data from \cite{Protopopescu:1973sh} and \cite{Estabrooks:1974vu}.
Figures taken from \cite{Pelaez:2015qba}, \cite{Oller:1998hw} and \cite{Pelaez:2015qba}, respectively.}
\label{fig:dataraw}       
\end{center}
\end{figure}

Unfortunately, it is hard to reach the kinematic region where the exchanged pion is almost on-shell and therefore there are other contributions that affect the extraction, plaguing the results with systematic uncertainties. As a consequence, the many experiments to determine meson-meson scattering are often incompatible among themselves and, moreover, even incompatible within the same experiment, since different extraction procedures can lead to different results incompatible within statistical uncertainties \footnote{On top of that even with the same method and data there were several ambiguities leading to different possible solutions, like for instance the so-called up or down solutions~\cite{Protopopescu:1973sh,Hyams:1973zf,Grayer:1974cr}. Over the years it has been possible to disentangle those with other input or dispersion relations \cite{Kaminski:2002pe}. We omit this discussion here and refer the reader to the review in \cite{Pelaez:2015qba} and references therein.}. In fact, there are other contributions from the exchange of other resonances, from reabsorption of other pions, corrections to the pole term, etc... (see \cite{Nys:2017xko}). This problem with the systematic uncertainties in the data is illustrated in Fig.~\ref{fig:dataraw}, where we show a sample of different $\pi \pi$ and $\pi K$ scattering data sets, some of them even coming from the same meson-nucleon experiment (note, for example, the five different solutions from \cite{Grayer:1974cr}). 

\begin{figure}
\begin{center}
\raisebox{-0.5\height}{%
  \includegraphics{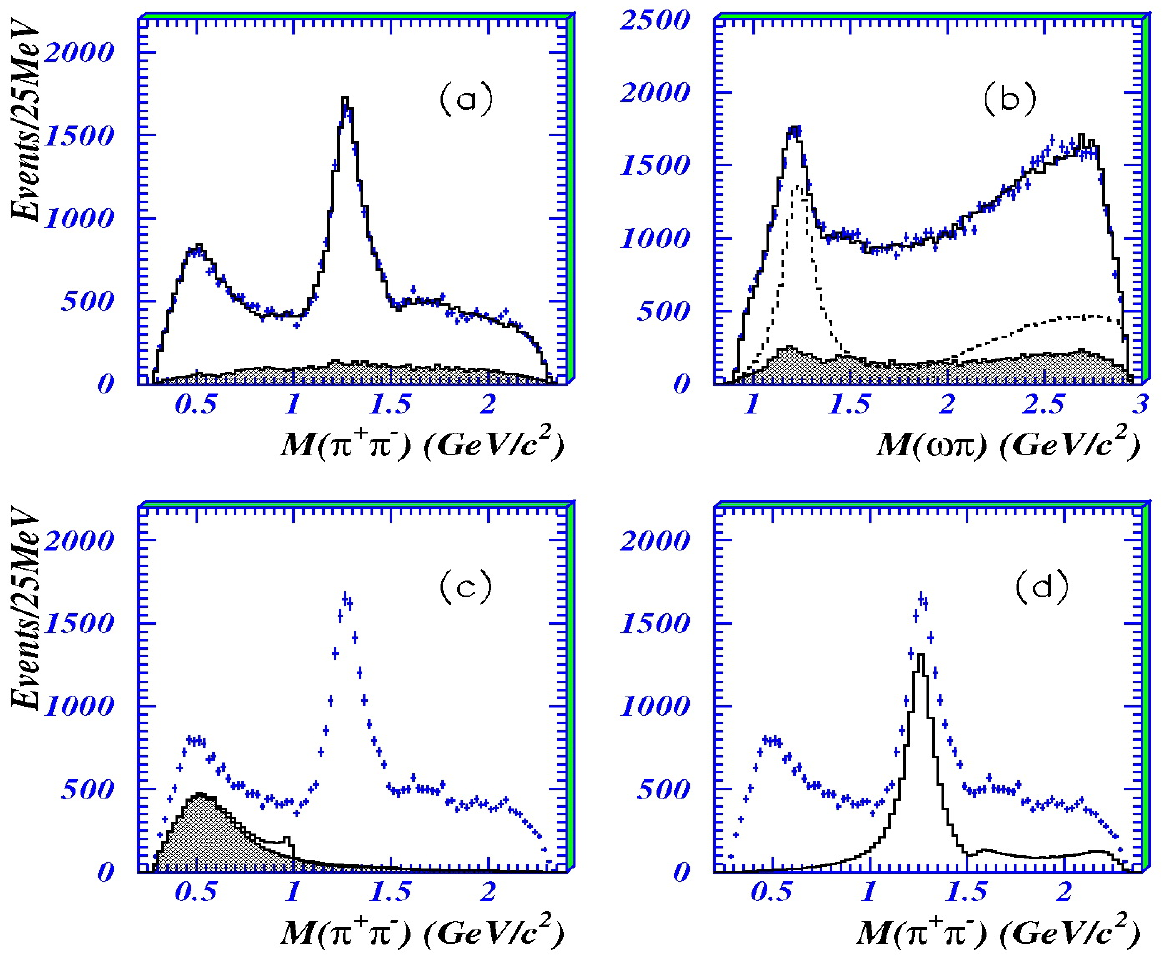} }
\raisebox{-0.5\height}{%
  \includegraphics{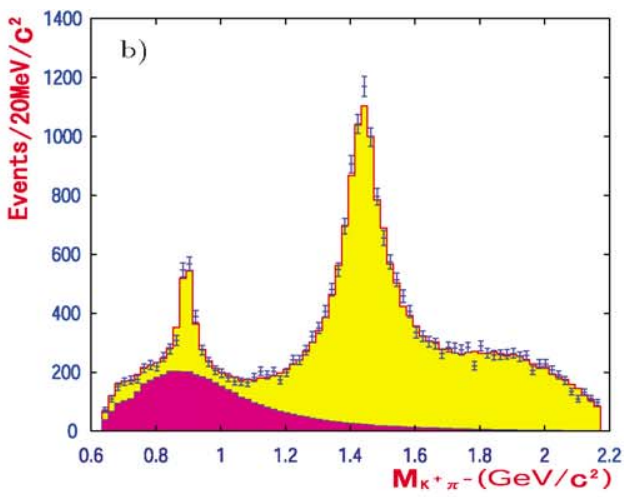} }
\caption{Example of the $\sigma$ and $\kappa$ conributions (as darker areas) in decay processes. 
{\bf Left:} The $\sigma$ in $J/\Psi\rightarrow\omega \pi^+\pi^-$. {\bf Right:} The $\kappa$ in 
$J/\Psi\rightarrow\bar K^*(892)^0 K^+\pi^-$. Figures taken from \cite{Ablikim:2004qna} and \cite{Ablikim:2005ni}, respectively.}
\label{fig:production}       
\end{center}
\end{figure}

Despite these clearly inconsistent data sets, some general trends were observed. First: no other coupled states, with additional pions, were detected until rather high energies, so that $\pi\pi$ was elastic up to the $K\bar K$ threshold, where the fast-rising shape of the $f_0(980)$ is clearly observed. Similarly, $\pi K$ scattering was also elastic in practice up to $K\eta$ threshold. Second: in this elastic region there is no evident sign of a resonance, nor for the $\sigma$ in the left panel of Fig.~\ref{fig:dataraw}, nor for the $\kappa$  in the central panel. There are no fast $180^\degree$ rises in either phase shift, as there is, for example, for the $f_0(980)$ in the left panel. Definitely, we cannot see the familiar Breit-Wigner shape associated to a well-isolated and relatively narrow resonance like the $\rho(770)$ that is easily identified in the $\pi\pi$ scattering data of the right panel, obtained by the same experiments using the same techniques. Note the $\pi K$ phase does not even reach $90^\degree$ in the elastic region.
It is worth mentioning that the $\kap$ debate has also raised interest in measuring $\pi K$ scattering in the recently accepted KLF proposal 
\cite{Amaryan:2020xhw} to use a neutral $K_L$ beam at Jefferson Lab with the Gluex experimental setup, to study strange spectroscopy and the $\pi K$ final state system up to 2 GeV.

rtant to mention that these states became more accepted once they were also observed in the decays of heavier hadrons (see Fig.\ref{fig:production}), which occurred around the turn of the millenium . As we commented in the introduction, there was a clear need for some light but very wide scalar resonance contribution in these processes. The relevance of these observations relied on the good definition of initial and final states, and the completely different systematic uncertainties from those that afflict scattering. Moreover, some sort of  ``peak"  or ``bump" can be seen with the naked eye in these processes around the nominal mass of these resonances, which seems to have made the acceptance of their existence more palatable. No doubt, these measurements were definitely very helpful for the general acceptance of the existence of these resonances. Nevertheless, these ``production" processes are also affected by the next problem, they rely strongly on the model used to extract a particular wave.

\subsection{The model-dependence problem}

The second feature that hindered the acceptance of the $\sigma$ and $\kappa$ for several decades was the extensive use of models for their determination and characterization. This is partly related to the previous problem with data, since for a long time, the lack of precise experimental results made acceptable many semi-quantitative or even just qualitative descriptions. Theoretical accuracy was not a demand. But it is now.

\begin{table}[!hbt] 
\label{tab:polessigma}
\caption{$\sig$ pole determinations using Roy-Steiner equations (three top rows), and the conservative dispersive average \cite{Pelaez:2015qba} which covers them, together with other  extractions using analytic techniques (three bottom rows)
using as input dispersively constrained input.}
\centering 
\begin{tabular}{l  c c} 
\hline
\hline
$\sig$ & \hspace{0.2cm} $\sqrt{s_{pole}}$\,\, (MeV) & $\vert g\vert$ (GeV)\\
\hline
Caprini, Colangelo, Leutwyler (2005) \rule[-0.2cm]{-0.1cm}{.55cm} \cite{Caprini:2005zr,Leutwyler:2008xd}& \hspace{0.1cm} $(441^{+16}_{-8})-i(272^{+9}_{-12.5})$ & $3.31^{+0.35}_{-0.15}$\\
Moussallam (2011) \rule[-0.2cm]{-0.1cm}{.55cm}  \cite{Moussallam:2011zg} & \hspace{0.1cm}  $(442^{+5}_{-8}) -i (274^{+6}_{-5})$ &-\\
Garc\'{\i}a-Mart\'{\i}n, {\it et al.} (2011)\rule[-0.2cm]{-0.1cm}{.55cm} \cite{GarciaMartin:2011jx}& \hspace{0.1cm} $(457^{+14}_{-13})-i(279^{+11}_{-7})$ & $3.59^{+0.11}_{-0.13}$\\
\hline
\rule[-0.2cm]{-0.15cm}{.45cm}
Pel\'aez (2015)  \rule[-0.2cm]{-0.1cm}{.55cm} \cite{Pelaez:2015qba}.  {\tiny Conservative dispersive average}& \hspace{0.1cm} $(449^{+22}_{-16})-i(275\pm15)$ \footnote{We have corrected a typo in \cite{Pelaez:2015qba} where the imaginary part reads $(275\pm12)$.} & $3.45^{+0.25}_{-0.29}$\\
\hline
\rule[-0.2cm]{-0.1cm}{.55cm} 
Caprini, {\it et al.} (2016) \cite{Caprini:2016uxy}& \hspace{0.1cm} $(457\pm 28)-i (292\pm 29)$ &-\\
Tripolt, {\it et al.} (2016) \rule[-0.2cm]{-0.1cm}{.55cm} \cite{Tripolt:2016cya}& \hspace{0.1cm} $(450^{+10}_{-11})-i (299^{+10}_{-11})$ &-\\
Dubnicka, S. {\it et al.} (2016) \rule[-0.2cm]{-0.1cm}{.55cm} \cite{Dubnicka:2016bhn}& \hspace{0.1cm} $(487\pm 31)-i (271\pm30)$ &-\\
\hline
\hline
\end{tabular} 
\hfill
\end{table}

\begin{table}[!hbt] 
\caption{ $\kap$ pole determinations using Roy-Steiner equations (two top rows), together with another extraction using analytic methods (bottom row) with dispersively constrained input.}
\centering 
\begin{tabular}{l  c c} 
\hline
\hline
$\kap$  & \hspace{0.2cm} $\sqrt{s_{pole}}$\,\, (MeV) & $\vert g\vert$ (GeV)\\
\hline
Descotes-Genon, Moussallam (2006) \rule[-0.2cm]{-0.1cm}{.55cm} \cite{DescotesGenon:2006uk}& \hspace{0.1cm} $(658\pm 13)-i (279\pm 12)$ &\\
Pel\'aez, Rodas (2020) \rule[-0.2cm]{-0.1cm}{.55cm}  \cite{Pelaez:2020gnd} & \hspace{0.1cm}  $(648\pm 7) -i (280\pm 16)$ & $3.81\pm0.09$\\
\hline
Pel\'aez, Rodas, Ruiz de Elvira (2016) \rule[-0.2cm]{-0.1cm}{.55cm} \cite{Pelaez:2016klv}& \hspace{0.1cm} $(670\pm 18)-i (295\pm 28)$ & $4.47\pm0.40$\\
\hline
\hline
\end{tabular} 
\label{tab:poleskappa}
\end{table}

Note that this is a different problem from the previous one, because {\it even using the same data} the determination of the pole associated to a resonance in the complex plane is a very delicate mathematical problem, and it becomes worse and worse as the resonance width is larger and the pole lies deeper in the proximal Riemann sheet of the complex plane. The same data fitted with naive models lacking the minimum fundamental requirements can yield a pole, {\it or not}, and the parameters of that pole can vary wildly. This was illustrated nicely in \cite{Caprini:2008fc} but as we will see is even more shocking for the $\kappa$ \cite{Pelaez:2020uiw,Pelaez:2020gnd}.
In particular, the incorrect use of Breit-Wigner shapes, often with some ad-hoc modifications that  violate the well-known analytic structure of partial waves, and sometimes even unitarity, was very frequent, both in scattering and production.

\begin{figure}
\begin{center}
\resizebox{0.5\columnwidth}{!}{%
  \includegraphics{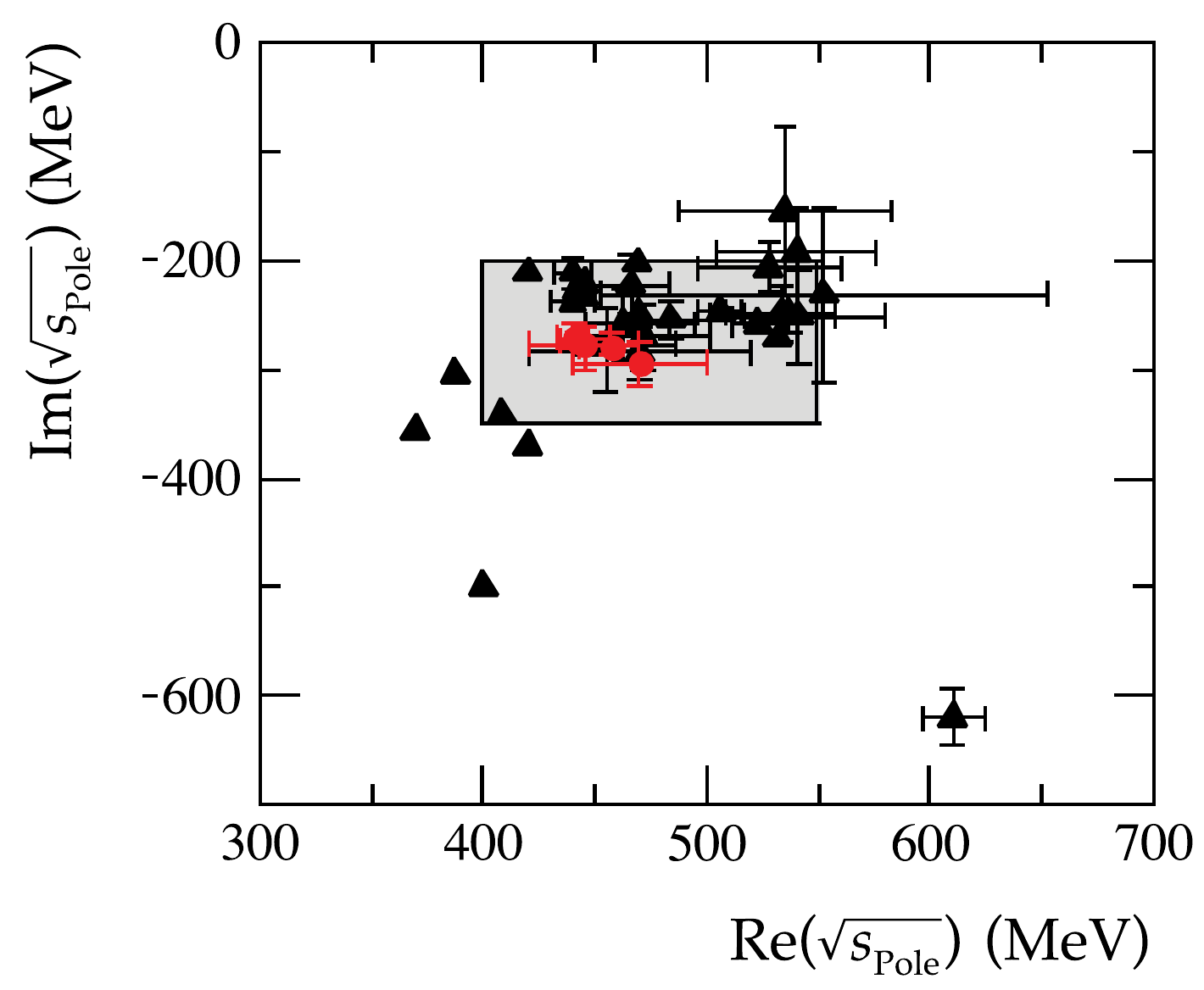} }
\caption{Status of the $\sig$ pole as presented in the 2020 edition of the RPP \cite{pdg}.
The shaded area is the RPP estimate for the pole mass, i.e., $M\equiv\re(\sqrt{s_{pole}})$, and the 
pole half width $\Gamma/2\equiv-(\im \sqrt{s_{pole}})$.
The red circles are considered the ``Most advanced dispersive analyses" of \cite{Ananthanarayan:2000ht,Caprini:2005zr,GarciaMartin:2011jx,Moussallam:2011zg}. For the rest of references see \cite{pdg}. Figure taken from the Note on ``Scalar mesons below 2 GeV" in \cite{pdg}. }
\label{fig:sigmapoles}       
\end{center}
\end{figure}

\begin{figure}
  \hfill
\begin{center}
\resizebox{0.5\columnwidth}{!}{%
  \includegraphics{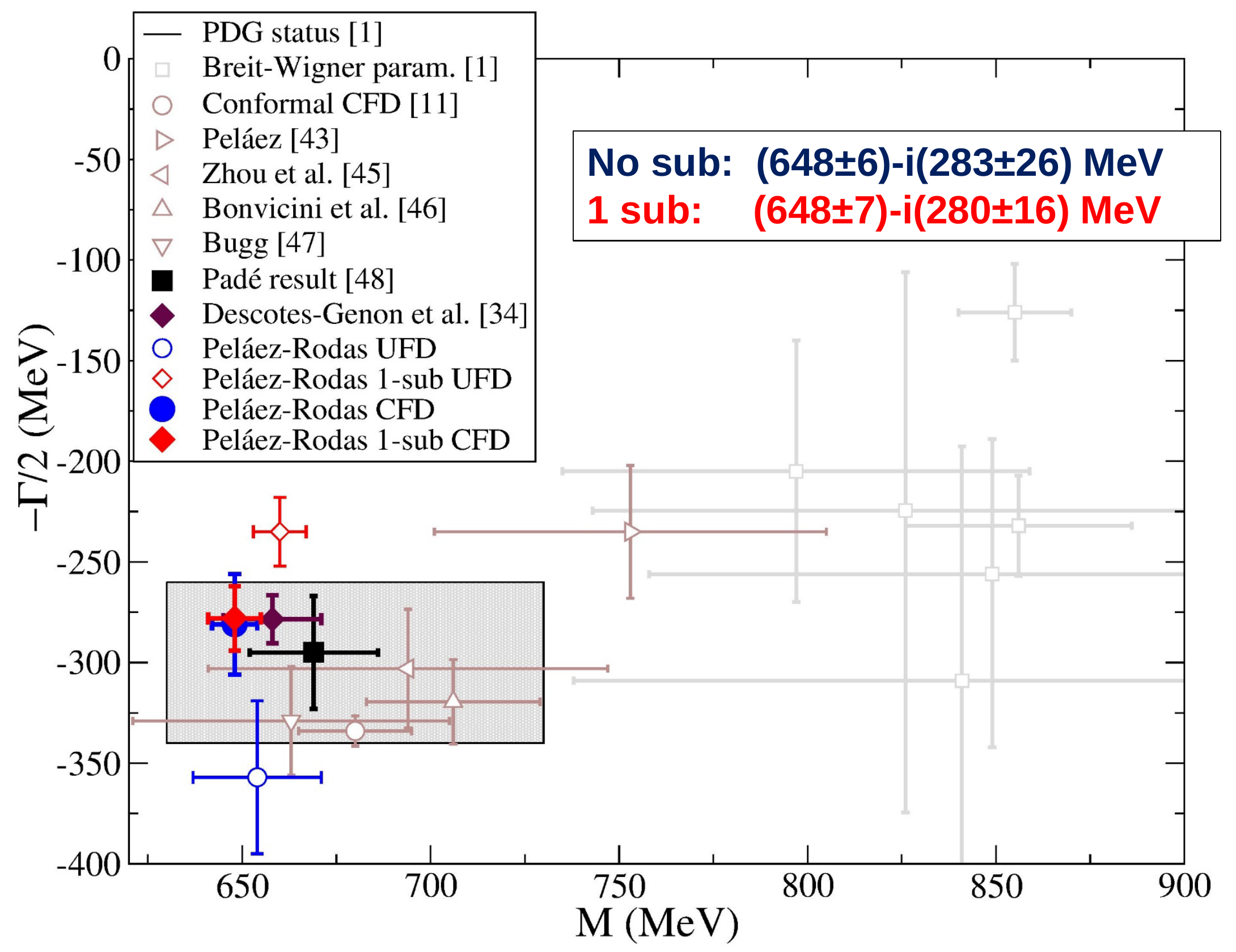} }
\caption{$K_0^*(700)$ pole positions. The RPP estimate (dark rectangle) and
 Breit-Wigner parameterizations are taken from \cite{pdg}. The rest are: Descotes-Genon et al. \cite{DescotesGenon:2006uk}, Bonvicini et al. \cite{Bonvicini:2008jw}, D.Bugg \cite{Bugg:2003kj}, J.R.Pel\'aez \cite{Pelaez:2004xp}, Zhou et al. \cite{Zhou:2006wm} and the ``Pad\'e Result'' \cite{Pelaez:2016klv}. The ``conformal CFD" is a simple analytic extrapolation of a conformal parameterization in \cite{Pelaez:2016tgi}.
 We also show results using Roy-Steiner dispersive equations, using as input
 the Pel\'aez-Rodas UFD or CFD parameterizations \cite{Pelaez:2020uiw,Pelaez:2020gnd}.
 Red and blue points use for the antisymmetric $\pik$ amplitude a once-subtracted or an unsubtracted dispersion relation, respectively. 
 This illustrates how unstable pole determinations are when using simple unconstrained fits to data (UFD).
 Only once Roy-Steiner Eqs. are imposed as a constraint (CFD), both pole determinations fall on top of each other. The final pole position is the main result of the dispersive analysis in \cite{Pelaez:2020uiw}, provided on the inset. Figure taken from \cite{Pelaez:2020uiw}.}
\label{fig:kappapoles}       
\end{center}
\end{figure}

In Figure \ref{fig:sigmapoles}, we show the present status at the RPP of the $\sigma$ poles. The wide spread of these poles is mainly due to the use of incorrect models and unreliable extrapolations to the complex plane. It should be noted that the RPP only keeps those models consistent with the lowest-energy $K_{\ell4}$ data \cite{Batley:2010zza} obtained in 2010. Before that date the spread was about a factor of 5 larger (see \cite{Pelaez:2015qba}). We also list in Table \ref{tab:polessigma}  the Roy-Steiner pole determinations, compared to other analytic extractions.
We also include the modulus of the coupling to meson-meson channel $\vert g\vert^2=-16\pi \vert Z\vert (2\ell+1)/(2q)^2$, where $\vert Z\vert $ is the residue of the associated pole in the partial wave $f_\ell(s)$ and $q$ is the CM momentum.
In Fig.~\ref{fig:kappapoles} (taken from \cite{Pelaez:2020uiw}) we show the present status of the $\kap$ pole, and as a dark rectangle the RPP estimate. The ``Breit-Wigner poles"
listed in the RPP are drawn explicitly as well, although the BW approximation is incorrect,
 to see the spread and disagreement of those poles with rigorous dispersive extractions (bold solid symbols), which are also listed in Table \ref{tab:poleskappa}.
 As an illustration of the present discussion, note the disagreement between the two ``UFD" determinations 
 which correspond to the same fit to data in the real axis, but continued to the complex plane with an unsubtracted or a once subtracted dispersion relation. This shows how unstable the analytic continuation is unless the data fit is consistent with all fundamental constraints and all contributions are calculated correctly. Actually, when the fit is constrained to do so, the resulting value is stable no matter what method of analytic continuation is used. These is the case of our very recent dispersive results  (both ``Pel\'aez-Rodas CFD" \cite{Pelaez:2020uiw}), whose values are given on the inset of Fig.\ref{fig:kappapoles}.


The model problem affects also our understanding of the classification of the $\sig$ and $\kap$ as well as their nature in terms of the underlying theory, QCD, and its degrees of freedom, quarks and gluons. Unfortunately, QCD and the confinement mechanism of quarks and gluons inside hadrons cannot be treated perturbatively below the 1.5-2 GeV region. Of course, we can resort to quark models. These take into account rather well the flavor symmetries but the hadronization has to be implemented by some ad-hoc mechanism. They have proved remarkable to describe semi-quantitatively ``ordinary" $\bar qq$ mesons, but are known to have difficulties to describe the QCD spontaneous chiral symmetry breaking and the light masses of the pseudo-NGB together with their decay constants. Moreover, the pure quark-model states are well-known not to describe easily the light scalars, particularly their huge widths, unless some unitarized  meson-meson interaction is taken into account \cite{vanBeveren:1986ea,Black:1998wt,Black:1998zc} and only then a reasonable description of meson-meson scattering is achieved. The essential role of unitarity for modern spectroscopy 
in the context of quark models has been illustrated in detail in the recent  review \cite{vanBeveren:2020eis}.

Alternatively, one can start from the rigorous effective field theory description of QCD at low-energies, i.e., ChPT, and use that information together with unitarity and analyticity to reduce the model dependence and obtain a connection with QCD through the common parameters between QCD and ChPT.
This is the idea of unitarized ChPT, which despite its name relies not only on unitarity and ChPT but also on the analitic properties of partial waves. We list in Table \ref{tab:polesuchpt}  some of the recent extractions using these techniques. Please note that some of these extractions may be more elaborated, or include dispersion relations to a higher degree than others. Finally, lattice QCD has been rather successful in generating hadron resonance masses, but widths remain a challenge. Nevertheless, meson-meson scattering is within the reach of realistic lattice calculations, in fair agreement with existing data. In these lattice calculations resonant shapes are easily recognized. However, for wide states or those overlapping with other analytic features, they suffer from the same problem as real experimental data, which is that so far they have to rely on models for the pole extraction. In this respect, the $\sig$ has been clearly identified on the lattice at various $m_\pi$ masses and both for $N_f=2+1$ \cite{Briceno:2016mjc,Briceno:2017qmb} and $N_f=2$ \cite{Guo:2018zss}. However the $\kap$ pole depends strongly on the model used to extract it from the lattice results \cite{Wilson:2019wfr,Rendon:2020rtw}. Once again, for a model-independent extraction from lattice data one would have to implement correctly the analytic features of meson-meson scattering and continue the results to the complex plane by means of dispersive techniques as those reviewed here.

\begin{table}[!hbt] 
\caption{Various $\sig$ and $\kap$ pole determinations using different approaches that can be considered unitarized ChPT. They are in fairly good agreement with 
those for the precise dispersive analyses in Table \ref{tab:polessigma},
although for the $\kap$ they tend to be somewhat more massive.
Let us emphasize that the uncertainties here should be interpreted with caution, since in most cases the error bars, when they exist,  do not include systematic uncertainties or model-dependent effects.  For instance, there are no estimations of higher order corrections and the left and circular cuts are absent or are calculated with NLO ChPT up to infinity or a large cutoff. Very often, the data fitted in these analyses may also not satisfy the dispersive representation.
\label{tab:polesuchpt}}
\centering 
\begin{tabular}{l  c c} 
\hline
\hline
$\sig$ & \hspace{0.2cm} $\sqrt{s_{pole}}$\,\, (MeV) & $\vert g\vert$ (GeV)\\
\hline
\rule[-0.2cm]{-0.1cm}{.55cm} 
Dobado, Pel\'aez (1996) \cite{Dobado:1996ps}& \hspace{0.1cm} $440-i 245$ &-\\
Oller, Oset (1997) \rule[-0.2cm]{-0.1cm}{.55cm} \cite{Oller:1997ti}& \hspace{0.1cm} $468.5 -i193.6$ &-\\
Oller, Oset, Pel\'aez (1998) \rule[-0.2cm]{-0.1cm}{.55cm} \cite{Oller:1998hw}& \hspace{0.1cm} $442 -i227$ &-\\
Oller, Oset (1998)\rule[-0.2cm]{-0.1cm}{.55cm} \cite{Oller:1998zr}& \hspace{0.1cm} $441 -i221$ & 4.26\\
Pel\'aez (2004) \rule[-0.2cm]{-0.1cm}{.55cm} \cite{Pelaez:2004xp}& \hspace{0.1cm} $(440\pm 8)-i (212\pm15)$ &-\\
Zhou {\it et al.} (2004) \rule[-0.2cm]{-0.1cm}{.55cm} \cite{Zhou:2004ms}& \hspace{0.1cm} $(470\pm 50)-i (285\pm 25)$ &-\\
Guo, Oller (2011)\rule[-0.2cm]{-0.1cm}{.55cm} \cite{Guo:2011pa}& \hspace{0.1cm} $(440\pm3)-i (258^{+2}_{-3})$ &$3.02\pm0.03$\\
Albaladejo, Oller (2012) \rule[-0.2cm]{-0.1cm}{.55cm} \cite{Albaladejo:2012te}& \hspace{0.1cm} $(440\pm 10)-i (238\pm 10)$ &-\\
Ledwig, {\it et al.} (2012)\rule[-0.2cm]{-0.1cm}{.55cm} \cite{Ledwig:2014cla}& \hspace{0.1cm} $(458\pm 2)-i (264\pm3)$ & $3.3\pm0.1$\\
Dai, {\it et al.} (2019)\rule[-0.2cm]{-0.1cm}{.55cm} \cite{Dai:2019zao}& \hspace{0.1cm} $(438\pm 52)-i (270\pm5)$ & $3.33\pm0.07$\\
Danilkin, Deineka, Vanderhaeghen (2020)\rule[-0.2cm]{-0.1cm}{.55cm} \cite{Danilkin:2020pak}& \hspace{0.1cm} $(457\pm 7)-i (249\pm5)$ &$3.17\pm0.04$\\
\hline
\hline
 $\kap$ & \hspace{0.2cm} $\sqrt{s_{pole}}$\,\, (MeV) & $\vert g\vert$ (GeV)\\
\hline
Oller, Oset, Pel\'aez (1998) \rule[-0.2cm]{-0.1cm}{.55cm} \cite{Oller:1998hw}& \hspace{0.1cm} $770 -i250$ &-\\
Oller, Oset (1998)\rule[-0.2cm]{-0.1cm}{.55cm} \cite{Oller:1998zr}& \hspace{0.1cm} $779 +i330$ &4.99\\
Pel\'aez (2004) \rule[-0.2cm]{-0.1cm}{.55cm} \cite{Pelaez:2004xp}& \hspace{0.1cm} $(754\pm 22)-i (230\pm27)$ &-\\
Guo, {\it et al.} (2005)\rule[-0.2cm]{-0.1cm}{.55cm} \cite{Guo:2005wp}& \hspace{0.1cm} $(757\pm 33)-i (279\pm 41)$ &-\\
Zhou, Zheng (2006)\rule[-0.2cm]{-0.1cm}{.55cm} \cite{Zhou:2006wm}& \hspace{0.1cm} $(694\pm 53)-i (303\pm 30)$ &-\\
Guo, Oller (2011) \rule[-0.2cm]{-0.1cm}{.55cm} \cite{Guo:2011pa}& \hspace{0.1cm} $(665\pm 9)-i (268^{+21}_{-6})$ &$4.2\pm0.2$\\
Ledwig, {\it et al.} (2012) \rule[-0.2cm]{-0.1cm}{.55cm} \cite{Ledwig:2014cla}& \hspace{0.1cm} $(684\pm 4)-i (260\pm4)$ &$4.2\pm0.1$\\
Danilkin, Deineka, Vanderhaeghen (2020)\rule[-0.2cm]{-0.1cm}{.55cm} \cite{Danilkin:2020pak}& \hspace{0.1cm} $(701\pm 12)-i (287\pm17)$ &$4.18\pm0.18$\\
\hline
\hline
\end{tabular} 
\end{table}

%% file: sections/dispre.tex
\subsection{Analytic structure of amplitudes}
\label{sec:anstruc}

In the right panel of Fig.~\ref{fig:mNmmN} we have depicted the
$M_1(p_1)M_2(p_2)\rightarrow M_1(p_3)M_2(p_4)$ scattering process, where $p_i$ are the meson four-momenta and in our case $M_1M_2=\pi\pi$ or $\pi K$. The corresponding amplitude is denoted as $F(s,t,u)$, where $s,t,u$ are the usual Mandelstam variables explained in the plot, with 
$ u=\sum_i m_i^2-s-t$. It is more convenient to use amplitudes with a given isospin, which we assume conserved, and since $u$ is redundant as a variable, to write the amplitudes as $F^I(s,t)$. 

We will be mostly interested in partial-waves of definite angular momentum and isospin since in that way we can identify the spin and quantum numbers of the resonances observed on each partial wave. 
These partial waves are obtained from the usual projection:
\begin{align*}
F^I(s,t)=16\pi N\sum_{\ell=0}^{\infty}(2\ell+1)P_\ell(z_s)f^I_\ell(s),\; f^I_\ell(s)=\frac{1}{32\pi N}\int_{-1}^{1} dz_s P_\ell(z_s) F^I(s, t(z_s)),
\end{align*}
where $P_\ell$ are the Legendre polynomials and $z_s$ is the scattering angle in the $s$ channel. Note that $N=1,2$ for $\pi K$ and $\pi\pi$, respectively, because, from the point of view of hadron interactions, pions are identical particles in the isospin limit that we are considering here. 

Let us also recall that it is more convenient to recast partial waves in terms of the phase shift $\delta_\ell^I$ and elasticity $\eta_\ell^I$ as follows\footnote{In the literature these are also written as $\delta_{I\ell},\delta^{(I)}_\ell$ and
 $\eta_{I\ell},\eta^{(I)}_\ell$ as we will see in several figures of this review.}:
\begin{equation}
f_\ell^I(s)=\frac{\eta_\ell^I(s) e^{i2\delta_\ell^I(s)}-1}{2i\sigma(s)},\quad \sigma(s)=\frac{2q(s)}{\sqrt{s}}
\label{eq:phasespace}
,\end{equation}
where $q$ is the CM momentum of the scattering particles. In the elastic regime $\eta_\ell^I=1$ and we can write:
\begin{equation}
f_\ell^I(s)=\frac{ e^{i\delta_\ell^I(s)}\sin \delta_\ell^I(s)}{\sigma(s)}.
\label{eq:felastic}
\end{equation}

\begin{figure}[ht]
\begin{center}
\resizebox{1.\columnwidth}{!}{%
  \includegraphics{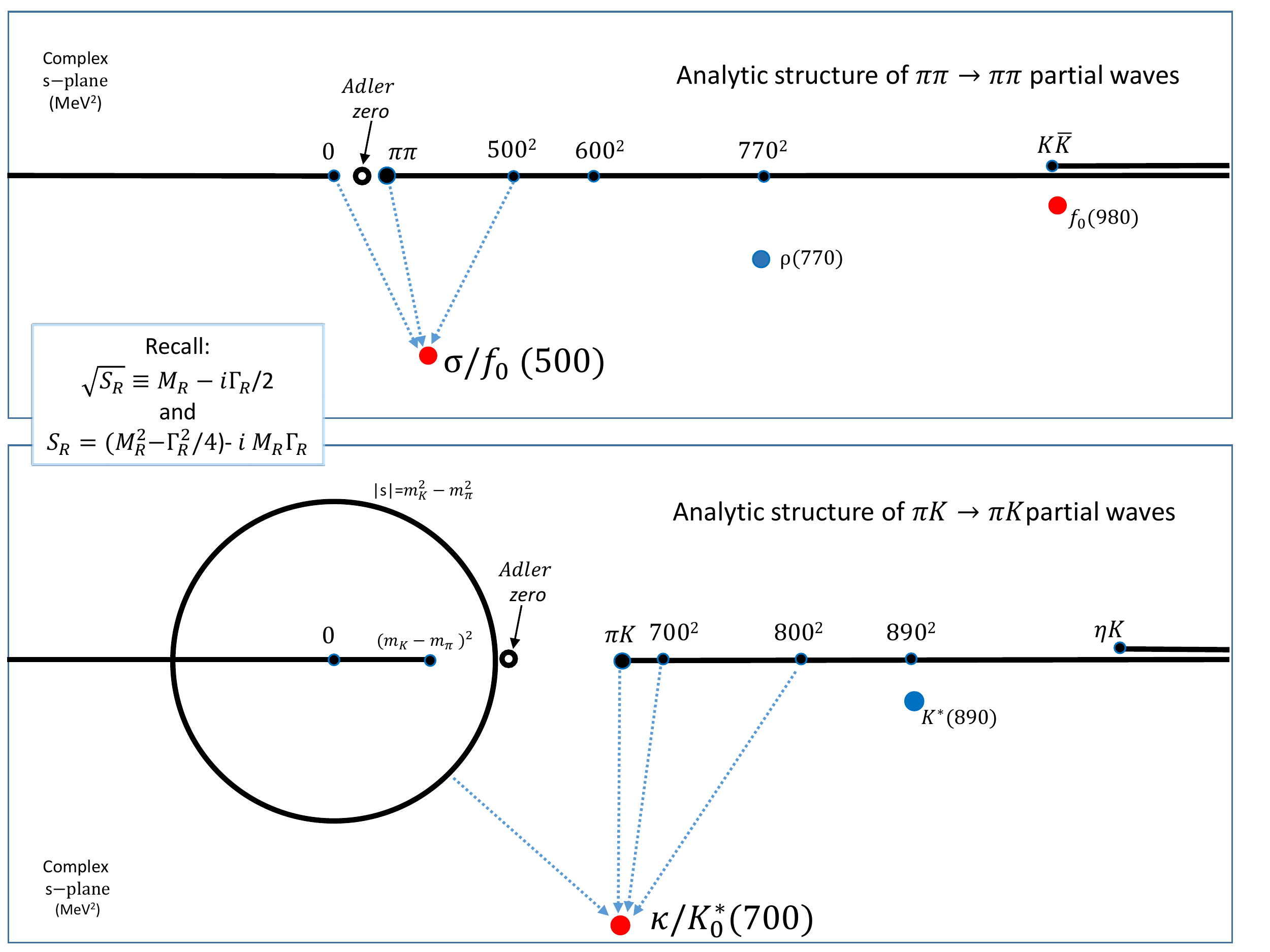} }
\caption{Analytic structure of meson-meson scattering partial waves in the complex-$s$ plane. They always have a right-hand cut, also called physical or unitarity cut, starting at the first two-body threshold and extending to $+\infty$, as well as a left cut extending from $s=0$ to $-\infty$. For the case of scattering of two mesons with different mass, there is also a circular cut and the left cut extends further from $s=0$ to the square of their mass difference. Top: For  $\pi\pi$ scattering  we illustrate the position of the known $\sig$, $\rho(770)$ and $f_0(980)$ resonance poles, in the second Riemann sheet. Note that the \sig pole is similarly close to the resonance nominal mass than to threshold, the left cut or the Adler zero required by chiral symmetry (which lies on the first Riemann sheet). Bottom: For $\pi K$ scattering we show the $K^*(892)$ and \kap poles in the second Riemann sheet. Note that the latter is roughly as close to the real axis around the resonance nominal mass than it is to other analytic features. The bottom figure is taken from  \cite{Pelaez:2020gnd}.}
\label{fig:anstruc}       
\end{center}
\end{figure}

In order to understand the relevance of analyticity for the determination of the $\sig$ and $\kap$ as poles in the complex plane, we first have to describe the analytic structure of meson-meson scattering partial waves (see \cite{Pelaez:2015qba,Pelaez:2020gnd} for a pedagogical detailed introduction and review). Thus, in the upper and lower panels of Fig.~\ref{fig:anstruc} we show the analytic structure of  $\pipi$ and $\pik$ scattering partial waves, respectively.  Note that we are showing the $s$-plane, which is the Lorentz invariant variable in which partial-waves are analytic in the whole complex plane, except for several cuts.

 Some features are common to the two cases \cite{Kennedy:1962}: 1) partial-waves satisfy the Schwartz reflection symmetry, $f(s^*)=f(s)^*$; 2) the right-hand cut on the real axis, also known as physical or unitarity cut, extending from threshold to $+\infty$, which can overlap with other cuts when additional states become available. The physical value of the partial wave is evaluated precisely over this cut as $f(s)=f(s+i\epsilon)$ with $\epsilon\rightarrow0^+$;  3) A left-hand cut coming from $-\infty$ and extending to the square of the mass difference between the incoming mesons. This cut is due to the presence of physical cuts in crossed channels, as seen from the $s$-channel. Finally, there is a circular cut centered at $s=0$,  whose radius is the difference of the squared masses of the incoming mesons (Note that it collapses to a point at the origin for $\pi\pi$ scattering). Poles in the real axis could appear below threshold, signaling the presence of bound states,  and causality guarantees (see \cite{smatrix}) that no other singularities exist in the so called first Riemman sheet of partial waves, where the CM momentum as a complex variable has a positive imaginary part.
 
 For the particular phenomenology of pion-pion and pion-kaon dynamics in the isospin limit and neglecting electromagnetic interactions, there are no bound  $\pi\pi$ or $\pi K$ scattering states (there are pionium bound states and their corresponding poles if electromagnetic interactions are included). In addition, in the massless limit, the spontaneous breaking of chiral symmetry requires the presence of so-called Adler zeros on the real axis below threshold. These are expected to appear in the massive case around zero, although slightly displaced by some amount of the order of the meson masses. This is actually the case in perturbative ChPT and in the dispersive analyses that we will review below.

However, when promoting the momentum to a complex variable, it can also have a negative imaginary part, and that defines a second Riemann sheet, accessible from the first by crossing continuously the unitarity cut. Actually, the number of sheets doubles every time a new accessible final state becomes available, which occurs when the $s$ variable reaches its center of mass-squared momentum.
In these ``unphysical sheets" other singularities can appear, like pairs of conjugates poles. In Fig.~\ref{fig:anstruc} we have thus signaled the position of the most relevant ones for our purposes here. Note that, out of their associated conjugate pair of poles, we are only showing the one in the lower half plane, which is the one that affects the most  the partial wave for real physical values.

\begin{figure}
\begin{center}
  \resizebox{\columnwidth}{!}{%
  \includegraphics{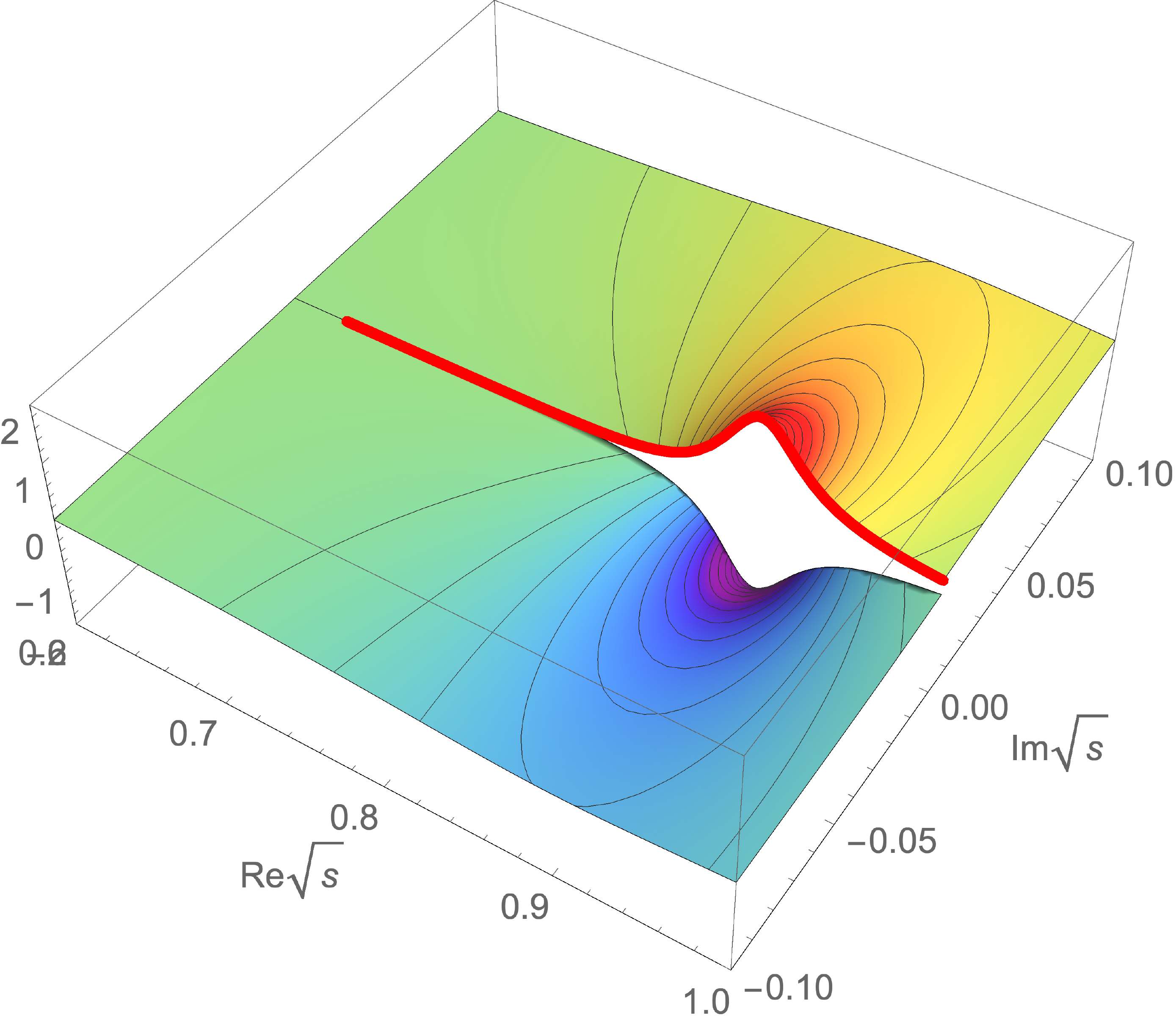} \includegraphics{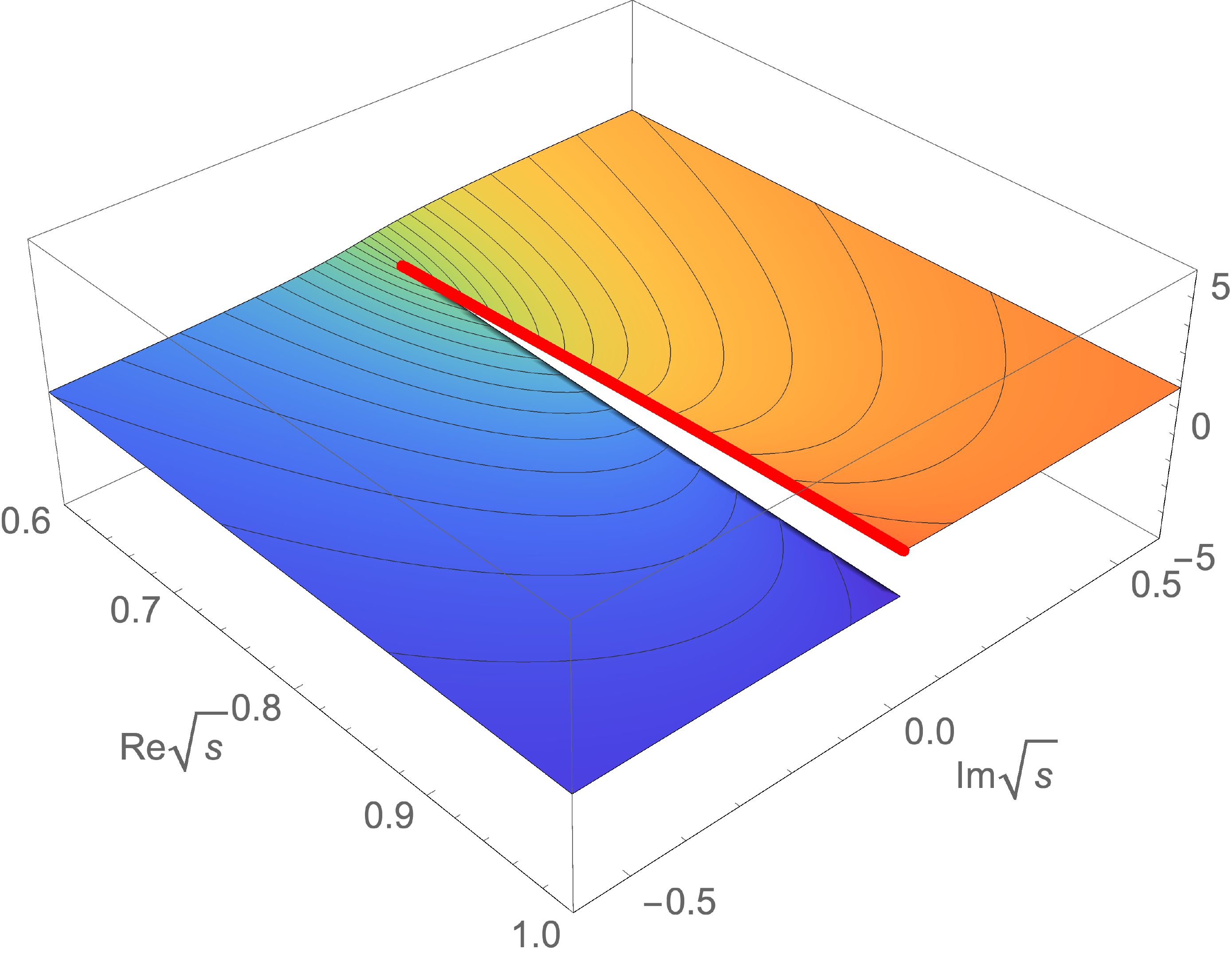} } \\
\resizebox{\columnwidth}{!}{%
  \includegraphics{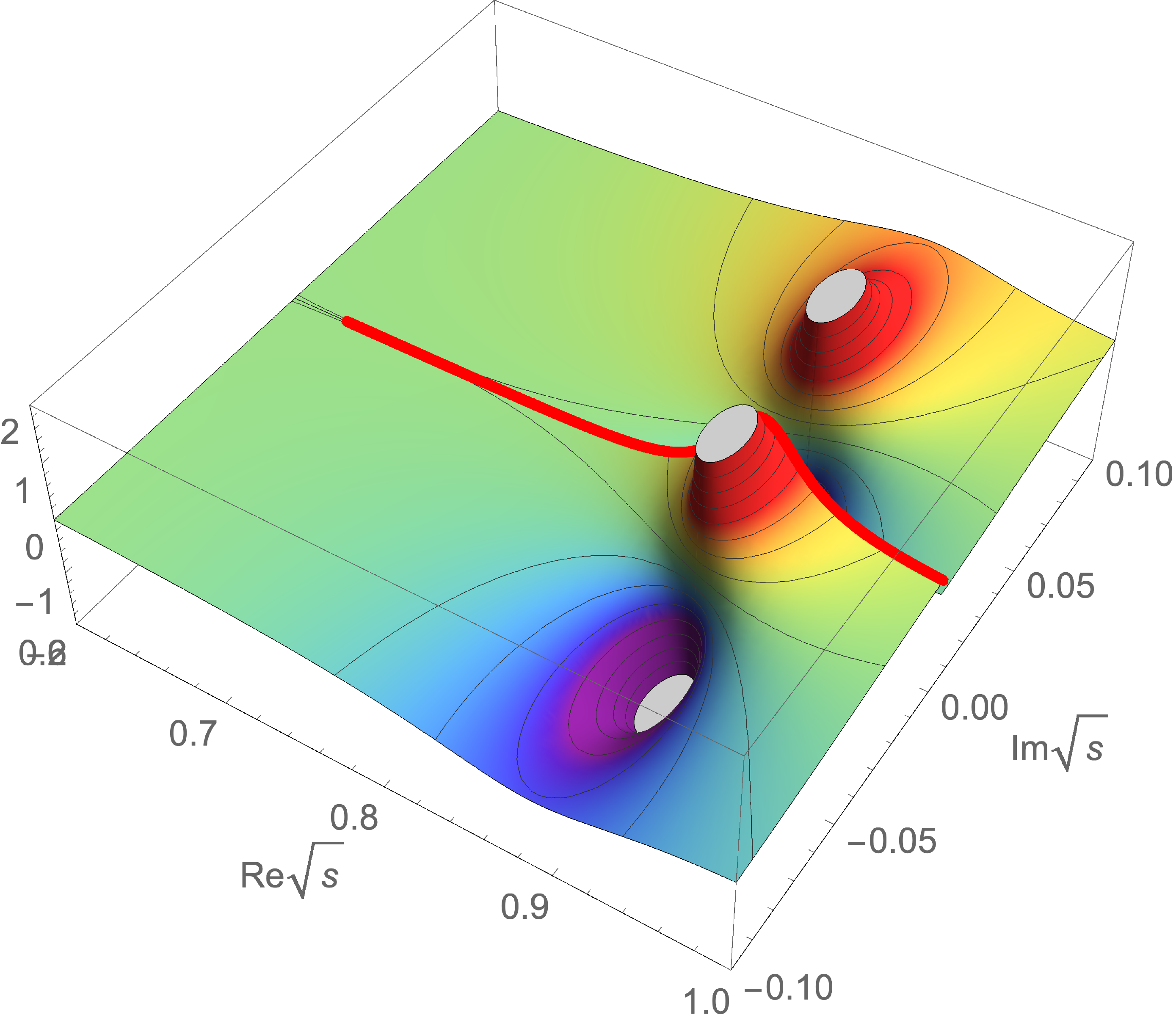} \includegraphics{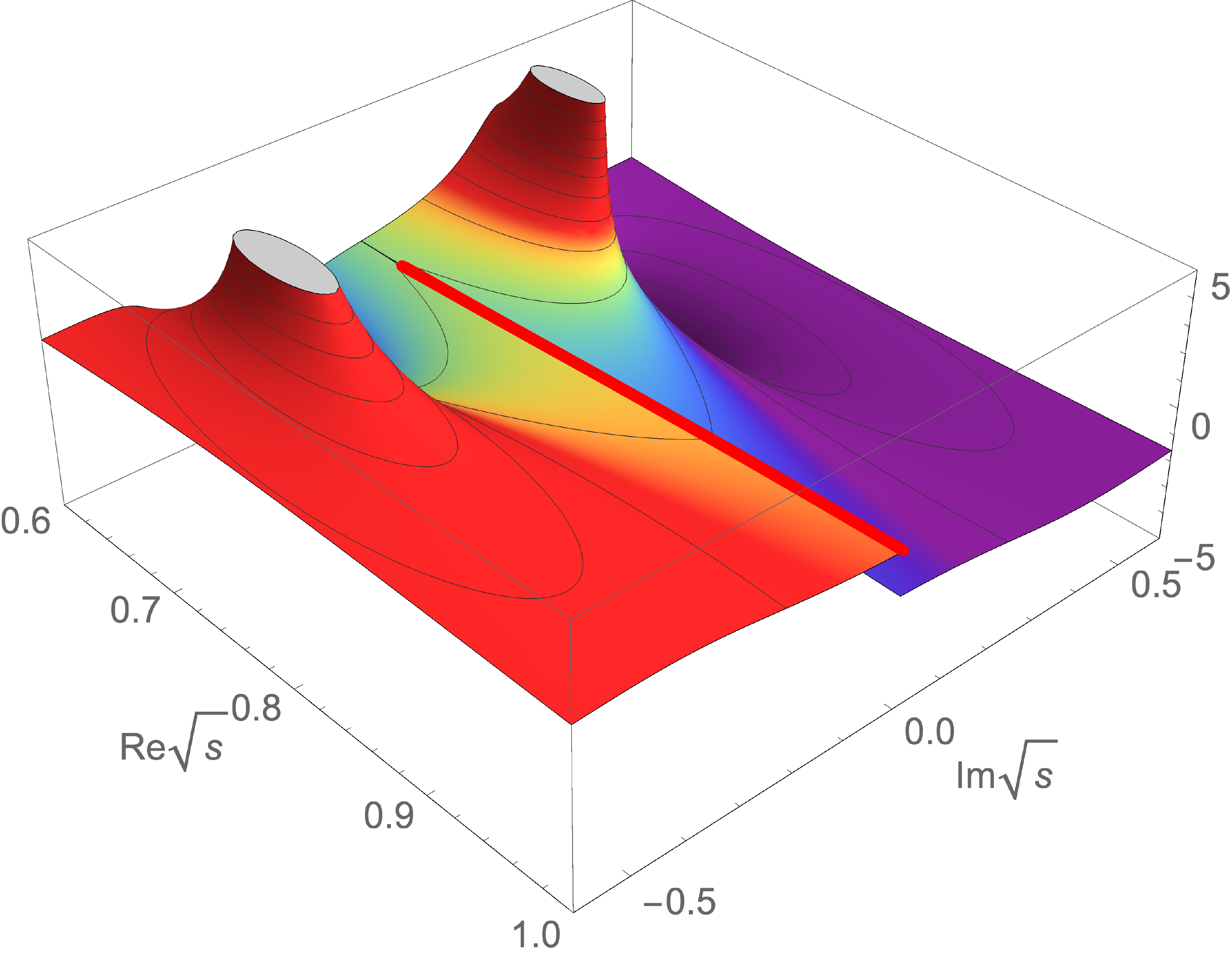} }
\caption{Imaginary parts of the isospin-1/2 $P$ (Left) and $S$ (Right) $\pik$ scattering partial waves from the UFD solution in \cite{Pelaez:2016tgi} in the complex plane (in GeV). Note the very different behaviors produced by the $K^*(892)$ and $\kap$ poles on the physical region, denoted by a thick red line, on the first (Top) and second Riemann sheets (Bottom).}
\label{fig:pwcomplex}       
\end{center}
\end{figure}

When these poles are very close to the real axis and isolated from other singular structures, they produce a very distinctive ``peak" shape
in the squared modulus of the partial wave and the cross section, as shown in the left panels of Fig.~\ref{fig:pwcomplex}. In those plots we are showing the imaginary part of the isospin-1/2 vector $\pi K$  partial wave in the first (top) and second (bottom) Riemann sheets.  In the second sheet we see for each pole a huge peak rise and an adjacent huge sink, which are the two-dimensional analog of the $1/x$ behavior that goes to $+\infty$ as $x\rightarrow 0^+$, but tends to $-\infty$ as $x\rightarrow0^-$.
As we will see below, unitarity tells us that for elastic waves the imaginary part is proportional to the square of the modulus and, therefore, the cross section. So the peak seen in the real axis translates into a peak in the partial-wave cross section. Simultaneously, they produce a huge rise of almost $180^\degree$ in the phase-shift around the mass value.
This is the case of typical resonances like the $\rho(770)$, whose phase-shift sharp rise in the real axis is shown in the right panel of Fig.~\ref{fig:dataraw}, reaching $90^\degree$ at $s=M_R^2$. The $K^*(892)$ has a similar shape. In such cases the familiar Breit-Wigner description is a fairly good approximation. 

In addition, in Fig.~\ref{fig:dataraw}, we can see that the $f_0(980)$ also presents a sharp increase on the isoscalar-scalar phase (left panel). In this case its associated pole is close to the real axis, but not isolated from other analytic structures. It is actually so close to the opening of the $K\bar K$ threshold that a naive Breit-Wigner shape is not valid. Moreover, it produces a dip in the cross section, not a peak, because the phase shift around 980 MeV is not $90^\degree$ but relatively close to $180^\degree$, due to the interference of the $\sig$ meson, which provides a rather large ``background" phase as a starting point for the sharp rise produce by the $f_0(980)$ pole.

The situation is also complicated for the $\sig$ and $\kap$ poles, which, as seen in Figs.~\ref{fig:anstruc} and \ref{fig:pwcomplex},
lie deep in the second Riemann sheet of the complex plane. Moreover, in Figs.~\ref{fig:anstruc} we can see that they lie as close to the real axis in their nominal mass region as they do to other analytic features like the threshold or the left and circular cuts. An analytic continuation deep in the complex plane is a delicate extrapolation from the real axis, but in these cases, it is even hindered by the presence of these other analytic structures, for which we need to have some reasonable control if we are to claim a precise determination of these pole parameters. Even the Adler zero, despite lying in the first Riemann sheet, is rather close to the pole and its absence can deform the whole amplitude.

As an example, we have recently shown that an inconsistent treatment of the unphysical cuts can move the pole-width  of the $\kap$ by 200 MeV, {\it even if using the same fit to data in the \kap region}. These are the two ``UFD" results in Fig.~\ref{fig:kappapoles}.  Moreover, being so deep in the complex plane, and not so close to their nominal mass, neither the $\sig$ nor the $\kap$ poles, shown in the right panels of Fig.~\ref{fig:pwcomplex}, produce the naively expected peak. As already commented, there is also no sharp rise of the phase shift seen in the left and central panels of Fig.~\ref{fig:dataraw}. Therefore, to have a precise determination of the pole we need both reliable data and a sound analytic formalism. Both goals can be achieved with dispersion relations.

\subsection{Dispersion relations}
\label{sec:DR}

\begin{figure}
\begin{center}
\resizebox{.5\columnwidth}{!}{%
  \includegraphics{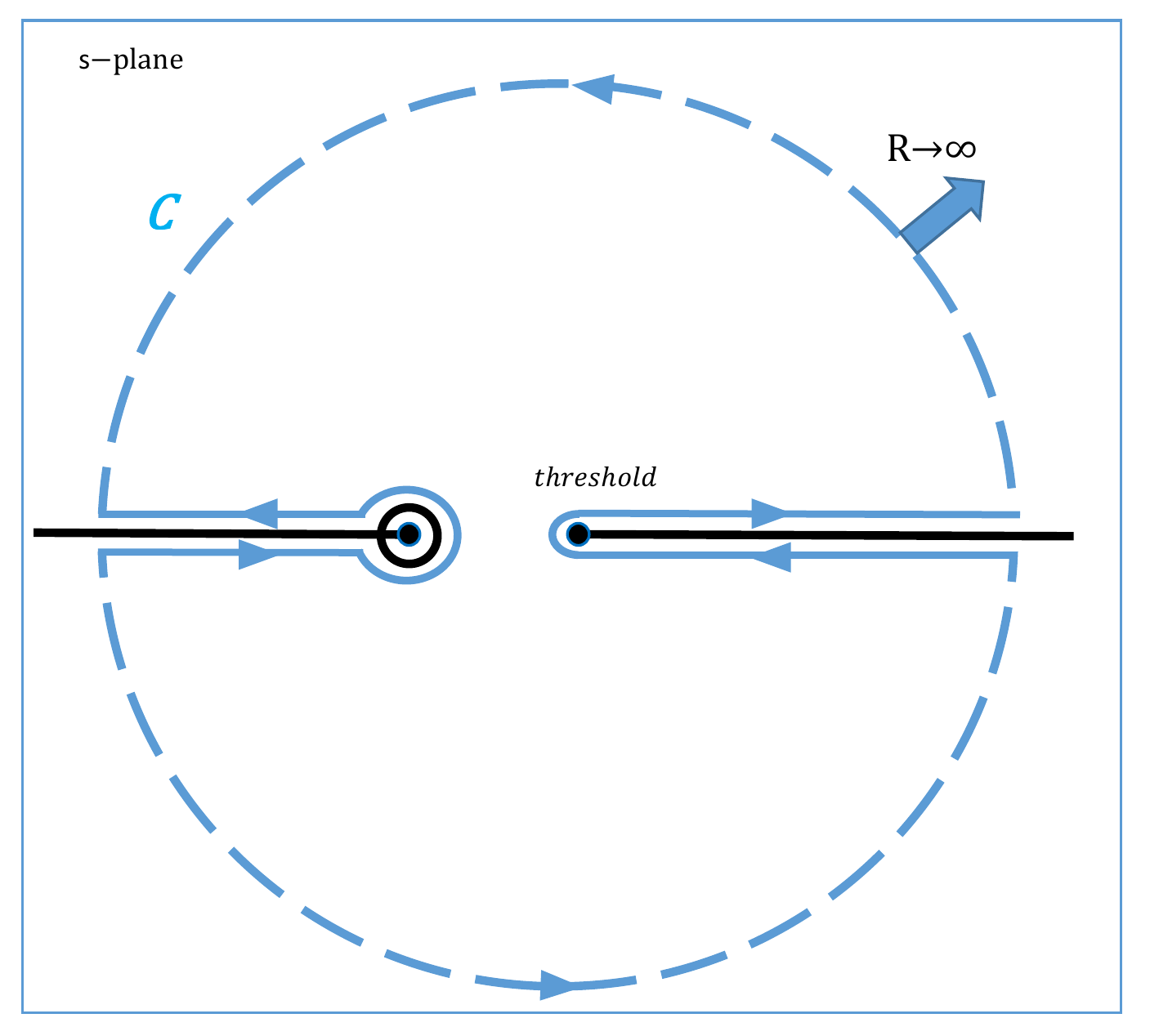} }
\caption{Integration contour $C$ (blue lines), surrounding the cuts 
  (black lines), used to obtain dispersion relations. The circular part of the contour (dashed curve) is taken to infinity. For all dispersion relations discussed in the text, the right cut starts at the corresponding threshold, whereas the left cut starts at different points: $-t$ for fixed-$t$ dispersion relations, $t=0$ for $\pi\pi$ partial waves and $(m_K-m_\pi)^2$ for $\pi K$ partial waves. The circular cut only exists for non-equal mass scattering partial waves.}
\label{fig:Cauchy}       
\end{center}
\end{figure}

These are nothing but the use of Cauchy's Integral Formula, which yields the value of an analytic function inside a given contour $C$ as an integral of the function over the contour, namely:
$$ f(z)=\frac{1}{2\pi i}\oint_C dz'\frac{f(z')}{z'-z}.$$
For amplitudes like the ones in $\pik$ we choose the contour as in Fig.~\ref{fig:Cauchy}. 
Of course, we need an amplitude of just one variable, so that, in practice, there are two kinds of dispersion relations: First, those for the full amplitude $F(s,t,u)$, fixing the $t$ variable (other variables could be fixed, but they are less useful), called fixed-$t$ dispersion relations (FTDR), 
or fixing them with another relation, like the hyperbolae $(s-a)(u-a)=b$, which lead to Hyperbolic Dispersion Relations (HDR). Second, partial-wave dispersion relations (PWDR), which in turn, can be obtained by integrating a family of fixed-$t$ dispersion relations (PWFTDR) \cite{Roy:1971tc} or a family of hyperbolic dispersion relations (PWHDR) \cite{Hite:1973pm,Johannesson:1976qp}, or other more complicated dispersion relations \cite{Mahoux:1974ej,Auberson:1974ai,Auberson:1974xu,Auberson:1975yx,Roy:1978tu}.

If the amplitude tends to zero sufficiently fast as the circular part of the contour is sent to infinity, the dispersion relation reduces to the integrals over the parts of the contour surrounding the cuts, infinitesimally separated from them.  
Let us recall now that both the amplitude and the partial waves satisfy the Schwartz reflection symmetry and thus their values in the upper and lower half-planes are conjugate to one another. Note also that there is one straight section of the contour infinitesimally above and another one infinitesimally below each one of the cuts, although they run in opposite senses. Thus, when integrating these pairs of straight sections together we are integrating the difference between  
the amplitude and its conjugate over each cut.
Therefore the straight sections of the contour become
a single integral for each cut for twice the imaginary part of the amplitude.

Finally, if the integral over the curved part of the contour does not converge sufficiently fast, one then applies Cauchy theory to the function $(f(s)-f(s_0))/(s-s_0)$, which obviously decreases faster.
This gives rise to a dispersion relation ``subtracted" at the $s_0$ point, which only determines $f(s)$ up to the $f(s_0)$ ``subtraction constant". If the function still does not decrease sufficiently fast at infinity, one can make one further subtraction at another point, which would require one more subtraction constant, etc... If one makes several subtractions at the same point, these constants correspond to the derivatives of the function at the subtraction point. Nevertheless, the existence of a Froissart bound for amplitudes\cite{Froissart:1961ux} guarantees that with two subtractions all dispersion relations will converge. Note that the more subtractions the less is weighted the high-energy region, which is usually less well known. Thus, one can make as many subtractions as deemed convenient for the problem.

In such case the singularity in the denominator has to be resolved using the Principal Value relation $1/(s'-s-i\epsilon)=PV[1/(s'-s-i\epsilon)]+i\pi\delta(s'-s)$. As a consequence, in the real axis above threshold dispersion relations do not provide the full amplitude, but just its real part. Of course, this means the full amplitude is also determined for physical values of $s$ because there the imaginary part is input.

Just for illustration and to fix ideas we show how a once-subtracted fixed-$t$ dispersion relation looks for $\pi\pi$ scattering:
$$F(s,t)=F(s_0,t)+\frac{s-s_0}{\pi}\int_{th}^\infty ds'\frac{\im F(s',t)}{(s'-s)(s'-s_0)}
+\frac{s-s_0}{\pi}\int_{-\infty}^{-t} ds'\frac{\im F(s',t)}{(s'-s)(s'-s_0)}.$$
Similarly, a partial-wave dispersion relations subtracted three times at $s_0=0$ would read, generically and suppressing the $I,\ell$ indices for brevity, as follows:
\begin{equation}
f(s)=f(0)+s f'(0)+\frac{s^2}{2}f''(0)+\frac{s^3}{\pi}\int_{RC}ds'\frac{\im f(s')}{s'^3(s'-s)}+LC(f)+CC(f),
\end{equation}
where, since the integrands look the same as for the right cut (RC),  we have abbreviated as $LC$ the integral over the left-hand cut  and $CC$ that over the circular cut. Remember the latter is only present for scattering of two unequal mass mesons. Recall also that, if the ``output" variable $s$ takes physical values on the real axis above threshold, the left hand sides are not the amplitudes but just their real parts and the right-hand cut integrals become just their principal value.

The above dispersion relations, and their different variants in terms on how we are left with just one variable (fixed-$t$, hyperbolic or partial waves), as well as in the number of subtractions, have to be satisfied by the scattering amplitudes. As a consequence, dispersion relations can address the two generic problems in the determination of the $\sig$ and $\kap$ that we described in the section~\ref{sec:problems}:

\begin{itemize}
\item {\it The data problem:} First, dispersion relations can be used to check the consistency of different data points or even full sets and possibly discard them. Second, they can also be used to obtain constrained fits to data. This will provide a consistent and model independent description of the data. Third, being integral representations, they tend to decrease the uncertainties and can yield more precise values at a given point than the direct measurement. This is particularly relevant right at threshold, where threshold parameters can be recast in terms of integrals, called sum-rules, thus avoiding dangerous extrapolations from regions where data exist.

\item{\it The model-dependence problem:} The analytic properties previously described rely only on fundamental principles like causality (which forbids the appearance of singularities for the full amplitude outside the real axis), Lorentz invariance and crossing symmetry. The only assumption is isospin conservation, and that pions and kaons are stable, which are both fairly good approximations within the realm of Hadron Physics. 
When solving or using these equations, it is assumed that amplitudes are reasonably smooth and that there are no wiggles or fast oscillations hidden between two data points, which, in a sense, is always implicit in Physics when describing data. Thus, for all means and purposes, dispersion relations are a  model-independent tool to study Hadron Physics. Moreover, they provide an analytic continuation to the complex plane in the first Riemann sheet that is unique.

\end{itemize}

Of course, for the $\sig$ and $\kap$ poles, we are also interested in the second Riemann sheet.
For this we have to know how to cross continuously through the physical cut. However, this is particularly simple for partial waves in the elastic case, since then the $S$ matrix in the second sheet is the inverse of the $S$-matrix on the first. Namely: 
\begin{equation}
S_\ell=1+2i\sigma(s)f_\ell,\quad  S_\ell^{(II)}=\frac{1}{S_\ell^{(I)}},\quad f^{(II)}(s)=\frac{f^{(I)}}{1+2i\sigma(s)f^{(I)}},
\label{eq:second}
\end{equation}
where $\sigma(s)$ was defined in Eq.~\eqref{eq:phasespace} and we have suppressed the isospin indices for convenience. Note, however, that in the second sheet we have to define $\sigma(s^*)=-\sigma(s)^*$. Since inelasticity has not been observed up to the $K\bar K$ threshold for $\pipi$ scattering or the $K \eta$ threshold for $\pik$ scattering in the $\sigma$ and $\kappa$ channels, respectively, and, as seen in Fig.~\ref{fig:anstruc}, their poles are very far from those first relevant inelastic thresholds, this elastic approximation to reach the second Riemann sheet is more than enough for our purposes.

Generically, the right or ``physical" contributions are simpler to calculate. In particular, it is usually possible to compare to data in the low-energy region, where the scattering is elastic, or the inelasticity is dominated by just a few channels, but not too close to threshold where data is scarce since the interaction becomes weaker. In contrast, the most difficult parts are the unphysical cuts, the very high energy region and the subtraction constants, which are taken at very low $s$, quite frequently in the subthreshold region.

The techniques used to deal with these kind of relations depend on the aim of the work.
For simplicity we have classified them in two classes, labeling them with a crude description:
 ``Precision dispersive analyses" and  ``Unitarized Chiral Perturbation Theory".
 Of course, this division in two types of approaches is not clear-cut and it is indeed possible to find works in the literature that, fitting better in one category, share features from the other approach.  
 This is the case of uses of Roy equations with input from ChPT \cite{Colangelo:2001df,Caprini:2005zr} or recent dispersive analysis with input from Roy-like equations where other part of the input is evaluated or constrained with ChPT \cite{Danilkin:2020pak}.
 Another recent example of interest, although applied to form factors instead of scattering, can be found in \cite{Shi:2020rkz}, where one of the most popular methods of unitarization (the IAM, to be explained below) is complemented with a full dispersive treatment.
The classification is nevertheless pedagogically useful for this introduction and we will next review the main features of each technique.

%% file: sections/dispanalysis.tex
The goal here is to evaluate as precise and rigorously as possible all contributions to the partial waves avoiding, as much as conceivable, model-dependence assumptions. These approaches yield the most robust values for 
$\pi\pi$ and $\pi K$ threshold and sub-threshold parameters and for the poles of resonances that appear within their applicability region. They also provide strong and rigorous constraints on data from threshold up to energies between 0.9 and 1.6 GeV, depending on the type of dispersion relation.

Since one wants to calculate precisely the unphysical-cut contributions, which come from crossed channels, crossing symmetry is a very important tool used to rewrite them in terms of amplitudes in their physical region. The $\pi\pi$ scattering case is particularly simple since all its crossed channels are $\pi\pi$ scattering again. However, the $t$ channel of $\pi K$ scattering is $\pi\pi\rightarrow K\bar K$ and, as a consequence, these two processes are coupled together in several dispersion relations when using  crossing.

The next issue of concern is that dispersive integrals extend to infinity. Of course, the high-energy contribution can be suppressed by increasing the number of subtractions. Nevertheless, we need data for these processes at high energies and, unfortunately, there are not many. Partial-wave data in terms of phase and elasticity exist for $\pi\pi$ scattering up to 1.8 GeV and for some $\pi K$  waves up to 2.4 GeV. However, it is known that higher partial waves become as large as lower ones at around those energies. Beyond that region, only data on total cross sections for $\pi\pi$ scattering are available, with huge uncertainties for the lowest energies. 
The high-energy behavior is thus obtained from Regge Theory, using the dominant Pomeron ($P$) and first Reggeon exchanges ($f_2$ or $P'$, $\rho$ and $K^*$) and assuming that the vertices coupling them to pions and kaons can be obtained from the factorization of meson-nucleon and nucleon-nucleon processes, for which there are abundant data. Factorization also provides the $t$ dependence of these processes. There are somewhat different determinations in the literature  \cite{Ananthanarayan:2000ht,Pelaez:2003ky,Buettiker:2003pp,GarciaMartin:2011cn,Caprini:2011ky,Pelaez:2016tgi,Pelaez:2018qny}, now agreeing within uncertainties for the total cross sections (forward direction) and the most relevant $t$ dependence, but with some differences, which are relatively minor once inside the dispersive integrals of interest (see \cite{Pelaez:2015qba} for a review on the $\pi\pi$ case). Taking into account that Regge Theory is a dual ``averaged" description~\cite{RuizdeElvira:2010cs}, it is well suited to be used inside integrals, but this implies that the output of the dispersion relations is only reliable locally at lower energies. For $\pi\pi$ and $\pi K$ scattering, in practice, this provides a first limitation of the dispersive approach below 2 GeV or less, depending on the process.

Note that, by crossing the unphysical cuts contributions to their respective energy regions, all partial waves of different angular momentum corresponding to the crossed contributions are coupled to the $s$-channel partial wave equations. This happens either if we use PWDR or if we use dispersion relations for the full amplitude, which at low energies is reconstructed from the partial-wave series.
These works~\cite{Caprini:2005zr,Ananthanarayan:2000ht,Pelaez:2003ky,Buettiker:2003pp,GarciaMartin:2011cn,Caprini:2011ky,Pelaez:2016tgi,Pelaez:2018qny} have been crucial to settle, once and for all, the controversy about the existence of the $\sig$ and $\kap$. 

At this point, let us remark that dispersion relations have been used in the literature in different ways. On the one hand, it is possible to solve  the relations for $S$ and $P$ waves, typically up to an energy around 1 GeV, assuming the input in all other regions and waves is known and fixed. This Roy-like approach has been followed for the $\sig$ and $\kap$ resonances in~\cite{Ananthanarayan:2000ht,Colangelo:2000jc,Caprini:2005zr,Moussallam:2011zg,DescotesGenon:2006uk}. In such case the resulting uncertainty comes from the input, which is kept fixed to phenomenological fits to data. Sometimes this is supplemented with theoretical information from ChPT, which helps decreasing the uncertainty. The relations are solved numerically, i.e., imposing that the dispersion relation is satisfied with a tiny relative numerical error of say $10^{-4}$, in the isospin limit and neglecting multi-pion states below 1 GeV. In a sense, this approach predicts the partial-wave in the region below 1 GeV for the $S$ and $P$ waves.
Generically, but not always, the results are found to describe well the data in those regions, even if they were not used as input.
On the other hand, dispersion relations can be used as constraints on fits to data. This approach is usually called a ``data-driven dispersive analysis" and for the $\sig$ and $\kap$ has been followed for instance in 
\cite{Pelaez:2004vs,Kaminski:2006qe,Kaminski:2006yv,GarciaMartin:2011cn,GarciaMartin:2011jx,Pelaez:2015qba,Pelaez:2016tgi,Pelaez:2017ppx,Pelaez:2018qny,Pelaez:2020gnd,Pelaez:2020uiw}.
This is done by introducing in the fit a penalty function that measures the distance squared between the input and the output for each dispersion relation, which is minimized together with the $\chi^2$ of the fit. This distance is calculated from the difference between input and output divided by the relative uncertainty at different energy values and averaged over the whole energy region. It can be used as a check of how well a dispersion relation is satisfied, or as a penalty function for a constrained fit. The source of uncertainty is the data, but now in all the energy regions. There is an uncertainty on how to weight the penalty function, but different possibilities are usually considered and the associated error is well below the uncertainty from the data. Note also that this procedure allows one for changes in the fits to data in regions that were considered fixed input in the previous approach. These two usages are therefore complementary and, as we will see, turn out to be remarkably consistent.

The most used dispersion relations for this purpose are the following:

\begin{itemize}

\item [$\bullet$]
{\bf Forward Dispersion Relations (FDR).} Likely the simplest ones, usually written in just one line. These are a particular case of fixed-$t$ dispersion relations for the whole amplitude $F(s,t,u)$, where $s\leftrightarrow u$ crossing is used to rewrite the left cut in terms of the same process. Since no partial-wave projection is needed for their derivation, one does not have to worry about the convergence of the partial-wave expansion. In principle, they are applicable up to arbitrary energies. In practice, only up to the energy where one starts using Regge Theory: 1.4 GeV for $\pi\pi$ \cite{Pelaez:2004vs,Kaminski:2006yv,Kaminski:2006qe,GarciaMartin:2011cn} and 1.8 GeV for $\pi K$ \cite{Pelaez:2016tgi}. In both cases, if their isospin-combination basis is well chosen, one of them can be calculated unsubtracted and the others with just one subtraction.

It has been shown that apparently good-looking fits to data fail to satisfy these equations \cite{Pelaez:2004vs,Pelaez:2016tgi} within uncertainties. Actually some $S$-wave data sets are so inconsistent with FDRs that they can be safely discarded \cite{Pelaez:2004vs}. It is fairly simple to impose FDRs as constraints to obtain constrained fits to data up to their maximum applicability region.

\begin{itemize}

\item {\it \underline{FDR Pros:}} Very simple. The only ones applied beyond 1.1 GeV for $\pi\pi$ scattering, or 1 GeV for $\pi K$. They yield several relevant sum rules for threshold parameters.

\item {\it \underline{FDR Cons:}} Their second Riemann sheet and the relevant poles there are not accessible using just dispersive techniques. But using their aforementioned constrained fits with other analytic techniques (to construct, for instance, sequences of Pad\'e approximants \cite{Masjuan:2013jha}) they determine the poles of all resonances up to their applicability energy \cite{Pelaez:2016klv,Masjuan:2014psa,Caprini:2016uxy}, with a very reduced model dependence.
\end{itemize}

\vspace{.3cm}
\item[$\bullet$]
{\bf Partial-wave projected fixed-$t$ dispersion relations (PWFTDR)}.
A family of fixed-$t$ amplitude dispersion relations is integrated to write the corresponding dispersion relations for partial-waves. The unphysical cuts are rewritten in terms of the physical region using $s\leftrightarrow u$ crossing. When projecting the output into partial waves with respect to the angle of the given channel, one finds in the input the tower of partial waves of the crossed channel with respect to its own angle. Thus, in principle, this approach leads to a dispersion relation for each partial wave coupled to the infinitely many other partial waves. In practice, since higher angular-momentum waves are suppressed at low energies, only the dispersion relations for the lowest partial waves are implemented and a few higher partial waves are  considered only as further input. At higher energies one uses Regge theory.

For $\pi\pi$ scattering, these equations are generically known as ``Roy equations", since their twice-subtracted version was first derived by Roy \cite{Roy:1990hw}, who also used $s\rightarrow t$ crossing to rewrite the subtraction constants for all waves, and each value of $t$, in terms of just the two $S$-wave scattering lengths. There is an extensive literature on Roy equations, extending from the early phenomenological applications in the 70's \cite{Basdevant:1972uu,Basdevant:1972uv,Basdevant:1973ru,Pennington:1973hs,Pennington:1973xv}, which later faded away with the advent of QCD, to the renaissance in the 2000's \cite{Ananthanarayan:2000ht,Colangelo:2001df,Caprini:2005zr,Moussallam:2011zg,GarciaMartin:2011cn,GarciaMartin:2011jx,Pelaez:2015qba} (and references therein).
One of the most relevant $\pi\pi$ scattering Roy analyses for our present knowledge of the $\sig$ come from a {\it solution} of $S$ and $P$-waves below 800 MeV in \cite{Ananthanarayan:2000ht}, later refined in \cite{Colangelo:2001df} with ChPT constraints at low energy. Note these poles were not extracted using the dispersion relation, but from a parameterization of the solution. Nevertheless,  these solutions were later calculated in the complex plane and, using Eq.~\eqref{eq:second}, a dispersive, robust and precise pole was found in \cite{Caprini:2005zr}. It should be emphasized that these are {\it solutions} of Roy equations where no data is used as input in the $S$ and $P$ waves below 800 MeV. The use of ChPT was relevant for precision, by constraining the boundary conditions at threshold. Later on, these solutions were obtained up to 1.1 GeV \cite{Moussallam:2011zg}, finding a similar pole for the $\sig$ and, at the same time, determining in a robust way the position of the $f_0(980)$ resonance.

More recently, it was shown that one subtraction is enough due to $s-u$ symmetry in the Pomeron exchange, leading to the so-called GKPY equations \cite{GarciaMartin:2011cn,Pelaez:2015qba}. When they are used as constraints and for the same data input, Roy equations are more accurate in the low-energy region and GKPY in the resonance region.
Not surprisingly, simple good-looking fits to data were shown to fail to satisfy these equations 
\cite{GarciaMartin:2011jx}. This is illustrated in the top left panel of Fig.~\ref{fig:consistency}. Notwithstanding, a set of constrained fits to data (CFD) were obtained in \cite{GarciaMartin:2011jx}, consistent within uncertainties with FDRs, Roy and GKPY equations, as seen in the bottom left panel of Fig.~\ref{fig:consistency}. 
In addition, the residual distributions of the CFD fulfill all normality requirements necessary for the standard error propagation~\cite{Perez:2015pea}. 
Note that these are fits  and not solutions in the mathematical sense as before. 

We show in the left panel of Fig.~\ref{fig:CFD}, the resulting CFD $S$-wave, which is dispersively constrained up to 1.42 GeV. Actually, we show our recent global parameterization up to 1.9 GeV, which is consistent with the CFD up to 1.42 GeV and, above that energy, includes three different phenomenological fits to data sets which are in conflict among themselves, which are matched  continuously to the dispersively constrained fits at 1.42 GeV.
These CFD, which also describe the precise low-energy $K_{\ell4}$ data \cite{Batley:2010zza},  allowed for a dispersive $\sig$ pole precise determination \cite{GarciaMartin:2011jx,Pelaez:2015qba} from data. 
Irrespective of the number of subtractions, the applicability regime of Roy or GKPY equations is limited by avoiding the region of support of the double spectral regions, which limits what values of $t$ can be fixed, and by the convergence of the partial-wave expansion, which is only guaranteed inside the so-called Lehmann ellipse.
For $\pi\pi$ scattering this means less than roughly 1.1 GeV in the real axis. In the complex plane the applicability region, shown in the left panel of Fig.~\ref{fig:app}  was found in \cite{Caprini:2005zr} and fortunately includes safely  the region where the $\sig$ pole is sitting.

These Roy and GKPY studies played a major role in the 2012 RPP revision of the sigma, reducing its mass uncertainty estimate by a factor of 5 and changing the name of the resonance to the present one $f_0(500)$ from $f_0(600)$.

\begin{figure}
\begin{minipage}{.44\columnwidth}
\resizebox{\columnwidth}{!}{%
  \includegraphics{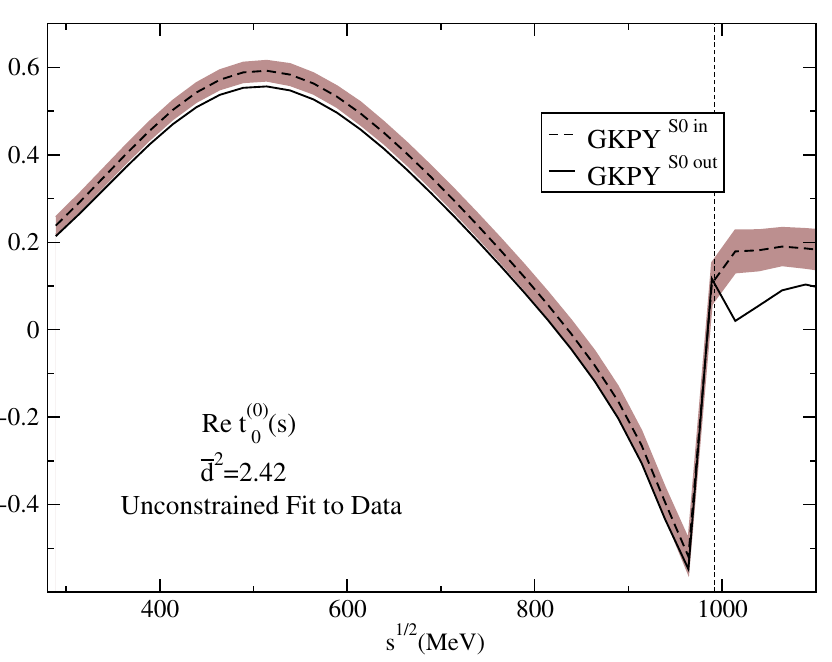} }
\resizebox{\columnwidth}{!}{%
  \includegraphics{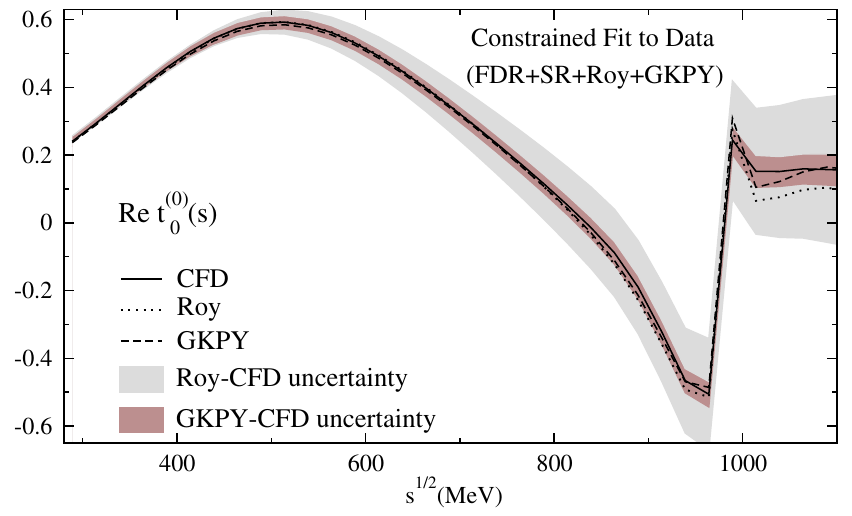} }
\end{minipage}
  \hspace{-.3cm}
\begin{minipage}{.57\columnwidth}
\resizebox{\columnwidth}{!}{%
  \includegraphics{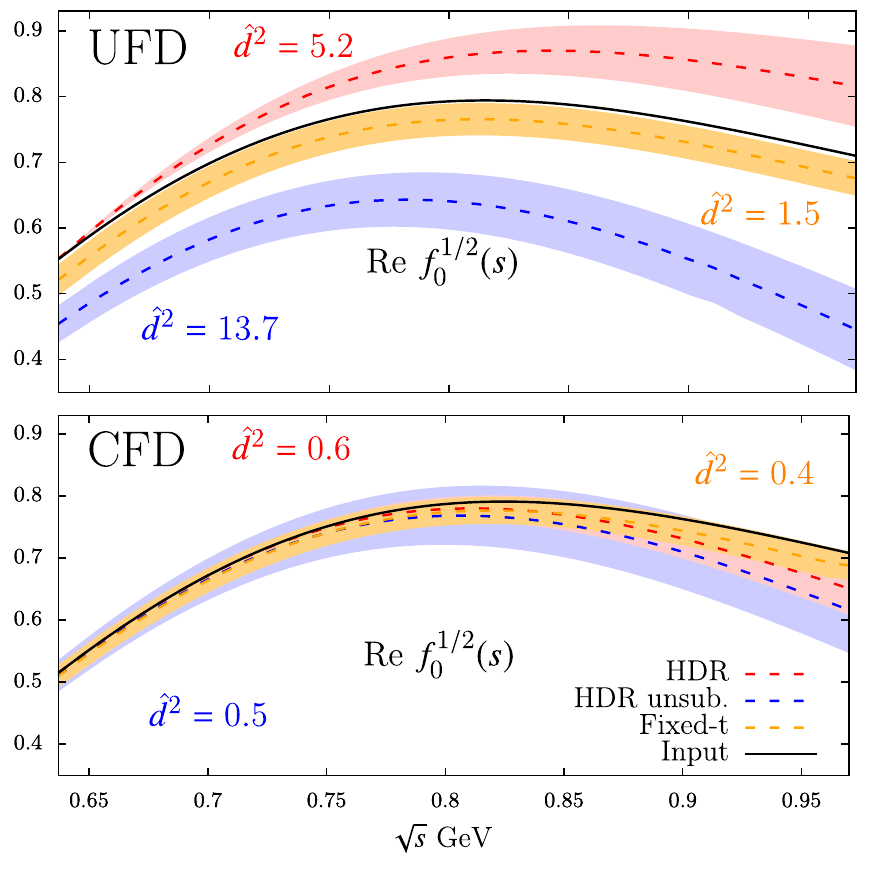} }
\end{minipage}
\caption{Fulfillment of some dispersion relations for $\pi\pi$ (Left) and $\pi K$ (Right) scattering in the $\sig$ and $\kap$ partial waves, respectively. Consistency with the dispersive representation requires input and outputs to agree within uncertainties. Note that unconstrained fits to data (UFD, top) do not fulfill well dispersion relations, whereas constrained fits to data (CFD, bottom) are very consistent.
Upper-left figure taken from \cite{GarciaMartin:2011cn}, bottom-left from \cite{GarciaMartin:2011jx} and
right from \cite{Pelaez:2020gnd}. 
}
\label{fig:consistency}       
\end{figure}

\begin{figure}
\begin{center}
  \hspace*{-.4cm}
  \begin{minipage}{0.48\columnwidth}
\resizebox{\columnwidth}{!}{%
    \hspace*{-.2cm}\includegraphics{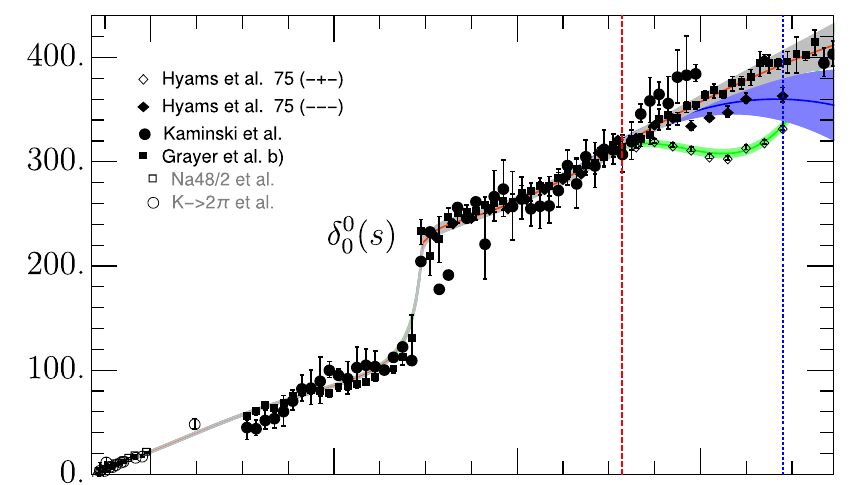} }
\resizebox{\columnwidth}{!}{%
  \includegraphics{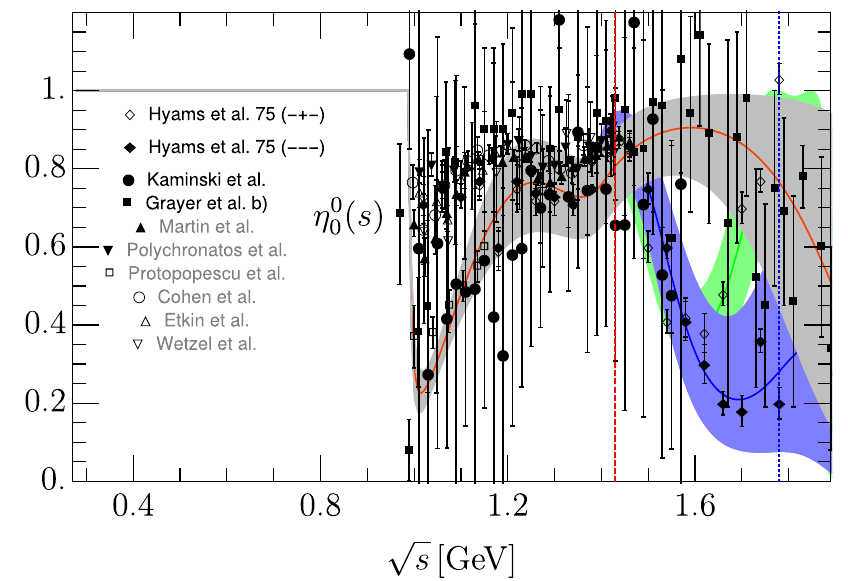} }
  \end{minipage}
  \hspace{-.3cm}
  \begin{minipage}{0.52\columnwidth}
   \resizebox{\columnwidth}{!}{%
  \includegraphics{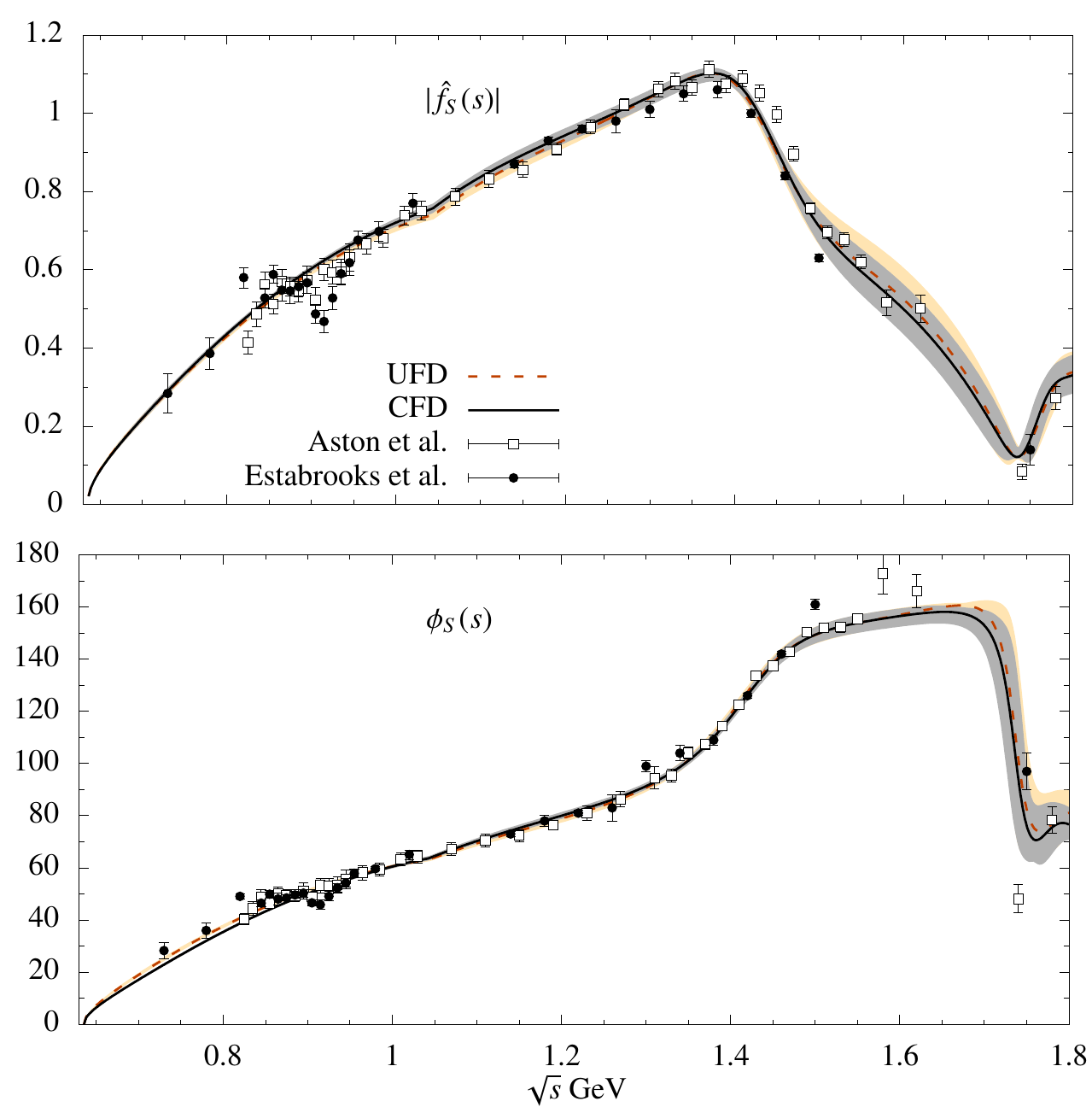} }
 
  \end{minipage}
\caption{$S$ waves, resulting from fits to data constrained with FDR and PWFTDR. 
{\bf Left:} For $\pi\pi$ scattering, we show the isospin-zero phase shift (top) and  elasticity (bottom). Figures taken from \cite{Pelaez:2019eqa}. The fit is constrained up to the red vertical line as in \cite{GarciaMartin:2011cn}, and beyond that there are several phenomenological fits to different data sets matched smoothly to the region constrained dispersively. {\bf Right:} 
For $\pi K$ scattering imposing also PWHDR. Figures taken from \cite{Pelaez:2020gnd}. We show the modulus (top) and the phase (bottom) of the $f_S\equiv f_0^{1/2}+f_0^{3/2}/2$ isospin combination. We also compare with unconstrained fits to data (UFD). 
}
\label{fig:CFD}       
\end{center}
\end{figure}

\begin{figure}
\begin{center}
  \hspace{-.2cm}
\raisebox{-0.5\height}{%
  \includegraphics[height=3.cm]{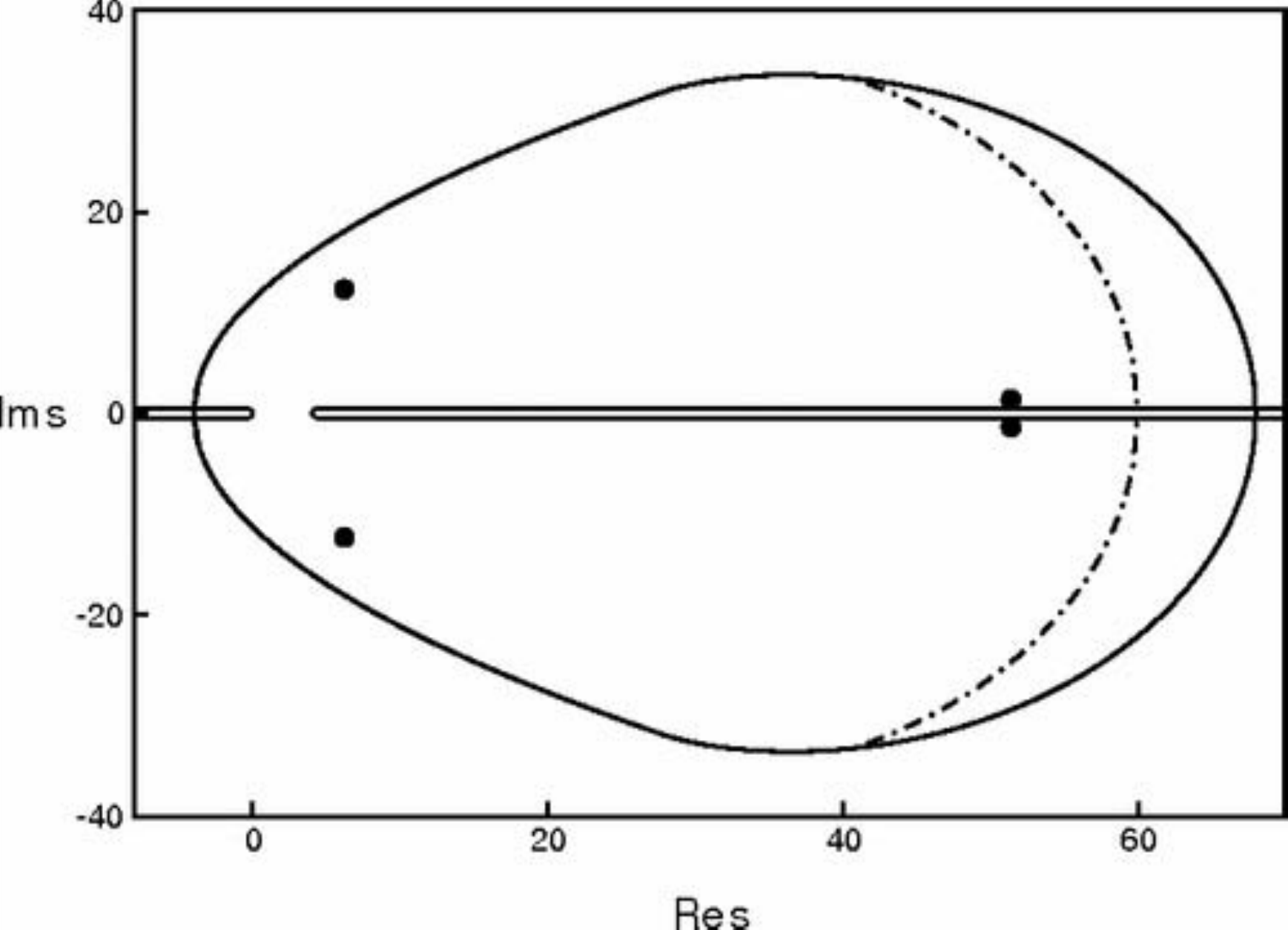} }
  \hspace{-.2cm}
\raisebox{-0.5\height}{%
  \includegraphics[height=3.2cm]{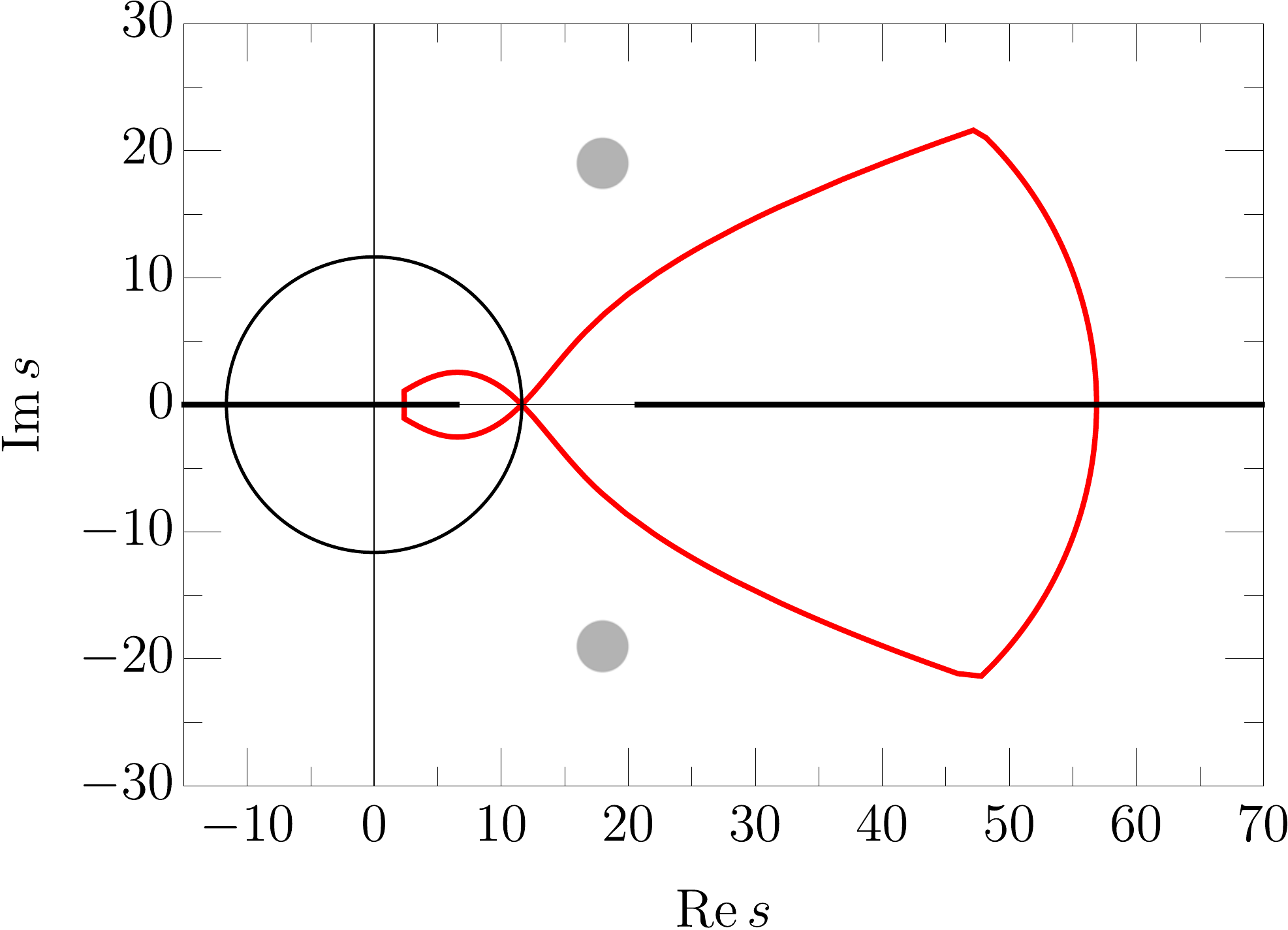} }
  \hspace{-.2cm}
\raisebox{-0.5\height}{%
  \includegraphics[height=3.3cm]{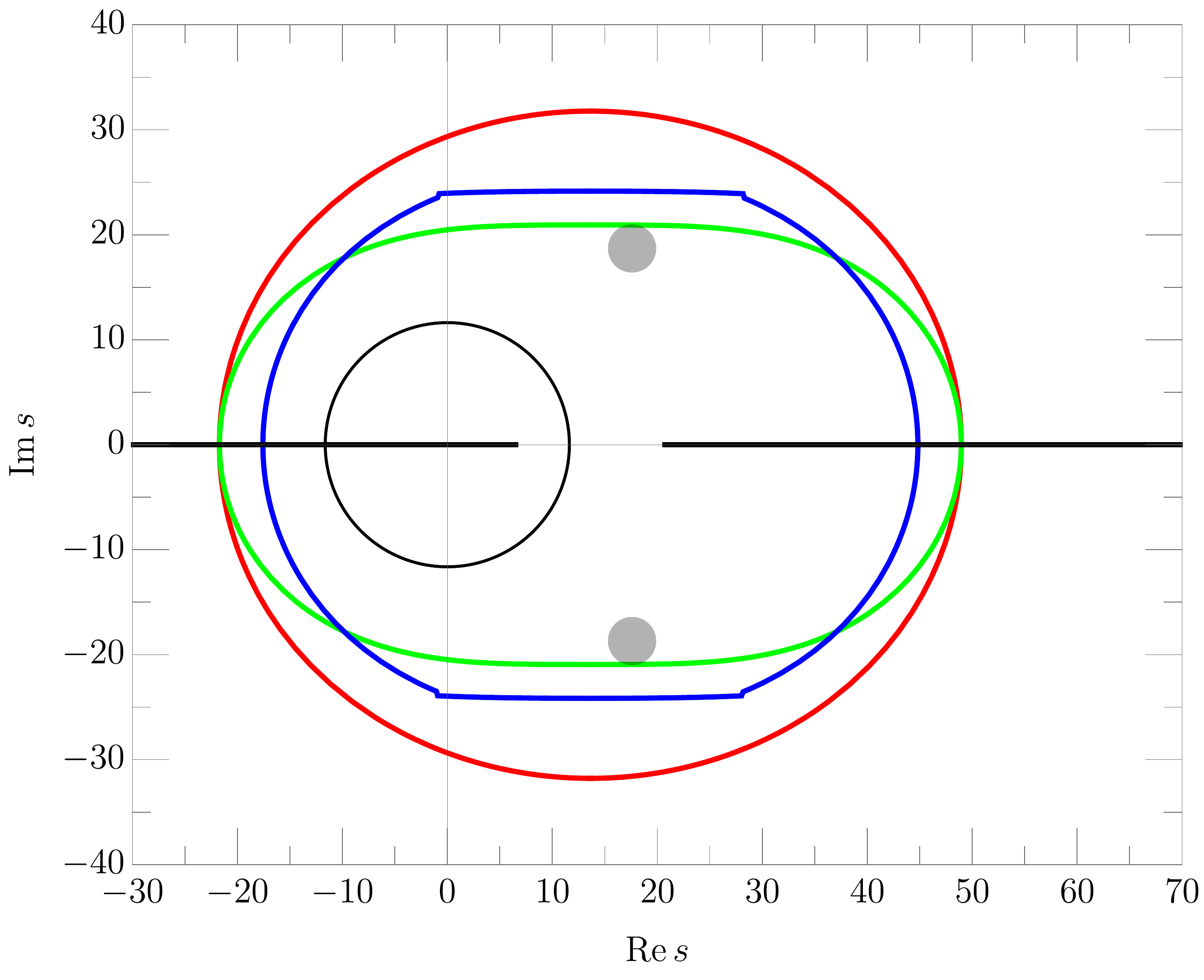} }
\caption{Applicability regions in the complex plane of partial-wave dispersion relations (all figures in $m_\pi^2$ units). We also show the conjugated pairs of $S$-wave poles.
{\bf Left:} For PWFTDR for $\pi\pi$ scattering.  Note the $\sig$ and $f_0(980)$ poles lie well inside the region.
{\bf Center:} For PWFTDR for $\pi K$ scattering. The $\kap$ pole lies outside.
{\bf Right:} For PWHDR for $\pi K$ scattering, with $a=0$ (blue line) and $a=-10m_\pi^2$ (green and red lines). Note the $\kap$ pole now lies inside.
Figures taken from \cite{Caprini:2005zr}, \cite{Pelaez:2020gnd} and \cite{Pelaez:2020gnd}, respectively.}
\label{fig:app}       
\end{center}
\end{figure}

Concerning $\pi K$ scattering, PWFTDR are sometimes called Roy-Steiner equations, since the unequal mass case was studied by Steiner and collaborators for $\pi N$ scattering \cite{Baacke:1970mi,Steiner:1971ms}. For our present interest on the $\kap$, let us remark that $S$ and $P$ wave {\it solutions to these equations} were obtained by the Paris group in a remarkable work \cite{Buettiker:2003pp}. Note that these are solutions that do not use data in the $\kap$ region as input, but just data at higher energies, including Regge asymptotics and other partial waves.
In principle, these equations would not require input from the $\pi\pi\rightarrow K\bar K$ crossed channel, but in practice it is needed for the determination of the subtraction constant for different values of $t$. Thus, there is a mild dependence on $\pi\pi\rightarrow K\bar K$. 
One should remark that this includes the ``pseudo-physical" region of 
$\pi\pi\rightarrow K\bar K$, between $\pi\pi$ and $K\bar K$ thresholds, where no $\pi\pi\rightarrow K\bar K$ data exists. Nevertheless it is possible to calculate dispersively the amplitude, since  $\pi\pi$ is in practice the only physically available state there and then its phase shift, well known from data and dispersion theory, is the same as the phase of  $\pi\pi\rightarrow K\bar K$, due to Watson's theorem \cite{Watson:1952ji}. The modulus is obtained dispersively with a Mushkelishvili-Omn\'es formalism, using this information and matching the physical region (see \cite{Buettiker:2003pp,Pelaez:2020gnd} for details).

No continuation to the complex plane was made since the applicability region for the PWFTDR does not reach the $\kap$ pole, as shown  in the center panel of Fig.~\ref{fig:app}. 

In our recent dispersive study \cite{Pelaez:2020gnd}, we have shown that simple fits to existing data do not satisfy well these equations, although the $f^{1/2}_0$ wave does not fare very badly, see upper right panel in Fig.~\ref{fig:consistency}. Nevertheless we provided a constrained fit to data (CFD), which, together with FDRs and PWHDR, also satisfies these PWFTDR within uncertainties, as seen in the lower-right panel of that same figure.  The resulting modulus and phase of the $S$-wave is shown in the right panels of Fig.~\ref{fig:CFD}.

\begin{itemize}
\item {\it \underline{PWFTDR Pros:}} Dispersion relations for individual partial waves. 
Rigorous treatment of unphysical cuts. Unique solution in elastic region. 
They are very accurate at low energies with two subtractions, and more accurate in the resonance region with one subtraction.
For $\pi\pi$, the $\sig$ pole lies within the applicability region. 
They were the key results to settle the $\sig$ controversy 
and provide a precise determination of its pole parameters.
For $\pi K$ the dependence on $\pi\pi\rightarrow K\bar K$ is mild.

\item {\it \underline{PWFTDR Cons:}} 
Coupled system of infinite partial waves, although at low-energies only the lowest ones are needed.
More complicated than FDRs. 
Only applicable in real axis up to roughly 
1.1 GeV for both $\pi\pi$ and $\pi K$.
Their applicability region in the complex plane does not reach the $\kap$.
For $\pi K$ it needs $\pi\pi\rightarrow K\bar K$ input, including its pseudo-physical region, which requires a Mushkelishvili-Omn\'es formalism with input from $\pi\pi$ scattering. This makes the $\pi K$ case much more complicated than for $\pi\pi$.
\end{itemize}

\vspace{.3cm}
\item[$\bullet$]
{\bf Partial-wave projected Hyperbolic Dispersion Relations (PWHDR).}
These were first derived for $\pi N$ scattering \cite{Hite:1973pm} although they were soon applied to 
 $\pi K$ scattering \cite{HedegaardJensen:1974ta,Johannesson:1974ma,Bonnier:1975ne,Johannesson:1976qp}. Note that we are interested in these equations for $\pi K$ scattering for the following reasons: First, as we have just commented, we need them to determine the subtraction constant of the symmetric PWFTDR. Second, in so doing we need input from $\pi\pi\rightarrow K\bar K$ and we would then like to constrain this process with dispersion relations, but only PWHDR reach the physical region of this channel. Finally, it was shown in \cite{DescotesGenon:2006uk} that these dispersion relations can reach the $\kap$ region, providing a rigorous mathematical tool for its determination. 
 
 None of the above motivations apply to $\pi\pi$ scattering and these equations have not been applied to the determination of the $\sigma$.
In contrast, these equations were first used in \cite{DescotesGenon:2006uk} to determine the $\kap$ using the previously obtained solutions of PWFTDR obtained in \cite{Buettiker:2003pp}. 
Despite this 2006 rigorous dispersive result, which  is shown as a dark solid diamond in Fig.~\ref{fig:kappapoles}, the $\kap$ pole was still classified in the RPP as ``Needs confirmation" until this year. 

With the aim of providing the required confirmation, we started a program of constraining fits to existing $\pi K$ scattering data, with these PWHDR as well as PWFTDR and FDR.  On a first step  $\pi K$ data fits constrained up to up to 1.75 GeV with FDRs were obtained in \cite{Pelaez:2016tgi}.
A naive continuation to the complex plane of the conformal parameterization we used for the fit in the elastic region already has a $\kap$ pole, shown as an empty circle in Fig.~\ref{fig:kappapoles}, although it has a larger width, not surprising, since this is still parameterization dependent.
Nevertheless, those FDR-constrained fits already allowed for a determination of the $\kap$ and other strange resonance poles using sequences of Pad\'es in \cite{Pelaez:2016klv}. The resulting $\kap$ pole, shown in Fig.~\ref{fig:kappapoles} as a solid black square, appears near but below 700 MeV, and was quite consistent with that of  \cite{DescotesGenon:2006uk}, triggering the change of name of the $\kappa$ from $K^*_0(800)$ to $K^*_0(700)$ in the 2018 RPP.

Nevertheless, to implement a fully dispersive determination of the $\kap$ pole {\it from data}, 
PWHDR were first used as constraints for  $\pi\pi\rightarrow K\bar K$ \cite{Pelaez:2018qny}, modifying the family of hyperbolae to maximize the applicability within the physical region of these equations up to 1.6 GeV. Note that in this way it was possible to check the data consistency, which was poor, and then constrain the data description of this channel, which otherwise was only fit and then used as input. Only very recently the simultaneous 
dispersive analysis of both $\pi K$ and $\pi\pi\rightarrow K \bar K$ data has been completed, using 16 different dispersion relations \cite{Pelaez:2020gnd}.

The unconstrained fits to data (UFD) fail to satisfy dispersion relations. 
For our purposes, the most relevant partial-wave is the scalar isospin-1/2 one, that we show here in the upper right panel of Fig.~\ref{fig:consistency}. As already explained, the PWFTDRdo not fare very badly, but the PWHDR deviate from the input by a large amount. Even worse, they deviate in different directions depending on whether one checks the subtracted or unsubtracted version of the antisymmetric $F^-$ amplitude used in the PWHDR derivation. Note that this happens even though the UFD fit to data, shown here in the right panels of Fig.~\ref{fig:CFD}, looks like a very nice fit.
Nevertheless, one should recall that there are other contributions to the dispersion relation.  For instance, for this wave the pseudo-physical region of $\pi\pi\rightarrow K \bar K$ in the vector channel (the $\rho$ exchange), is very relevant. Remember there is no data there and the amplitude has to be calculated with a dispersive Mushkelishvili-Omn\'es method, using the UFD sets and the $\pi\pi$ phase shift as input. However, in \cite{Pelaez:2020gnd} it has been shown that with unconstrained fits this prediction can vary dramatically from
 the subtracted to the unsubtracted case, as shown in the left panel of Fig.~\ref{fig:MO}. As a consequence, we have shown in Fig.~\ref{fig:kappapoles} two different UFD poles, with one or no subtractions, but the same nice-looking UFD input from Fig.~\ref{fig:CFD}. This is a remarkable illustration that just fitting the data with any model does not ensure the extraction of a consistent $\kap$ pole. It is necessary to check that other contributions besides that wave are also well described.
 
 For this reason, a dispersively constrained  fit to data (CFD) set has been obtained \cite{Pelaez:2020gnd}, describing data fairly well but being consistent within uncertainties with 16 dispersion relations, including PWHDR sets with two different subtractions. Indeed, we show in the lower right panel of Fig.~\ref{fig:consistency} how the output of all dispersion relations agree when using this CFD as input. We also show in the right panel of Fig.~\ref{fig:MO}, that for the CFD the prediction for the pseudo-physical region is the same irrespective of whether the subtracted or unsubtracted Mushkelishvili-Omn\'es method is used.

With this fully consistent CFD set,  the PWHDR can be used to find the $\kap$ pole, since it lies within their applicability region, finding  \cite{Pelaez:2020uiw} a remarkably consistent and accurate result with either the unsubtracted or the subtracted case. We show both poles in Fig.~\ref{fig:kappapoles}, together with their numerical values. These results, although remarkably consistent with that from \cite{DescotesGenon:2006uk}, are independent from it, since in the $\kap$ region the CFD is a fit to data, not a solution, and also because some other inputs have been checked against the dispersive representation, or updated (like the $K^*(892)$, Regge and $\pi\pi\rightarrow K\bar K$ descriptions). It is for this reason that, as already commented in the introduction, the ``Needs confirmation" 
for the $\kap$ will be removed in the next RPP version. 

\begin{itemize}
\item {\it \underline{PWHDR Pros:}}
Their applicability region reaches the $\kap$ pole. 
They can be applied to $\pi\pi\rightarrow K \bar K$ in the physical region.

\item {\it \underline{PWHDR Cons:}} 
In the physical region only applicable up to $\sim 1.1$ GeV.
The most complicated expressions so far. 
They require $\pi\pi\rightarrow K \bar K$ in the pseudo-physical region, but this can be treated
with a Mushkelishvili-Onm\'es method.
\end{itemize}

\begin{figure}
\begin{center}
  \hspace{-.2cm}
\resizebox{.45\columnwidth}{!}{%
  \includegraphics{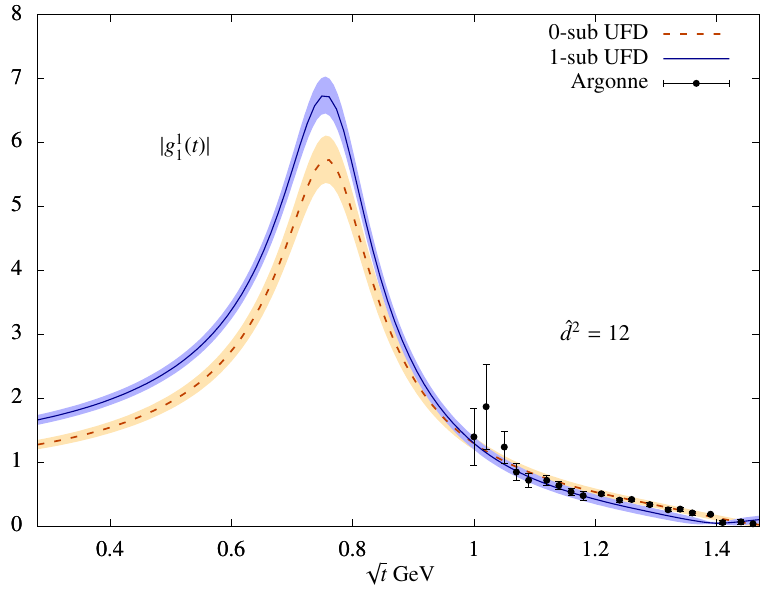} }
  \hspace{-.2cm}
\resizebox{.45\columnwidth}{!}{%
  \includegraphics{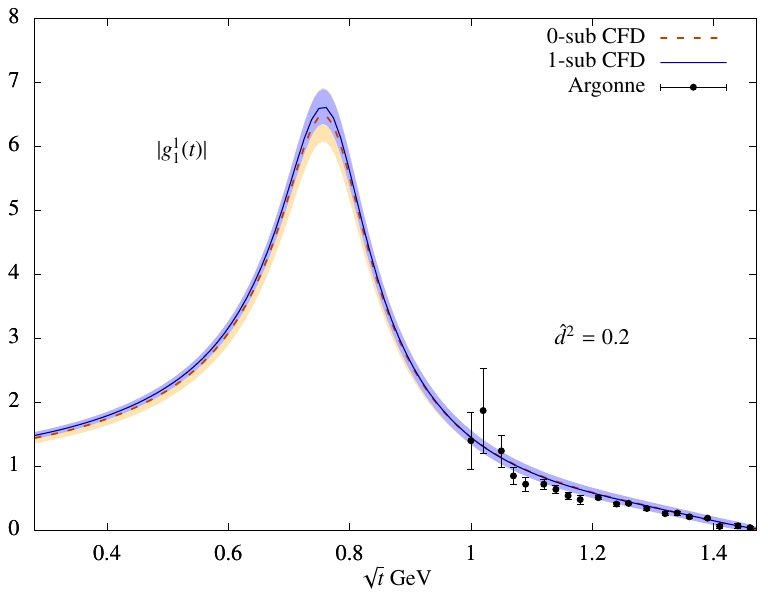} }
\caption{Vector partial wave in $\pi\pi\rightarrow K\bar K$ scattering. 
Note that there is only data above the $K\bar K$ threshold. In the ``unphysical" region below, 
the amplitude is calculated dispersively using a Mushkelishvili-Omn\'es approach. This region yields a relevant contribution to the $\kap$ channel in $\pi K$ scattering. Using unconstrained fits to data (UFD, left) the predictions with one or two subtractions are inconsistent among themselves. They come out consistent with constrained fits to data (CFD, right).
Figures taken from \cite{Pelaez:2020gnd}.}
\label{fig:MO}       
\end{center}
\end{figure}

%

\end{itemize}

In summary, the fully dispersive approach is well suited for precision studies, since it deals as accurately as possible not only with the physical scattering regions, but also with the contributions from analytic structures in the so-called unphysical regions. To achieve this precision, the dispersive integrals are evaluated by means of partial waves up to a region where Regge models are used. In practice, the output of these dispersion relations has to be calculated numerically.
These dispersive integrals can be used as checks to discard data and, therefore, contribute to the solution of the ``data problem" explained above. In addition, they can be used to constrain the data description, or,  for certain energy regions of the lowest waves, they can be solved using input from the other regions and waves. 

The main advantage is the model independence, since these dispersion relations are derived just from first principles like causality and crossing, although unitarity can also be implemented as an additional constraint on the solutions. In principle, no further dynamical input is needed, although it is possible to add further constraints or to match them with Chiral Perturbation Theory.
Depending on the number of subtractions, it is possible to attain more accuracy on the threshold or the resonance region and suppress regions where our experimental knowledge is limited. 

The price to pay for this precision and rigor is that expressions are rather complicated and the numerical treatment is heavy. One may be dedicating a lot of effort to contributions which are numerically small. This has been a price well-worth paying for in the case of $\pi\pi$ and $\pi K$ scattering, in order to settle the controversy of the $\sig$ and $\kap$ mesons, as well as for an accurate description of low-energy $\pi\pi$ and $\pi K$ scattering. Accurate data analyses on the threshold behavior are of relevance since ChPT provides there a systematic and precise theoretical approach to compare with. 

Similar dispersive analyses have also been carried out for $\pi N$ \cite{Baacke:1970mi,Steiner:1971ms,Hite:1973pm,Ditsche:2012fv,Hoferichter:2012wf,Hoferichter:2015hva}, $\gamma^{(*)} \gamma^{(*)} \to \pi \pi$ \cite{GarciaMartin:2010cw,Hoferichter:2011wk,Moussallam:2013una, Danilkin:2018qfn, Hoferichter:2019nlq}, $e^+ e^- \to \pi^+ \pi^-$ \cite{Colangelo:2018mtw}, $\gamma \pi\to \pi\pi$~\cite{Hoferichter:2012pm,Hoferichter:2017ftn} and $\gamma K\to \pi K$ \cite{Dax:2020dzg}. The approach for obtaining rigorous and precise poles from dispersively constrained data fits, is also well-suited for the recent lattice calculations that we hope may shed further light on these issues.

Of course, if high precision is not a requirement, there are many contributions that can be neglected or approximated in simple ways, making it simple to introduce further information about existing symmetries and associated dynamics, but still providing a fairly good description of data and poles. This is the subject of the next section.

%% file: sections/uchpt.tex
\subsection{Two-body elastic unitarity}
In this approach the emphasis is made on two-body unitarity for partial waves.
Let us concentrate on the elastic case, $\eta_{I\ell}=1$, which is much simpler and the relevant one for the $\sig$ and $\kap$ resonances. Then, recalling  Eq.~\eqref{eq:felastic}, {\it in the physical elastic region} we can write: 
\begin{equation}
\im f(s) =\sigma(s) \vert f(s) \vert^2 \hspace{0.2cm} \Rightarrow \hspace{0.2cm} \im \frac{1}{f(s)}=-\sigma(s) \hspace{0.2cm} \Rightarrow \hspace{0.2cm}
f(s) =\frac{1}{\re (1/f(s))-i\sigma(s)},
\label{eq:elunit}
\end{equation}
where we have omitted the $I,\ell$ indices for clarity 
and the second step follows trivially from the fact that, for any complex number $z$, $\im 1/z= \im z^*/(z z^*)=\im z^*/\vert z\vert^2=-\im z/\vert z\vert^2$. In other words, Eq.~\eqref{eq:elunit} means that we only need to know the real part of the inverse of a two-body scattering partial wave, since its imaginary part is given by unitarity.

This is not applicable to the general inelastic case, although, {\it in the particular case when all available states at a given energy are two-body states}, it is possible to generalize the above Eq.~\eqref{eq:elunit} to a matrix form:
\begin{equation}
\im F_{ij}(s)=F_{in}(s)^\dagger \Sigma_{nn}(s) F_{nj}(s) \quad \Rightarrow\quad F_{ij}(s)=[K^{-1}(s)-i \Sigma(s)]^{-1}_{ij},
\label{eq:coupledunit}
\end{equation}
where $K^{-1}(s)=\re F(s)^{-1}$. Note that we have suppressed the isospin and angular momentum indices for simplicity, $F=F_{ij}$ is a symmetric matrix whose elements are the partial waves between the $k=1...n$ coupled two-body states,  $\Sigma(s)={\rm diag}[\sigma_1(s)...\sigma_n(s)]$ and $\sigma_k(s)$ is the phase-space of each accessible physical state at that energy. 
As usual a summation is understood over repeated indices. This is the so-called ``$K$-matrix" method for two-body scattering and ensures two-body unitarity and the correct physical or ``right-hand" cuts for all the two-body states coupled via scattering to some given isospin and angular momentum.

This $K$-matrix method is a very simple, successful and popular way of describing scattering in partial waves in Hadron Physics. It can be useful when multiple-body states are available, as long as their contribution is negligible or they can be approximated by a quasi-two-body scattering. Moreover, by using simple real functions for $K$ it is possible to generate a wide variety of shapes, including those of resonances and, since the $\sigma_i(s)$ functions are easy to continue to the complex plane, look for their associated poles in unphysical sheets.

However, it is clear that this method is just a model that does not have all the analytic structures that we have described in Section~\ref{sec:anstruc} and illustrated in Fig.~\ref{fig:anstruc}. In particular it lacks the left and circular cuts. In addition, due to the singularity of $\sigma(s)$, it imposes a zero in the partial waves at $s=0$.
Of course, when far from these structures, the $K$-matrix can be versatile enough and provide a good description of data.
But, as we have seen, the ``unphysical" singularities, the dynamical Adler zero and the low-energy constraints dictated by the spontaneus chiral symmetry breaking, are essential for a rigorous and precise determination of $\pi\pi$ and $\pi K$ scattering in the region below 1 GeV and even more so for the determination of the $\sig$ and $\kap$ poles.

The caveats we have just described can be addressed by combining dispersion theory and chiral QCD constraints. The latter will be implemented by means of Chiral Perturbation Theory that we briefly sketch next.

\subsection{Chiral Pertubation Theory for meson-meson scattering}

Since the $u,d,s$ quarks are much lighter than the hadrons they form and the other quarks, QCD in the chiral limit, i.e., with massless $u,d,s$ quarks, should be a good approximation to Hadron Physics for energies well below the charm quark mass. In such case, the QCD Lagrangian is invariant under $SU(3)$ transformations 
among the three lightest quarks with definite right ($R$) or left ($L$) chirality. Namely, in the chiral limit for the lightest three quarks the QCD Lagrangian is $SU(3)_L\times SU(3)_R$ symmetric, implying that the vector $(V=L+R)$ and axial $(A=L-R)$ quark currents should be conserved up to relatively small quark mass corrections. As a consequence, mesons with a given spin should appear in almost degenerate pairs of nonets with opposite parity.  We have already shown the lightest nonets in Fig.~\ref{fig:multiplet}. The different masses within the same nonet are explained, assuming these mesons are quark-antiquark states, because the heavier strange quark contributes 100-200 MeV more to the meson mass than the $u$ and $d$, which also differ between themselves by a few MeV. As we have explained, the  only exception is the scalar $0^+$ nonet, whose ``inverted" hierarchy is more similar to the one expected from a ``tetraquark" or ``molecular" composition, although 100-200 MeV mass differences are once again due to the larger strange mass.
In any case, we can see that the expected chiral degeneracy does not occur, since the lightest states with the same flavor and $J$ but opposite parity $P$ typically differ by 300-500 MeV. For instance, the mass difference between the $\rho(770)$ and $a_1(1260)$, which have no strangeness, is too large to be explained with the few MeV $u$ and $d$ quark masses. 

Therefore, chiral symmetry must be broken spontaneously by the vacuum from $SU(3)_L\times SU(3)_R$ down to $SU(3)_V$. In other words, the axial currents do not annihilate the vacuum, so that {$\langle 0\vert A^i_\mu \vert \phi^k(p_\mu)\rangle=if_0 p_\mu\delta^{ik}$}, where $f_0$ is a constant to be defined below. Therefore,
Goldstone's Theorem implies the appearance of eight massless pseudo-scalar Nambu-Goldstone Bosons (NGB), $\phi^k$, and a mass gap between them and the other massive hadrons. These are the pions, kaons and the etas. 
Note that this means that when interacting between states ``1" and ``2", the axial current can interact directly as usual but can also become a massless NGB, namely, $\langle 1\vert A_\mu^j(0)\vert 2\rangle=R^j_\mu+f_0\,p_\mu T^j/p^2$, where the last term includes the propagator of a massless particle $1/p^2$ and the NGB interaction $T^j$. But then current conservation implies: $0=\langle 1\vert \partial^\mu A_\mu^j(0)\vert 2\rangle=p^\mu R^j_\mu+f_0\, T_j$. Thus, at low energies $p_\mu\rightarrow 0$, the NGB interactions $T_j$ must also vanish, which means that these are derivative interactions. 
Of course, in real life, quarks have a small mass and NGB states are not massless but just much lighter than their counterparts in other nonets \footnote{The only exception is the $\eta'$ whose huge mass is due to another well-understood effect, which is the chiral anomaly, beyond the scope of this tutorial review}. As a consequence, not only they can have derivative interactions but also
interactions containing powers of their masses, which vanish in the chiral limit. Since these masses are relatively small compared to other hadrons, this allows  for the construction of the QCD low-energy effective theory as a Lagrangian, built 
out of NGB only, as a systematic expansion ${\cal L}= {\cal L}_2+ {\cal L}_4+ {\cal L}_6...$ in powers of the NGB momenta (derivatives) and masses. The NGB, which form and $SU(3)_V$ multiplet, are gathered in an $SU(3)$ matrix 
\begin{equation}
U=\exp(i\sqrt{2}\Phi^k\lambda^k/f_0),\quad 
\Phi (x) \equiv
\left(
\begin{array}{ccc}
\pi^0/\sqrt{2}  + \eta/\sqrt{6}  & \pi^+ & K^+ \\
\pi^- & - \pi^0/\sqrt{2} + \eta/\sqrt{6}  & K^0 \\
K^- & \bar{K}^0 &  - 2\eta/\sqrt{6}
\end{array}
\right),
\end{equation}
where $\lambda_k$ are the Gell-Mann matrices. When coupled to electroweak forces, which we do not need for our purposes here,  $f_0$ is identified with the NGB decay constant in the chiral limit. Only those terms consistent with the spontaneous chiral symmetry breaking of QCD are allowed. This approach is known as Chiral Perturbation Theory (ChPT)
\cite{Weinberg:1978kz,Gasser:1983yg,Gasser:1984gg}. For example, the leading order (LO) is:
\begin{equation}
{\cal L}_2 = \frac{f_0^2}{4} Tr( \partial_{\mu} U^{\dagger}
\partial^{\mu} U +
M_0^2 (U + U^{\dagger})),\quad M_0^2=2 B_0 \,{\rm diag}(\hat m,\hat m, m_s).
\label{ec:Lag2}
\end{equation}
A similar formalism can be implemented with just the $u$ and $d$ quarks and $SU(2)$ symmetry groups, but we have preferred to present the $SU(3)$ case since we want to discuss the $\kap$, which has strangeness.

Note that ${\cal L}_2 $ only depends on the meson-mass matrix $M_0$ and the decay constant $f_0$, which at leading order is the same for $\pi, K$ and $\eta$. Therefore the form of this term is universal for any theory sharing the same chiral symmetry breaking pattern as QCD and its predictions are known as Low Energy Theorems. Each LO vertex in a Feynmann diagram adds two powers of mass/momentum. The NLO has 10 terms with four derivatives, or a squared mass and two derivatives or a mass to the fourth power \cite{Gasser:1984gg}. Each term is multiplied by a low energy constant $L_i$ (LEC), whose value is not fixed from symmetry, but from the QCD underlying dynamics. They are known with different levels of precision from data (see \cite{Bijnens:2014lea}) or lattice QCD (see \cite{Aoki:2019cca} and references threin). 
Note that, with these lagrangians, each additional loop in a Feynmann diagram increases its expansion order by one. However, having all possible terms consistent with symmetries, the one-loop divergences with LO vertices can be absorbed in the NLO LECs, and similarly occurs with more loops and higher order constants; thus rendering all calculation finite up to a given order \cite{Weinberg:1978kz}. These ``chiral" loops give rise to 
the so-called chiral logarithms, which are very relevant at low energies, as well as to the analytic structure of cuts for the amplitude that we described above.

\begin{figure}
\begin{center}
  \hspace{-.2cm}
\resizebox{\columnwidth}{!}{%
  \includegraphics{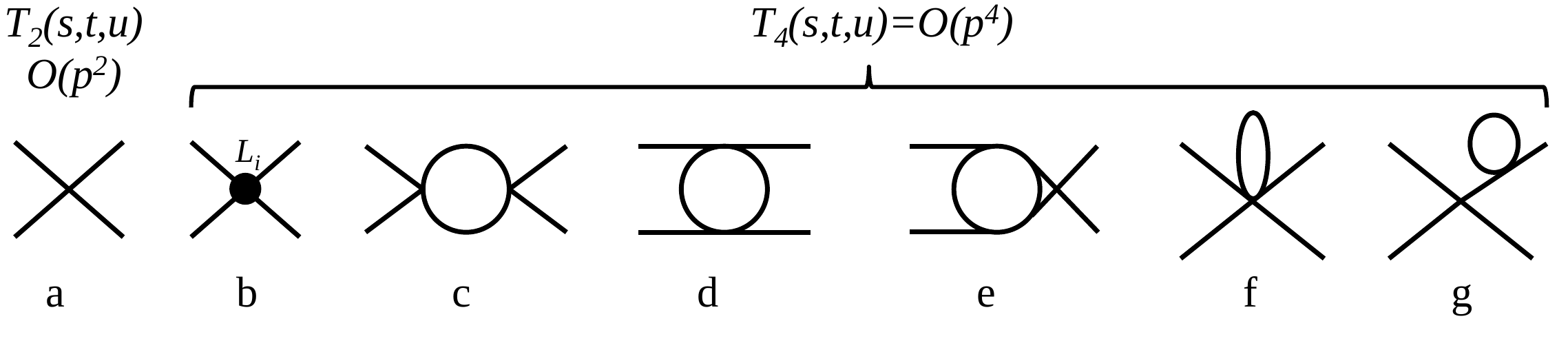} }
\caption{Classes of diagrams that appear in the
  NLO ChPT calculation of meson-meson scattering. }
\label{fig:ChPTdiagrams}
\end{center}
\end{figure}

In particular, meson-meson scattering partial waves are obtained as a series expansion $f(s)=f_2(s)+f_4(s)+...$, where for simplicity we have dropped the isospin and angular-momentum indices. When several channels are
coupled, these partial-wave amplitudes are collected in a $T$ matrix.  The LO is just a tree level calculation with the lagrangian in Eq.~\eqref{ec:Lag2} and therefore real. Let us provide, for illustration and later use, the LO results for the $(I,\ell)=(0,0),(1,1)$ $\pi\pi$ and $(1/2,0),(1/2,1)$ $\pi K$ scattering partial waves:
\begin{eqnarray}
&&f_0^{0}(s)=\frac{2s-M_\pi^2}{32\pi f_0^2},\quad
f^{1/2}_{0}(s)= \frac{5s^2-2(M_K^2+M_\pi^2)s-3(M_K^2-M_\pi^2)^2}{128\pi f_0^2 s}, \label{eq:LET0}\\
&&f_1^{1}(s)= \frac{s-4 M_\pi^2}{96\pi f_0^2},\quad 
f^{1/2}_{1}(s)= \frac{M_K^4+(s-M_\pi^2)^2-2M_K^2(s+M_\pi^2)}{128\pi f_0^2 s},\label{eq:LET1}
\end{eqnarray}
where $M_\pi$ and $M_K$ denote the pion and kaon mass, respectively.
To LO all NGB decay constants are equal to $f_0$, and in principle we could take any of them, although for the pion case it seems natural to use $f_0^2=f_\pi^2$. 
 Up to NLO, the different kinds of Feynmann diagrams that contribute to $T$ are shown in Fig.~\ref{fig:ChPTdiagrams}. All the one-loop divergences are absorbed through renormalization into the $L_i$ as well as the masses and decay constants, $f_i$, which at NLO have different values for $f_\pi$, $f_K$ and $f_\eta$.

A very relevant aspect of ChPT for the $\sig$ and $\kap$ is that it provides a model-independent link with QCD parameters. 
We have already seen the appearance of the quark masses $m_q$ in Eq.~\eqref{ec:Lag2}, 
but the leading order behavior in terms the $1/N_c$ expansion, where $N_c$ is the number of colors in the QCD Lagrangian, is also known.  For instance, meson masses behave as $O(1)$, the decay constants as $f_i^2\sim O(N_c)$ and the NLO LECS behavior varies between $O(N_c)$ and $O(1)$ \cite{Gasser:1984gg,Pelaez:2015qba}.

The drawback of ChPT is that it is limited to very low energies, well below resonances, whose associated poles it cannot describe. Moreover, as for any other perturbative treatment, unitarity is only satisfied order by order. In the elastic regime we are interested in, this means that instead of Eqs.\eqref{eq:elunit}, the ChPT series satisfies:
\begin{equation}
\im f_2(s)=0, \quad \im f_4(s)=\sigma(s) f_2(s)^2,\quad  \im f_6(s)= 2  \sigma(s)  f_2(s)\re f_4(s), ...
\label{eq:pertunit}
\end{equation}
However, whereas for other perturbative series the expansion parameter may be small, so that neglecting higher orders may be justified and unitarity may be well approximated, 
in this case we are expanding in powers of momentum, and therefore unitarity is badly violated as soon as the NGB momentum grows enough to reach the resonance region. In brief we will use that $f_2(s)$ is real and that the imaginary part for a given order is fixed from the previous orders.

Nevertheless, as we will review next, the ChPT series can still be used as input and ``unitarized" to describe resonances while respecting the low-energy chiral symmetry constraints.

\subsection{Unitarization: naive approach}
\label{sec:naive}
For more comprehensive and detailed accounts of unitarization of chiral interactions we refer the reader to the reviews in \cite{Pelaez:2015qba,Oller:2020guq,Yao:2020bxx}.

Let us first provide a very naive derivation of the unitarization equations, reviewing their usefulness, paying particular attention to the $\sig$ and $\kap$ case. In the next subsection we will provide a more sound justification using dispersion theory.

We just saw in Eq.~\eqref{eq:elunit} that, for physical values of $s$ in the elastic regime, the imaginary part of the inverse partial-wave amplitude is fixed by unitarity. 
What we are  calling here {\it the ``naive approach" to unitarization is just to use directly ChPT to calculate} $\re(1/f)=\re \,1/(f_2+f_4+...)\simeq (1/f_2)(1-\re f_4/f_2+...)$, where we have profited from the fact that $f_2$ is real, and write a unitarized elastic partial wave at different orders as:
\begin{equation}
f^U_{LO}(s)=\frac{1}{1/f_2(s)-i\sigma(s)}, \quad f^U_{NLO}(s)=\frac{1}{1/f_2(s)-\re f_4(s) /f_2(s)^2-i\sigma(s)}, ...
\label{eq:naiveu}
\end{equation}
and similar expressions for NNLO, etc... Obviously, by re-expanding again one recovers  at low-energies the ChPT series up to the order that has been used as input, but in addition one gets the correct  imaginary part of the next order. These expressions are unitary and can be analytically continued to the complex plane and the second sheet 
using Eq.~\eqref{eq:second}. 

\begin{figure}
\begin{center}
  \hspace{-.2cm}
\resizebox{.48\columnwidth}{!}{%
  \includegraphics{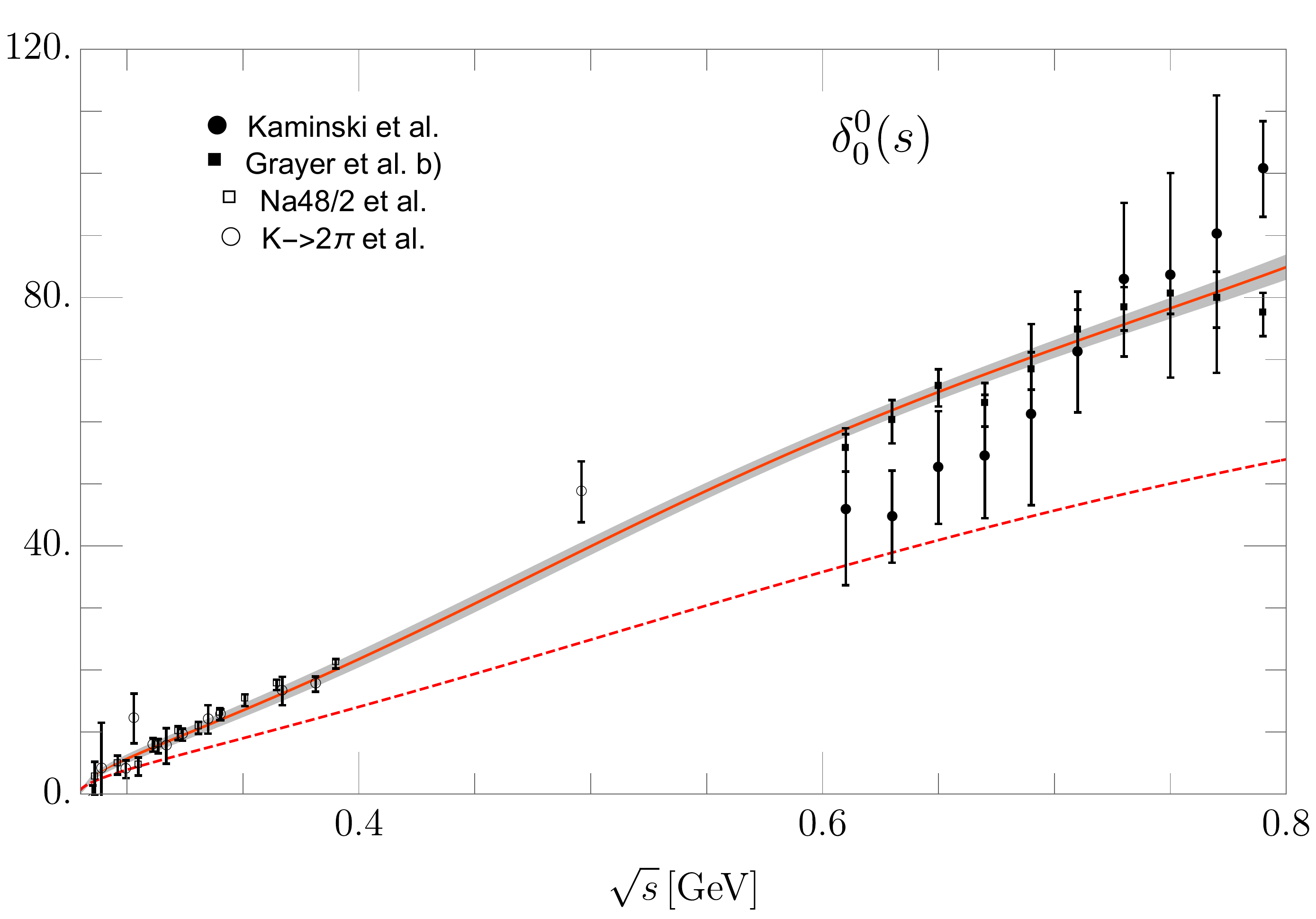} }
  \hspace{-.1cm}
\resizebox{.48\columnwidth}{!}{%
  \includegraphics{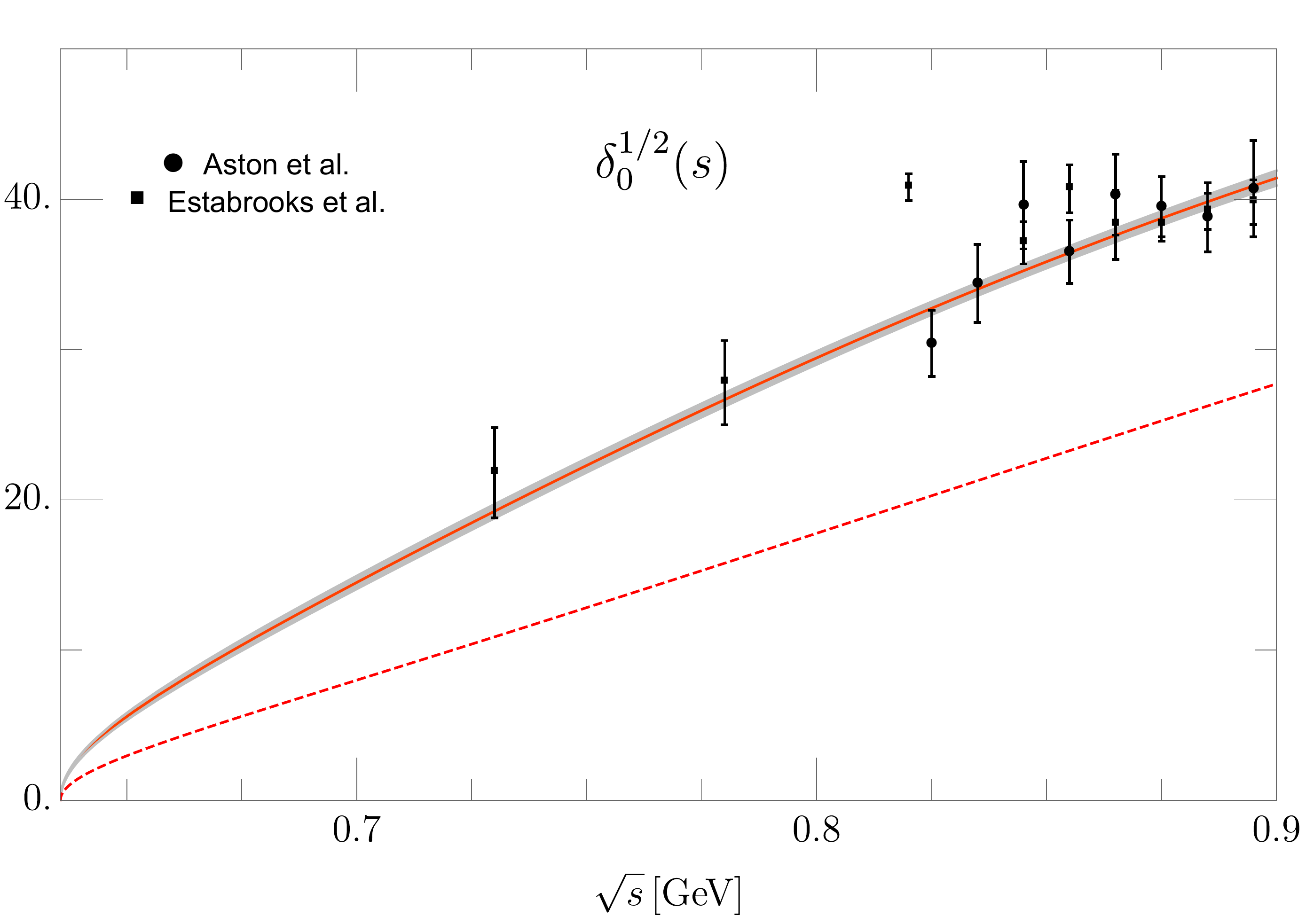} }
\caption{Unitarized LO (dashed lines) for both $\pipi$ (left) and $\pik$ (right) as described in Eq.~\eqref{eq:naiveu}, compared to the CFD parameterizations of \cite{Pelaez:2019eqa} and \cite{Pelaez:2020gnd} respectively. Note that the there is a qualitative resemblance between both shapes, even though the LO is incapable of describing the data. Both LO parameterizations contain poles that can be related to the $\sig$ and $\kap$ resonances, as shown in Eqs.~\eqref{eq:signaive} and \eqref{eq:kapnaive}.}
\label{fig:LOvsCFD}
\end{center}
\end{figure}

All unitarization schemes can be reduced in the real axis to the equations above in the elastic case, or to their matrix generalization  when several two-body channels are coupled. Unitarization is often criticised for its non-uniqueness, but the various unitarization schemes just correspond to different approximations to $\re f$. Significant differences in the predictions of different methods are only caused by the use of very crude approximations to $\re 1/f$. However, once the basic ingredients are taken into account, predictions are rather robust, particularly regarding the appearance of resonances and their poles. Note that, contrary to the fully dispersive approaches explained in previous sections, here the description of left and circular cuts for partial waves is not implemented exactly, as it is the unitarity cut (for the inverse amplitude). Therefore, there will always be some amount of crossing violation, since the cuts due to crossed channels are treated differently. With few exceptions, the left cut, if not simply neglected, is approximated perturbatively up to the desired order in ChPT. One possible way of quantifying crossing violation is by means of Roskies sum rules \cite{Roskies:1970uj}, but, as done in~\cite{Nieves:2001de}, one has to carefully avoid fitting data, since otherwise one would be measuring the crossing violation in the data, not due to the unitarization procedure.

In what follows we will first illustrate unitarization within the simplest possible case and then  we will discuss the most popular implementations.

\begin{itemize}
\item [$\bullet$] {\bf The simplest unitarized LO amplitude}. 
It is illustrative to take a look at the simplest possible calculation, whose results are shown in Fig.~\ref{fig:LOvsCFD}. Thus, we are using $f^U_{LO}(s)$, taking as input the LO partial waves in  Eqs.\eqref{eq:LET0} and \eqref{eq:LET1} in the chiral limit $M_\pi,M_K\rightarrow 0$, which then allows for very simple analytic calculations. In so doing, we find the following poles in the second Riemann sheets of the partial waves
where the $\sig$ and $\kap$ are seen:
\begin{eqnarray}
&&f^0_0:\quad \sqrt{s_{\sigma}}=(1-i)\sqrt{8\pi}f_0\simeq (463-i463)\,{\rm MeV},\quad \textrm{ \footnotesize{(LO chiral limit)}} \label{eq:signaive}\\
&&f^{1/2}_0:\;\sqrt{s_{\kappa}}=(1-i)\sqrt{64\pi/5}f_0\simeq (638-i638)\,{\rm MeV},\quad \textrm{ \footnotesize{(LO chiral limit)}}\label{eq:kapnaive}
\end{eqnarray}
where for the numerical values of $f_0^2$ we have taken $f_\pi^2$ for $\pi\pi$ and $f_\pi f_K$ for $\pi K$ scattering. The corresponding physical values are $f_\pi \simeq 92.3\,$ MeV  and $f_K\simeq1.19 f_\pi$~\cite{pdg}.
Despite the crudeness of the LO and chiral approximations, the masses of the 
$\sig$ and $\kap$ are surprisingly close to the rigorous values we so painstakingly obtained from dispersion relations. Although the widths are larger, they have the correct order of magnitude.

In contrast, for the vector partial waves we find the poles 
\begin{eqnarray}
&&f^1_1:\quad \sqrt{s_{pole}}=(1-i)\sqrt{48\pi}f_0\simeq (1133-i1133)\,{\rm MeV},\quad \textrm{ \footnotesize{(LO chiral limit)}}\\
&&f^{1/2}_1:\: \sqrt{s_{pole}}=(1-i)\sqrt{64\pi}f_0\simeq (1428-i1428)\,{\rm MeV},\quad \textrm{ \footnotesize{(LO chiral limit)}}
\end{eqnarray}
which are very far from the masses of the $\rho(770)$ and $K^*(892)$, respectively, with widths that are orders of magnitude larger than the physical ones. Restoring the physical values of the pion and kaon masses changes very little this scenario. In summery, the simplest unitarized LO approximation yields crude estimates for scalar poles that can already be identified with the $\kap$ and $\sig$, but the poles in the vector channels are completely off, although they will be well reproduced at NLO. The fact that the unitarized LO ChPT description for the $\sig$ and $\kap$ is reasonably good and so bad for the  
$\rho(770)$ and $K^*(892)$, which are dominated by the NLO contributions, is already telling us that the nature of light scalars is very different from that of the vector resonances.  In particular, the dynamics that generates the light-scalar mesons is dominated by the LO Lagrangian and, therefore, by the scale of $f_\pi, f_K$, which is
related to the scale of the spontaneous chiral symmetry breaking of QCD. This explains the problem already commented in the introduction, namely, why the $\sig$ and $\kap$, despite being the lowest lying resonances above the NGB, do not saturate the values of the LECs. They are dominated by different dynamics.

It should also be noted that this simplest unitarization scheme can also be applied to a chiral model with an explicit $\sigma$ or $\kap$ meson \cite{Black:2000qq}.
In such case one also gets a fair description of  scalar $\pi\pi$ and $\pi K$ scattering in the elastic region and their poles are once again not of the BW form.

\begin{itemize}
\item \underline{\it Simplest LO unitarization Pros:} Extreme simplicity. Most calculations can be done analytically. The lightest scalar resonances come out surprisingly well, even in the chiral limit. 

\item \underline{\it Simplest LO unitarization Cons:} 
It is just good for illustration purposes, since it only contains
information about the scales of the problem, i.e., the NGB masses and the scale of spontaneous chiral symmetry breaking. When re-expanded at low energies it only recovers the LO and the imaginary part of the NLO in the physical region. However, it misses the rest of genuine QCD dynamics, particularly the well-known quark-antiquark resonances in the vector channel.
Of course, it is limited to two-body scattering  and has no unphysical cuts. 
\end{itemize}

\vspace{.3cm}
\item [$\bullet$]
{\bf The Inverse Amplitude Method.} 
A much better description is obtained by unitarizing the full NLO ChPT partial waves, whose fit to data we show in Fig.~\ref{fig:IAMphase}. The calculations are more complicated, of course, and now the LECs introduce the genuine QCD dynamics beyond just the scale of the spontaneous chiral symmetry breaking. This improves the previous LO description of the lightest scalars, which now appear with an acceptable width, but also provides a remarkable description for the vectors. Their corresponding poles can be found in \cite{Pelaez:2004xp}. Note also the remarkable enlargement of the region where data can be described compared to standard NLO ChPT, also shown in Fig.~\ref{fig:IAMphase}.

\begin{figure}
\begin{center}
  \hspace{-.2cm}
\resizebox{\columnwidth}{!}{%
  \includegraphics{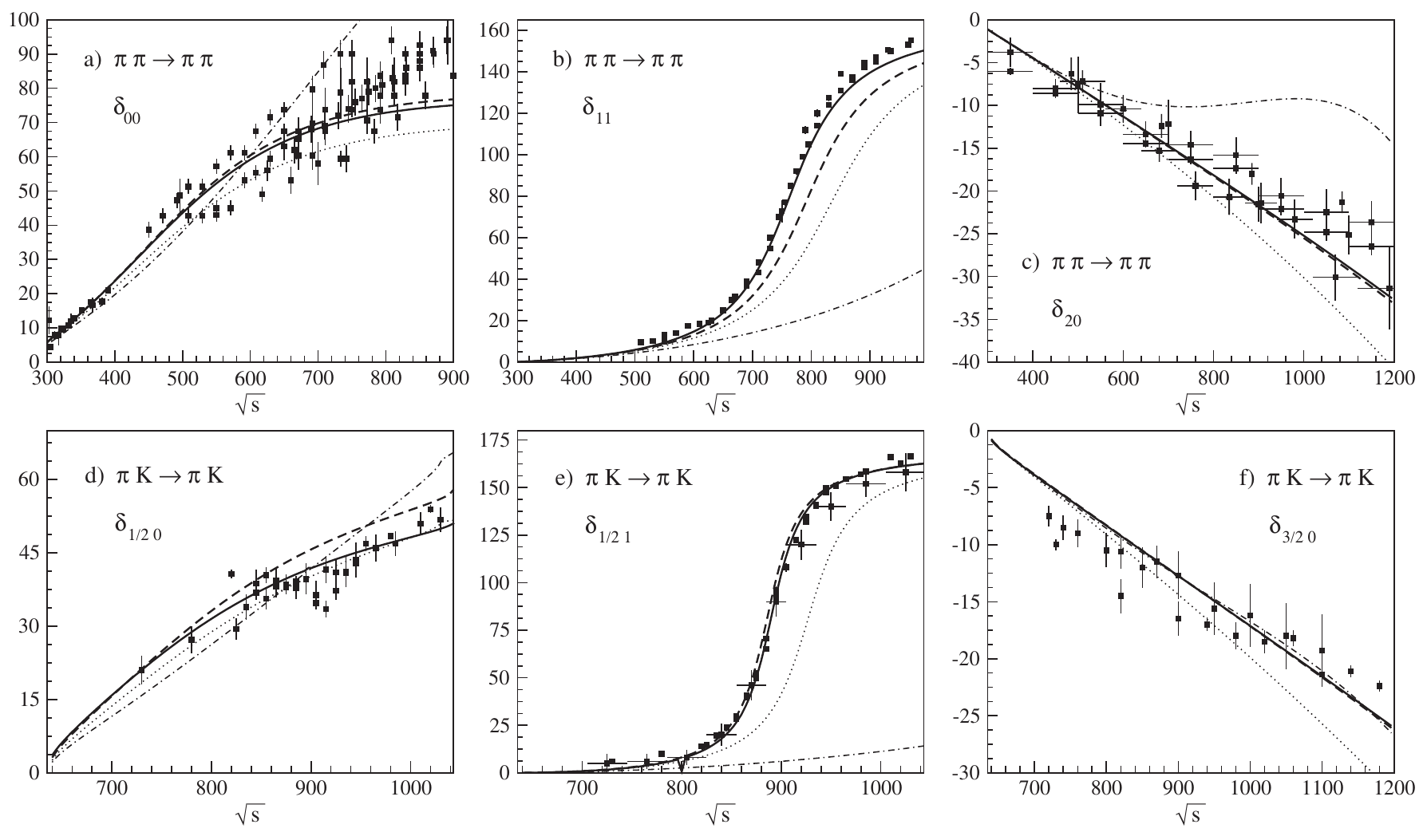} }
\caption{Phase shifts $\delta_{I\ell}$ obtained unitarizing NLO $SU(3)$ ChPT with the elastic IAM (continuous, dotted and dashed lines correspond to different LECs sets). We also show the pure NLO ChPT result as a dot-dashed line, which is almost straight in most cases.
Figures taken from \cite{Nebreda:2010wv}.}
\label{fig:IAMphase}       
\end{center}
\end{figure}

At this point, we want to ease the reader's concern about a possible caveat, which is that in the unitarized LO and NLO expressions in Eqs.~\eqref{eq:naiveu}, we are separating the real and imaginary parts of the inverse partial wave in the real axis. But the real and imaginary parts of a function are not necessarily analytic functions. Thus, one could be concerned about the  analyticity properties of the unitarized amplitude. However, the unitarized NLO can be rewritten as the NLO ``Inverse Amplitude Method" (IAM) \cite{Truong:1988zp,Dobado:1989qm,Dobado:1992ha}, namely:
\begin{equation}
f^U_{\rm IAM}(s)=\frac{f_2(s)^2}{f_2(s)-\re f_4(s)-i\sigma(s) f_2(s)^2}=\frac{f_2(s)^2}{f_2(s)-f_4(s)},
\label{eq:IAM}
\end{equation}
where in the last step we have used the perturbative unitarity relation for $f_4(s)$ that we gave in Eq.~\eqref{eq:pertunit}. Since both $f_2(s)$ and $f_4(s)$ are analytic functions in the whole complex plane except for the required cuts that we studied in the previous section, which come from $f_4(s)$, the Inverse Amplitude Method inherits the correct analytic structure.

Unitarized formulas for the elastic IAM at NNLO \cite{Dobado:1996ps,Nieves:2001de}  and  NNNLO (in the chiral limit) \cite{Dobado:2001rv} also exist. With the latter, the $f_2(1270)$ tensor resonance is reproduced, although with a somewhat small width, since inelasticity is neglected. In addition, although we have been dealing only with the elastic case here, since it suffices to describe the $\sig$ and $\kap$, it is possible to extend the IAM to the inelastic case, as long as the new accessible states are two-body states, or they can be approximated by a quasi-two-body formalism. This coupled-channel IAM derivation is completely similar to the elastic one, but now using the unitarity conditions in matrix form that we gave in Eq.~\eqref{eq:coupledunit}. The coupled channel IAM was first applied to the $\pi\pi$, $K\bar K$ system \cite{Guerrero:1998ei} and then to all possible two-meson states made of pions, kaons and etas \cite{GomezNicola:2001as}. Using LECs reasonably close to the perturbative values, the description is remarkable again, even though data is fitted up to much higher energies of the order of 1.2 GeV \cite{GomezNicola:2001as}.
The vectors $\rho(770)$, $K^*(892)$ are still nicely reproduced but now the scalars $\sig$ and $\kap$,
are accompanied by the $f_0(980)$ and $a_0(980)$, which also appear as poles in the I=0 and I=1 scalar partial waves, respectively \cite{Pelaez:2004xp}. 

The IAM has been very successful not only for low-energy  low-energy meson-meson scattering, but it has also  been applied to the thermal description of the $\sig$, $\kap$ behavior and chiral restoration \cite{Dobado:2002xf,GomezNicola:2010tb,GomezNicola:2012uc,Nicola:2013vma,Nicola:2020wxy},
or meson-nucleon interactions \cite{GomezNicola:1999pu}, as well as in the description of the pion-nucleon interaction \cite{GomezNicola:1999pu}. Moreover, it has been widely used for the electroweak symmetry breaking sector \cite{Dobado:1989gr}, which shares a similar symmetry breaking pattern but with $SU(2)$ groups, but differs in the choice of several scales. In particular, in the Higgs mechanism, where the NGBs become the longitudinal components of the massive gauge bosons, the Higgs field is narrow contrary to the $\sig$ and is a fundamental field whose mass $\simeq 125\,$ GeV is less than the scale of spontaneous symmetry breaking $v\simeq 250\,$GeV parameter, which plays the same role as $f_0$ here.

\begin{itemize}
\item \underline{\it IAM Pros:} 
It can be formulated up to any order. It contains information on the spontaneous chiral symmetry breaking  scale through the LO but also the additional genuine QCD dynamical information encoded in the LECs at higher orders. It is fully renormalized and, therefore, contains no spurious parameters. 
As we will see below, this is very relevant to establish a connection to fundamental QCD parameters like the quark masses and the number of colours, which allows us to  study the nature of resonances.
When re-expanded at low energies it recovers the ChPT series up to the order that has been used as input as well as, in the physical region, the imaginary part of the next order.
Since it is expressed in terms of the ChPT expressions for the partial waves, in the elastic case, it has the correct analytic structure in terms of cuts in the complex plane. 
It can be easily extended to the coupled channel case.
It is able to reproduce the scalar nonet and those resonances in the lightest vector multiplet that decay to two mesons.
We will see below that the elastic case can be derived from dispersion theory. This dispersive derivation has an additional term, easy to implement, in general negligible numerically, except in the vicinity of the Adler zeros, below the physical region of the scalar partial waves.

\item \underline{\it IAM Cons:} 
It requires the fully renormalized ChPT amplitudes. The left and unphysical cuts  are only perturbatively correct up to the order used in ChPT. For this reason, it is not so good for precision as the fully dispersive formalism. Nevertheless,
despite the general belief that this crossing violations may be large, 
it was found in \cite{Nieves:2001de} using the Roskies sum rules discussed above for the IAM that:  ``Generally speaking, they are not very large in percentage terms", when uncertainties allowed to draw conclusions. 
Finally, there is no dispersive IAM derivation in the coupled channel case, for which it should only be used in the physical region of all the opened coupled channels. 
\end{itemize}

\vspace{.3cm}
\item[$\bullet$]
{\bf The Chiral Unitary Approach.}
Note that, in principle, the IAM requires the fully renormalized ChPT calculation up to a given order. 
In practice, however, for specific processes, regions or channels, many contributions can be numerically negligible or can be absorbed in additional  parameters. This is no surprise,  after all we have just seen that  only with the LO the light-scalar poles were 
not too far from their values. This is due to the fact that these are wide states whose formation dynamics is dominated by the LO vertices and unitarity. Sometimes it is said that these states are ``dynamically generated", meaning that  for this kind of processes the relevant numerical contribution comes from meson loops in the s-channel.
Actually, in \cite{Oller:1997ti} it was shown that meson-meson scattering in scalar partial waves and the first nonet of scalars could be very nicely described using a unitary coupled channel formalism and 
just LO vertices. The authors showed that, for scalar waves,  the vertices  of the 
s-channel one-loop diagrams (Fig.~\ref{fig:ChPTdiagrams}, c) had an off-shell part that could be absorbed in the masses and decay constants together with tadpole diagrams, as well as an on-shell part that could be factorized outside the momentum integral of the loop. The crossed loops could be neglected numerically.
This means that they could approximate the chiral expansion as $F\simeq F_2+ F_2 G F_2...$, where G is a diagonal matrix whose elements are:
\begin{equation}
 G_{ii}(P)=i \int \frac{d^4 q}{(2 \pi)^4} \frac{1}{q^2-m^2_{1i}+i\epsilon}
\frac{1}{(P-q)^2-m^2_{2i}+i\epsilon},
\label{Gii}
\end{equation}
i.e., the loop function  with mesons 1 and 2 without the vertices, 
whose imaginary part is $\im G= \Sigma=- \im F^{-1}$
Here $P$ is the CM four-momentum.
Recalling that $F_2$ is real above all thresholds in the open channels of $F$, we find that 
$\re F^{-1}\simeq F_2^{-1}- F_2 {\text Re} G F_2...$. All together, we can thus write the unitarized partial-wave matrix of Eq.~\eqref{eq:coupledunit} as:
\begin{equation}
F=F_2[F_2-F_2 G F_2]F_2=[1-F_2 G ]^{-1} F_2,
\label{eq:ChUA}
\end{equation}
which is known as the Chiral Unitary Approach (ChUA). 
The huge advantage is its simplicity, because  only the LO terms, which are usually tree-level calculations, are needed. 

Observe that, once again, all the ingredients are analytic functions, since by replacing $\sigma(s)$ in the denominator by $G_{ii}$, we have avoided taking real and imaginary parts. Moreover, the spurious zero at $s=0$ that appears with the simplest unitarization when using the $\sigma(s)$ function is gone now by using  $G_{ii}(s)$ instead. 
Actually it was well known since the sixties that it is very simple to write an analytic function whose imaginary part on the real axis is 
$\sigma_i(s)$. For example, it can also be achieved with the Chew-Mandelstam function \cite{Chew:1960iv}
\begin{equation}
J_i(s)=\frac{s-s_i}{\pi}\int_{s_{th}}^\infty ds'\frac{\sigma_i(s)}{(s-s_i)(s-s')}=\frac{-\sigma_i(s)}{\pi}\log\frac{\sigma_i(s)-1}{\sigma_i(s)+1}.
\end{equation}

The caveat is that  $G_{ii}$ is divergent and has to be regularized, but since many diagrams and the NLO terms have been dropped in the calculation, this regulator cannot be reabsorbed in the parameters of the theory and is kept as an additional parameter. However this caveat ends up becoming a phenomenological advantage since it can be tuned in the fit to improve the performance of the simpler pure LO formalism in Eq.~\eqref{eq:naiveu}.
Any regularization method is in principle acceptable, along with the inclusion of subtractions as we have done for $J_i(s)$ above, which generate constants that become parameters in the real part.
However, when the regulator is a simple cutoff, it was shown in \cite{Oller:1997ti} that, in order to describe the scalar nonet, it had a value of several hundred MeV in momenta or $\simeq 1\,$GeV in energy, which once again suppresses the dynamics beyond that of the scale of spontaneous chiral symmetry breaking. This is why resonances generated this way are dominated by meson-meson dynamics. 
Additional dynamical information beyond the scale of symmetry breaking can only be taken into account by the ChUA once the LECS are reintroduced back (through diagrams b in Fig.~\ref{fig:ChPTdiagrams}). When this is done \cite{Oller:1998hw} the ChUA reproduces the scalar nonet as well as the $\rho(770)$, the $K^*(892)$ 
and their associated poles \cite{Oller:1998hw}, or even the octet part of the $\phi(1020)$ \cite{Oller:1999ag}, responsible for its two-body decay.

It is also interesting to notice that Eq.~\eqref{eq:ChUA} can be reinterpreted as a {\it resummation}
of the geometric series $F\simeq F_2+ F_2 G F_2+F_2 G F_2 G F_2+F_2 G F_2 G F_2 G F_2+...$.
This would be a re-summation of s-channel ``bubble diagrams'', factorizing on-shell the LO vertices from each bubble made of two-mesons in a loop.
Therefore, the ChUA could be reinterpreted as a solution of the Bethe-Salpeter equation $T=T_2GT$, obtained assuming on-shell factorization. For this reason, very often unitarization methods are called re-summations, but this is only the case of some particular approximations neglecting left and circular cuts.
 A detailed Bethe-Salpeter treatment of LO and NLO ChPT unitarization can be found in \cite{Nieves:1998hp,Nieves:1999bx}.
 
 This approach, or variants of it, has  become the most popular unitarization method for chiral Lagrangians. The reason is its simplicity, which allows for it to be applied even when no such a rigorous expansion and counting as that of meson ChPT exists. The leading order is enough and, when the dominant dynamics is due to the ``bubble loops", the rest of the effects can be easily mimicked by the presence of the cutoff or other regulators or subtraction constants, which would crudely approximate the role of the LECs. The method is thus particularly well suited to identify or describe the so-called ``dynamically generated" states, meaning states where the meson-meson interaction scale is the most  relevant in their formation. 
 Moreover, it is possible to extend its applicability beyond the pure ChPT Lagrangian and include explicitly other resonances in the Lagrangian, as long as their formation dynamics is not that of meson-loops.
 Sometimes these are called ``pre-existing" resonances, in the sense that they would exist even before turning the unitarization of meson-meson dynamics.
 Unitarization will then give these resonances a width and will provide the other resonances, which, like the $\sig$ and $\kap$ are ``dynamically generated" from meson loops. This allows to extend the unitarized chiral description to even higher energies, as long as they are dominated by these additional ``preexisting" states
 \cite{Oller:1998zr,Albaladejo:2008qa}.
 
 The ChUA has thus been applied, with different variants, in numerous systems besides meson-meson interactions. These are beyond the scope of this tutorial review, but we refer to \cite{Oller:2000ma,Oller:2020guq} and references therein.

\begin{itemize}
\item \underline{\it ChUA Pros:}
 Remarkably simple. It only requires tree-level calculations. It generates nicely the resonances which are dominated by the meson-loop interaction dynamics, i.e., the spontaneous chiral symmetry breaking scale. These are the relevant ones for the usually called ``dynamically generated" states, which do not appear directly from naive quark models. It can be applied even if no chiral-counting is well defined, by using chiral models or just the leading order in an effective chiral lagrangian. Hence it has become very popular even beyond the realm of NGB scattering. It also provides the link with Bethe--Salpeter-like equations and can be easily understood as a resummation of s-channel bubble diagrams. Additional non-``dynamically generated" states can easily be added to the description.

\item  \underline{\it ChUA Cons:}
 It is not fully renormalized, so that there is always some regulator (cut-off, scale or subtraction constant) dependence left. But that regulator comes together with a mild energy-dependent function that approximates the higher energy dynamics. For systems dominated by chiral loops such a cutoff momentum comes out with  a natural size of several hundred MeV up to 1 GeV.
 Also, the ChUA neglects the left and circular cuts. For the $\sigma$ and $\kap$ it therefore cannot be used for precision studies, although their poles come reasonably close to the physical values.
 \end{itemize}

\end{itemize}

Of course, the reader may have already noted an apparent inconsistency between the naive derivation 
we have provided in this section 
and the uses of unitarized ChPT. Namely, we have happily used the chiral expansion to approximate $\re F^{-1}$, which is only justified at low energies below the resonance region, but then we have used the unitarized expression in the resonance region. Why would the chiral expansion of $\re F^{-1}$ be reliable beyond the ChPT applicability regime?. Of course, as such it does not make sense. However, it works, and the reason is that the unitarized ChPT expressions can be derived in a more rigorous way from dispersion theory, making it clear why it is used beyond the convergence region of ChPT.

\subsection{Unitarization: the dispersive justification.}

In this subsection we will not review other unitarization methods, but justify them 
using dispersion relations for elastic partial waves. For simplicity let us then suppress the isospin and angular momentum indices and recall that for a 
partial wave $f(s)$, we can write a
dispersion relation (we subtract three times,
since we will also use it for the ChPT NLO term $f_4$(s), that grows with $s^2$)
\begin{equation}
f(s)=C_0+C_1s+C_2s^2+
\frac{s^3}\pi\int_{s_{th}}^{\infty}\frac{\im f(s')ds'}{s'^3(s'-s-i\epsilon)} + UC(f).
\label{disp}
\end{equation}
Note we have only written explicitly the integral over the physical cut,
extending from threshold, $s_{th}$, to infinity, 
whereas  the integrals over the
unphysical cuts ( left cut from the square of the mass difference of the incoming mesons to $-\infty$ and the circular cut when it exists) 
are abbreviated by $UC$.
The same cut structure is inherited by the inverse of the amplitude.
Following \cite{Dobado:1992ha}, let us define an auxiliary function $G(s)=f_2(s)^2/f(s)$
that should then satisfy a dispersion relation similar to \ref{disp}, since $f_2(s)$ is real and has no cuts, namely:
\begin{equation}
G(s)=G_0+G_1s+G_2s^2+     \\   \nonumber
\frac{s^3}\pi\int_{s_{th}}^{\infty}
\frac{\im \,G(s')ds'}{s'^3(s'-s-i\epsilon)}+UC(G)+PC,
\label{Gdisp}
\end{equation}
where now $PC$ stands for possible pole contributions in $G$ coming from 
zeros in $f(s)$.  Now, the subtraction constants terms come from the expansion of the function at $s=0$, where it is justified to expand them using the ChPT series.
In addition, as already seen in Eq.~\eqref{eq:elunit} in the elastic physical region $\im (1/f(s))=-\sigma(s)$,
so that by definition $\im G= -\sigma f_2^2$.
However, we already saw in Eq.~\eqref{eq:pertunit} that $\im f_4(s)=\sigma f_2^2(s)$, which produces as a result
$\im G(s)= -\im f_4(s)$. In addition,
 up to NLO $UC(G)\simeq -UC(f_4)$,
whereas $PC$ is of higher order and we will momentarily neglect it.
Thus, we can rewrite the previous dispersion relation as:
\begin{eqnarray}
\frac{f_2^{2}(s)}{f(s)}&\simeq& a_0+a_1s-b_0-b_1s-b_2s^2 
-\frac{s^3}\pi\int_{s_{th}}^{\infty}\frac{\im f_4(s')ds'}{s'^3(s'-s-i\epsilon)}-LC(f_4)
\nonumber  \\   
\frac{f_2^{2}(s)}{f(s)}&\simeq& f_2(s)-f_4(s), \quad \Rightarrow f(s)\simeq
\frac{f_{2}^2(s)}{
f_{2}(s)-f_{4} (s)}, \nonumber
\label{preIAM}
\end{eqnarray}
where the second step follows because the $a_i, b_i$ are the coefficients of the ChPT expansion of $G_i(s)\simeq f_2(s)-f_4(s)...$ 
around $s=0$. They  are the subtraction terms of a dispersion relation for
$f_2(s)-f_4(s)$.
Thus we have re-derived the elastic IAM expression already presented  in Eq.~\eqref{eq:IAM}.

The $PC$ contribution has been calculated  explicitly \cite{GomezNicola:2007qj} and is not just formally suppressed, but numerically negligible
except near the Adler zeros, away from the physical region.
Only when the energy is close to the Adler zero one should use
a slightly modified version of the IAM \cite{GomezNicola:2007qj} taking care of that term.

Thus, as we saw in the previous subsection, naively, the IAM 
looks like replacing $\re  f^{-1}$ by its $O(p^4)$ 
ChPT expansion in Eq.~\eqref{eq:elunit}, but 
 Eq.~\eqref{eq:elunit} is only valid in the real axis, whereas our derivation allows us to consider the amplitude in the complex plane and look for poles associated to  resonances.
Note that {\it we have not used ChPT on the right, physical cut, which is exact for $1/f$ in the elastic approximation.} We have used ChPT only to expand the subtraction constants, which is justified since it is a calculation at $s=0$,
and to approximate the unphysical cuts. 
This later approximation is also not a problem for the circular cut, which for $\pi K$ scattering  lies at relatively low values of $s=m_K^2-m_\pi^2$, and hence it is always less than the absolute value of the threshold energy.
However, we have used the NLO approximation of the integrand of the let cut, which extends to $s=-\infty$.
Nevertheless, we should note that these are very subtracted integrals, heavily weighted on the low-energy region. In addition, the left cut integrand has another $1/(s'-s)$ factor
which suppresses the contribution when the dispersion relation is integrated far from it.
This is actually the case of the resonance region, which generically lies quite far from the left cut, although the $\sig$ and $\kap$ will be the states more affected by this approximation. 

Therefore, although it may look like we are using the ChPT series inside the IAM in the resonance region, we are actually just
using it to approximate the low energy constants, or to evaluate the unphysical contributions to the integrals. These were anyway dominated by the low-energy region, and are also suppressed far from it. This means that it is well justified to use the IAM expression in the physical region and in search for poles, although maybe not for precision studies. Of course, one should avoid calculations near the left cut using unitarization.  Recent attempts to systematize the uncertainties in the use of the IAM and modifications to circumvent special cases can be found in~\cite{Molina:2020qpw,Niehus:2020gmf,Salas-Bernardez:2020hua}. 
According to \cite{Salas-Bernardez:2020hua} in the scalar case for meson-meson interactions ``the IAM seems to be faring even better than we have a right to state with the uncertainty analysis that we have carried out".

Unfortunately, there is not such a dispersive derivation for the coupled channel IAM. As a matter of fact, all coupled channel unitarization methods in which the unphysical cuts, or an approximation to them, are introduced in the denominator of T through $\re T^{-1}$  in Eq.~\eqref{eq:coupledunit}, yield an incorrect analytic structure for the other cuts. This is independent on their possessing the correct two-body unitarity cuts.
The reason is easy to understand with the simplest example of the $\pi\pi$ and $K \bar K$ systems.
All possible scattering partial-waves between these two states have the same right-hand cut singularity structure. We have already seen in Fig.~\ref{fig:anstruc} that the left cut in $\pi\pi\rightarrow\pi\pi$ partial waves extends from $-\infty$ to 0, and similarly occurs for $\pi\pi\rightarrow K\bar K$. However, for the $K \bar K\rightarrow K\bar K$ amplitude the left cut extends to $4(m_K^2-m_\pi^2)$ \cite{Guerrero:1998ei,GomezNicola:2001as}. Therefore, in the calculation of the determinant needed to obtain the inverse in Eq.~\eqref{eq:coupledunit}, the $K \bar K$ scattering left-hand-cut singularity propagates to the whole unitarized matrix and appears in the resulting partial waves for $\pi\pi\rightarrow\pi\pi$ and $\pi\pi\rightarrow K\bar K$, where it should not have been. For the case at hand, the numerical effect of this cut is not very large,  and, with appropriate care, one can still get fairly reasonable results \cite{Guerrero:1998ei,GomezNicola:2001as} below the $K\bar K$ threshold.
The appearance of these spurious singularities is not characteristic of the IAM only, but also of the Bethe-Salpeter method as shown in~\cite{Ledwig:2014cla}.

Nevertheless, if one wants to get a dispersive justification of coupled channel unitarization procedures in matrix form, one can still implement the two-body unitarity cuts exactly in the inverse of a matrix, say as a ``denominator" $D$, 
but the left hand cuts should appear multiplying, not in an inverse, say as a ``numerator" $N$.
This is the reason why for the coupled channel formalism the so-called $N/D$ method \cite{Chew:1960iv}, which makes explicit such a separation, is more appropriate. However, it is beyond our scope, since here we are interested in the $\sig$ and $\kap$ for which the elastic formalism is more than enough. Suffices to say that the $N$ and $D$ pieces satisfy each a dispersion relation that has the other one as input and there is some freedom in multiplying both by analytic functions, in the number of subtractions used, etc.... Despite being devised to implement the left cut through $N$, the method is still fairly complicated and very often the discontinuity over the left cut is set to zero, leading to a coupled channel unitary approach relatively similar to the ChUA, with very similar results for the $\sig$ and $\kap$ \cite{Oller:1998zr,RuizdeElvira:2018hsv}.
Notwithstanding, this method is more versatile due to the factors that can be used to multiply both $N$ and $D$. Furthermore, note that the N/D representation can be easily obtained in configuration space through a decomposition of the wave-equation solutions in terms of Jost functions, which in turn can be computed by coarse-graining the short-distance interactions \cite{RuizdeElvira:2018hsv}. In this way the left cut is implemented automatically from the spectral representation of the corresponding potential.
For a recent review on the N/D method in the context of unitarized ChPT we refer to the recent review by Oller \cite{Oller:2020guq}.

\subsection{Mass and $N_c$ dependence of the $\sig$ and $\kap$ from unitarized ChPT}

A very relevant feature of unitarized ChPT compared to the fully dispersive approach, is that it is possible to explore the dependence of the $\sig$ and $\kap$ on the quark masses and the number of colors.
These dependences shed light on the nature or composition of these states in terms of quarks and gluons~\cite{Pelaez:2003dy,Pelaez:2006nj,Hanhart:2014ssa,RuizdeElvira:2017aet}.

\begin{figure}
\centering
  \hspace{-.2cm}
\raisebox{-0.5\height}{\includegraphics[height=4.4cm]{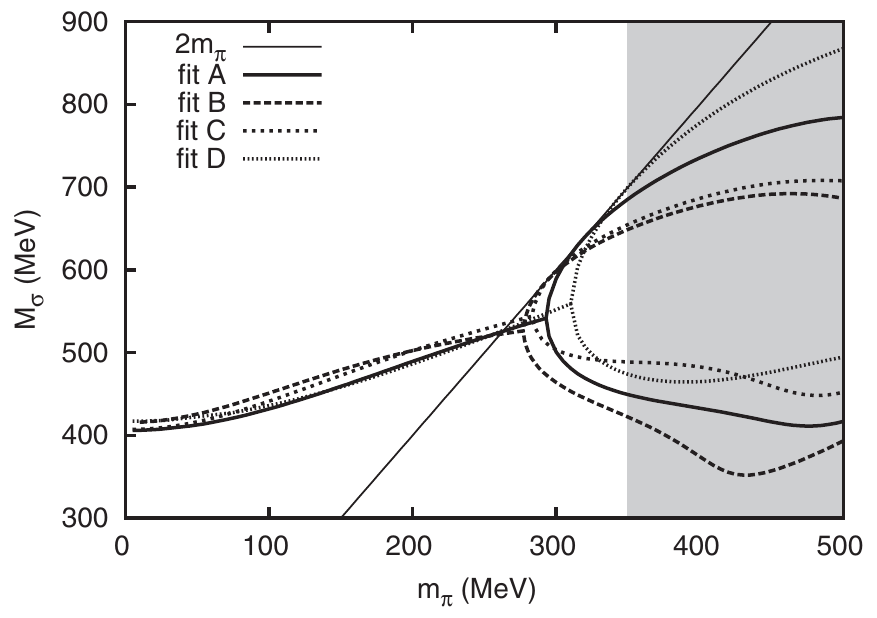}}
 \hspace{-.1cm}
\raisebox{-0.5\height}{\includegraphics[height=5.cm]{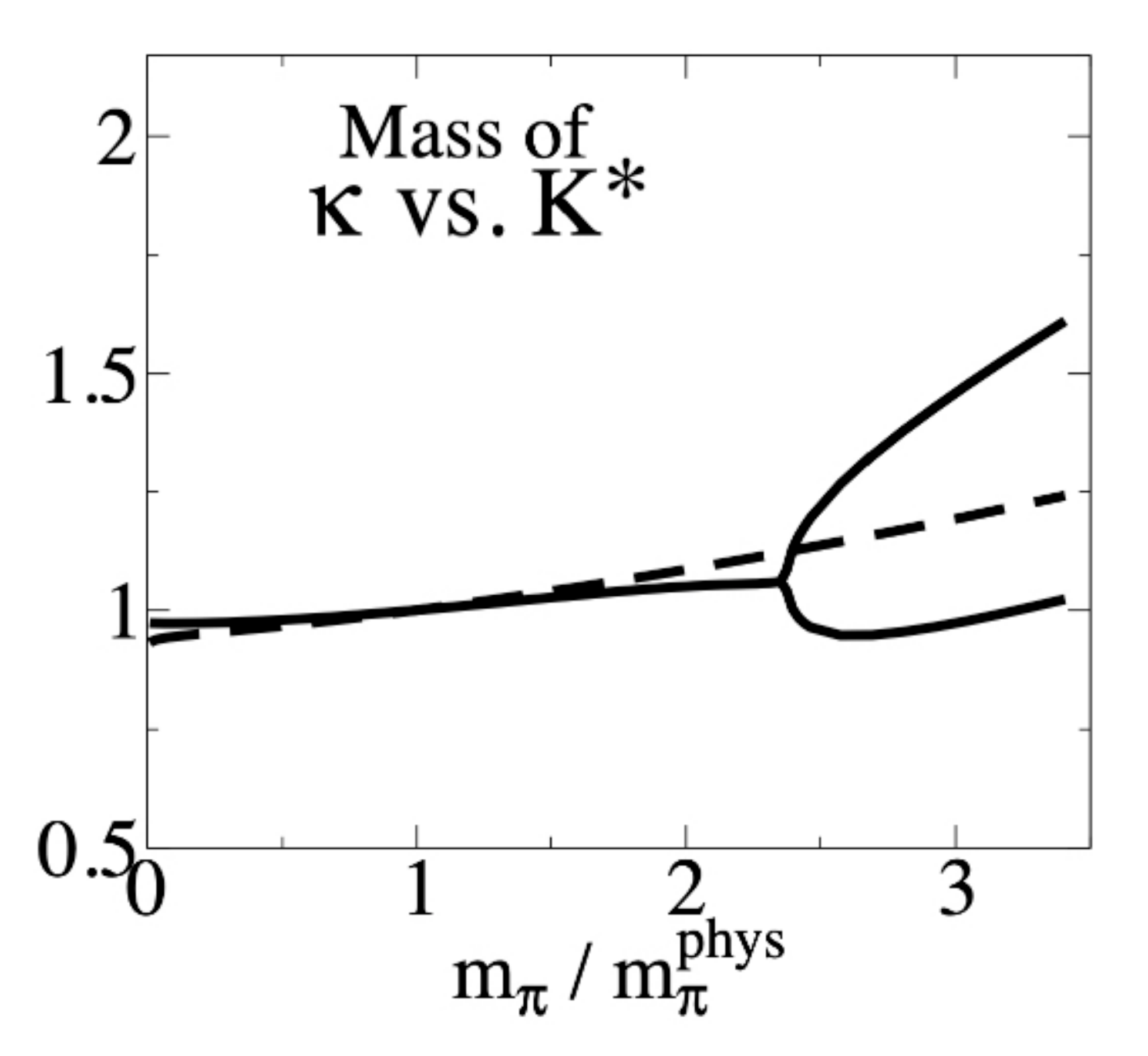} }
\caption{Left: Dependence of the sigma mass $M_\sigma$ on the pion mass,
from the NNLO (two-loops) IAM \cite{Pelaez:2006nj}. Different curves represent different 
fits on \cite{Pelaez:2006nj}. The thin continuous line shows the $2m_\pi$ threshold.
Right: $m_\pi$ dependence of the 
 $\kap$ (solid line) and $K^*(892)$ (dashed line) masses \cite{Nebreda:2010wv}.
All masses and widths are defined from the pole positions
as obtained from NLO IAM fits. 
Figures taken from \cite{Pelaez:2010fj} (left) and \cite{Pelaez:2010je} (right).
\label{fig:NNLOykappa}}
\end{figure}

Changing the quark masses in the ChPT Lagrangian is rather simple and it translates into a change of the meson masses. At leading order this is trivial, as can be seen in Eq.~\eqref{ec:Lag2}. To higher orders, the decay constants  and the masses both have additional contributions that depend perturbatively on the LECS and the masses themselves. Note that  the LECs do {\it not} carry any dependence on the quark or meson masses. As an illustration of this approach, in Fig.~\ref{fig:NNLOykappa} we can see the expected behavior for the $\sig$ (left, obtained with the $SU(2)$ NNLO IAM) and $\kap$ (right, obtained from the $SU(3)$ NLO IAM) as a function of the pion mass. 
First of all, one should recall that by making the pion or kaon masses larger, the ChPT convergence necessarily worsens. It has been estimated that beyond 350 MeV the results are at most qualitative~\cite{Niehus:2020gmf}, but it will probably stop being precise much before, providing just semi-quantitative results. 

There are, however, some generic features of this quark-mass dependence observed for the $\sig$ at NLO \cite{Hanhart:2008mx} and NNLO \cite{Nebreda:2011di}
and $\kap$ at NLO \cite{Nebreda:2010wv}.  The mass of these two scalar resonances grows with the pion mass, although slower than the two-pion threshold, which means that their width, or better the imaginary part of their pole position, becomes narrower and narrower. At some value of the pion mass between roughly twice and three times its physical value, the two-pion threshold lies above the pole-mass of these resonances.
At this point the behavior of scalar poles differs dramatically from poles in other waves. 
It can be shown that, generically, the width of resonances in waves with angular momentum higher than 0,
tends to zero at that point, and their pair of conjugated poles meet at threshold \cite{Hanhart:2008mx,Hanhart:2014ssa}. This is the case of the $\rho(770)$ and $K^*(892)$, 
which are Breit-Wigner like and their width is therefore proportional to some power of the available momentum. If the pion mass keeps increasing, one of the poles jumps to the first sheet, whereas the other one remains at a symmetric position in the second sheet, both below threshold. This is a bound state. However, this is not what happens for the $\sig$ and $\kap$. Since they appear in scalar waves, their pairs of conjugated poles can, and actually do, meet in the second sheet below threshold. After that, one of them approaches threshold, whereas the other one moves away from it, both staying always in the real axis. These are the two branches that are observed in Fig.~\ref{fig:NNLOykappa}. For some values of the pion mass, we therefore have two poles in the real axis of the second Riemann sheet in very asymmetric position with respect to threshold. The closest one to threshold, which influences the most the physical region, is known as a ``virtual" or quasi-bound state.  It keeps on moving towards threshold, now sufficiently fast to reach it and jump to the first Riemann sheet. When it does so, it also becomes a bound state, but now its counterpart in the second sheet is in a rather different position. The more asymmetric their positions the more predominant is the ``molecular" nature. i.e., a lightly bound state whose constituents are the particles involved in the scattering process. As a technical remark, note that since we are dealing with poles below threshold, in this case it is important to use the modified-IAM \cite{GomezNicola:2007qj} including the $PC$ contribution that we momentarily neglected after Eq.~\eqref{Gdisp}, which takes into account the Adler zero properly.

In conclusion, unitarized ChPT seems to indicate that, at high pion masses, both the $\sig$ and $\kap$ are closer to molecular states made of two mesons than to ordinary mesons whose formation dynamics are quark interactions.

It is also worth noticing that calculations at unphysical masses are of relevance for lattice QCD.  Note that lattice does not provide results on the complex $s$ plane, but the result for physical energies have to be continued once again to the complex plane with different methods.
The first full lattice calculation to find a $\sigma$ used pion masses well above the ones reviewed here \cite{Kunihiro:2003yj} but the same group had a later work \cite{Wakayama:2014gpa}, supporting the molecular picture for the $\sigma$ at those very large pion masses. A bound isoscalar state is also found for $m_\pi\simeq 325$ and $391\,$MeV in \cite{Prelovsek:2010kg} and \cite{Briceno:2016mjc}, respectively, qualitatively consistent with unitarized ChPT. For $m_\pi=236\,$MeV \cite{Briceno:2016mjc} the lattice phase shifts described with different parameterizations are consistent with a pole now in the second Riemann sheet. 
  A similar lattice calculation \cite{Guo:2018zss}, analyzed using unitarized ChPT, found a 
  compatible pole result. Therefore, it  is likely that for pion masses between 200 and 300 MeV in the lattice the  $\sig$ resonance
  should pass through the virtual-state stage. Actually, such a virtual state was found  for the $\kap$ pole  when analyzing $\pi K$ scattering on the lattice at $m_\pi=391$ and $230\,$MeV in \cite{Dudek:2014qha}
  and \cite{Brett:2018jqw}, respectively. These results are again in qualitative agreement with unitarized ChPT. Note, however, that as the pion mass becomes lighter, the determination of the $\sig$ and $\kap$ pole is plagued again with similar instabilities \cite{Briceno:2017qmb,Wilson:2019wfr,Guo:2018zss} as in the case with physical data,  since it requires an extrapolation deep in the complex plane, where once again the unphysical singularities and Adler zeros may play a significant role. For an illustration of the relevance of Adler zeros in the $\kap$ determination from lattice, see \cite{Rendon:2020rtw}. It is therefore likely that a dispersive ``data driven" approach of the kind explained in section \ref{sec:DR} above,  may be relevant for a robust extraction of the pole from lattice results.

\begin{figure}
\begin{center}
  \hspace{-.2cm}
\resizebox{.6\columnwidth}{!}{%
  \includegraphics{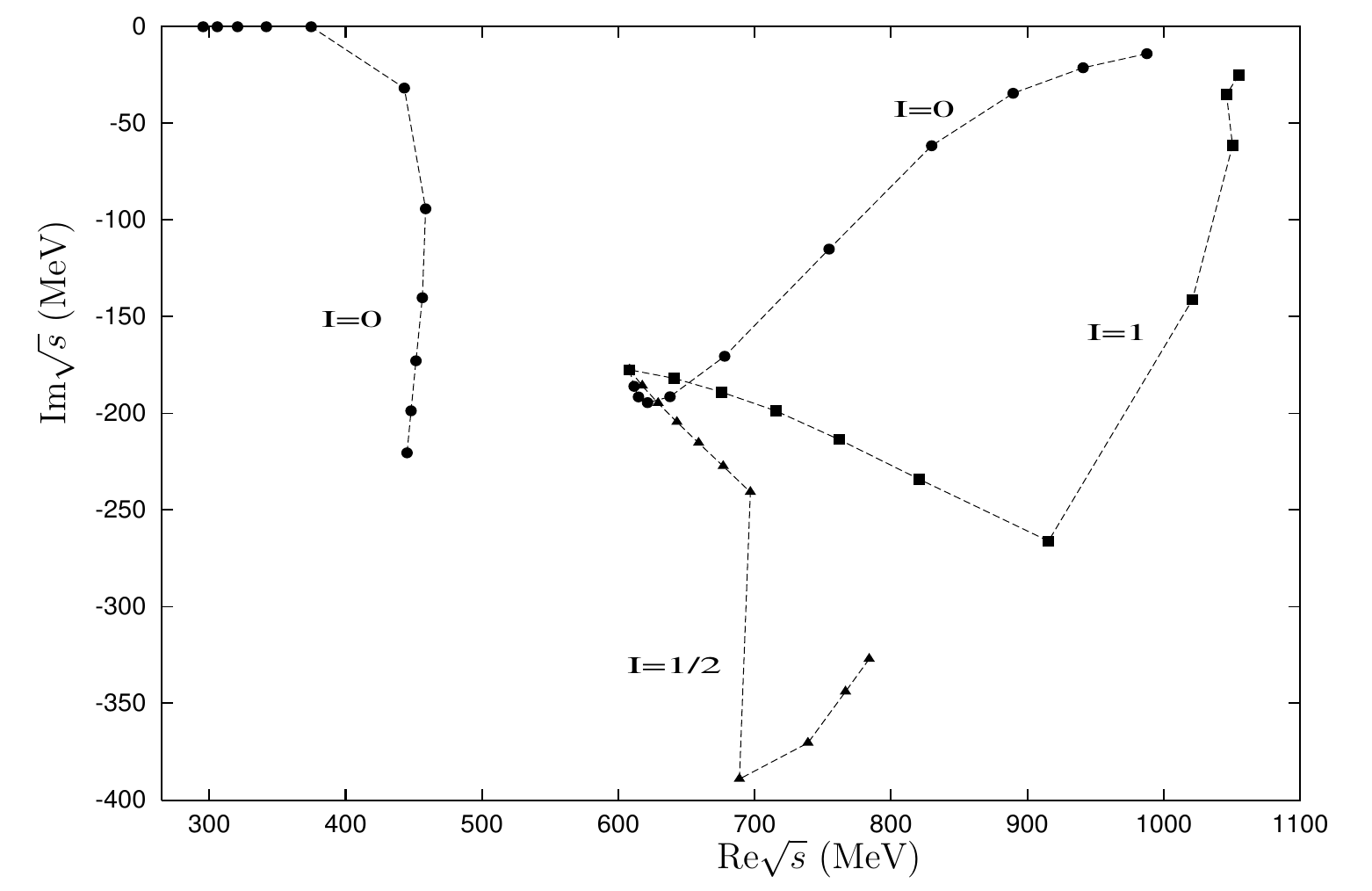} }
\caption{Trajectories of the poles that appear in coupled channel unitarized amplitudes of different isospin
as the pion, kaon and eta masses are varied from their physical values to a common value of 350 MeV \cite{Oller:2003vf}. This shows that the 
lightest scalars actually belong to a nonet in the $SU(3)$ limit.
The two trajectories with $I=0$ correspond to the singlet and octet states, not directly to the poles of the
$\sigma$ or $f_0(980)$ resonances, which are a mixture of these two.
 Figure taken from \cite{Oller:2003vf}. }
\label{fig:equalmasslimit}
\end{center}
\end{figure}

It is also possible to change the strange-quark mass, but since it is already relatively large, it cannot be made much larger due to the convergence of the series. In addition, it cannot be made too light if one wants to use the elastic IAM, which is the one well justified from dispersion theory and has no spurious left cuts. Within this regime, the evolution of the $\sig$ and $\kap$ poles is rather smooth \cite{Nebreda:2010wv}. Of course, since we already know that left cuts only provide a correction, it is possible to use the ChUA and change the pion and kaon masses so that they tend to a common value.
One also has to assume the cutoff to remain constant, which is probably a good approximation, since we have seen it mimicks the high-energy and LEC effects. Within this approximation, it was possible to show that \cite{Oller:2003vf} the $\kap$ pole, as well as a combination of the $\sig$ and $f_0(980)$ become degenerate with the $a_0(980)$ poles, as illustrated in Fig.~\ref{fig:equalmasslimit}. This strongly supports the assignment of these states to the same
octet and a clear identification of the members of the lightest scalar nonet.

Finally, another relevant feature of the $\sig$ and $\kap$ revealed by unitarized ChPT is their behavior under changes in the number of colors of QCD. This is of relevance since, at leading order in the $1/N_c$ expansion \cite{tHooft:1973alw,Witten:1980sp},  the mass of ordinary mesons behaves as  $M\sim O(1)$ whereas their width goes as $\Gamma \sim O(1/N_c)$. Customarily  $q \bar q$ states are understood as ordinary mesons, but it was recently shown \cite{Weinberg:2013cfa,Knecht:2013yqa} that genuine tetraquark states also have at least that same $N_c$ behavior, which is even more suppressed for glueballs or special tetraquark configurations~\cite{Cohen:2014vta} \footnote{While this work was in the referee process a review has appeared, where large-$N_c$ behavior of tetraquarks is reviewed in great detail \cite{Lucha:2021mwx}.}
So, here by ordinary mesons we mean those with the same $N_c$ behavior as $q \bar q$ states. 

Back to ChPT, the $1/N_c$ leading order of its parameters is known from a model-independent analysis: pion and kaon masses are $O(1)$, the decay constants scale like $f_0\sim \sqrt{N_c}$ and the NLO LECs behavior, compiled in \cite{Pelaez:2015qba}, is either $O(1)$ or $O(N_c)$ \cite{Gasser:1984gg,Peris:1994dh}. It is fairly simple to change the $N_c$ behavior in the ChPT Lagrangian: let us call $p$ one of these parameters, whose behavior is $O(N_c^k)$ and simply change its value to $p\rightarrow p (N_c/3)^k$. As long as we are using ChPT, this is model independent. We can now use those new parameters into the unitarized ChPT formalism and follow the movement of the poles associated to each resonance in the complex plane as $N_c$ is increased.
This provides the leading $1/N_c$ behavior of resonances.

For illustration we can just look at the simplest unitarization scheme, with only LO input, that we 
presented in \ref{sec:naive}. In Eqs.~\eqref{eq:signaive} and \eqref{eq:kapnaive} we found that in the chiral limit, the masses and widths of the $\sig$ and $\kap$ only depend on $f_0\sim O(\sqrt{N_c})$ and are proportional to it. Therefore, within this approximation, none of these poles behave like an ordinary meson. This is not too surprising, since the LO only depends on the scale of chiral symmetry dynamics, and is just governed by meson-meson physics through their  loops, which are suppressed with respect to the dominant LECS that provide the information on ordinary resonances.

\begin{figure}
\begin{center}
  \hspace{-.3cm}
\raisebox{-0.5\height}{%
  \includegraphics[height=3.cm]{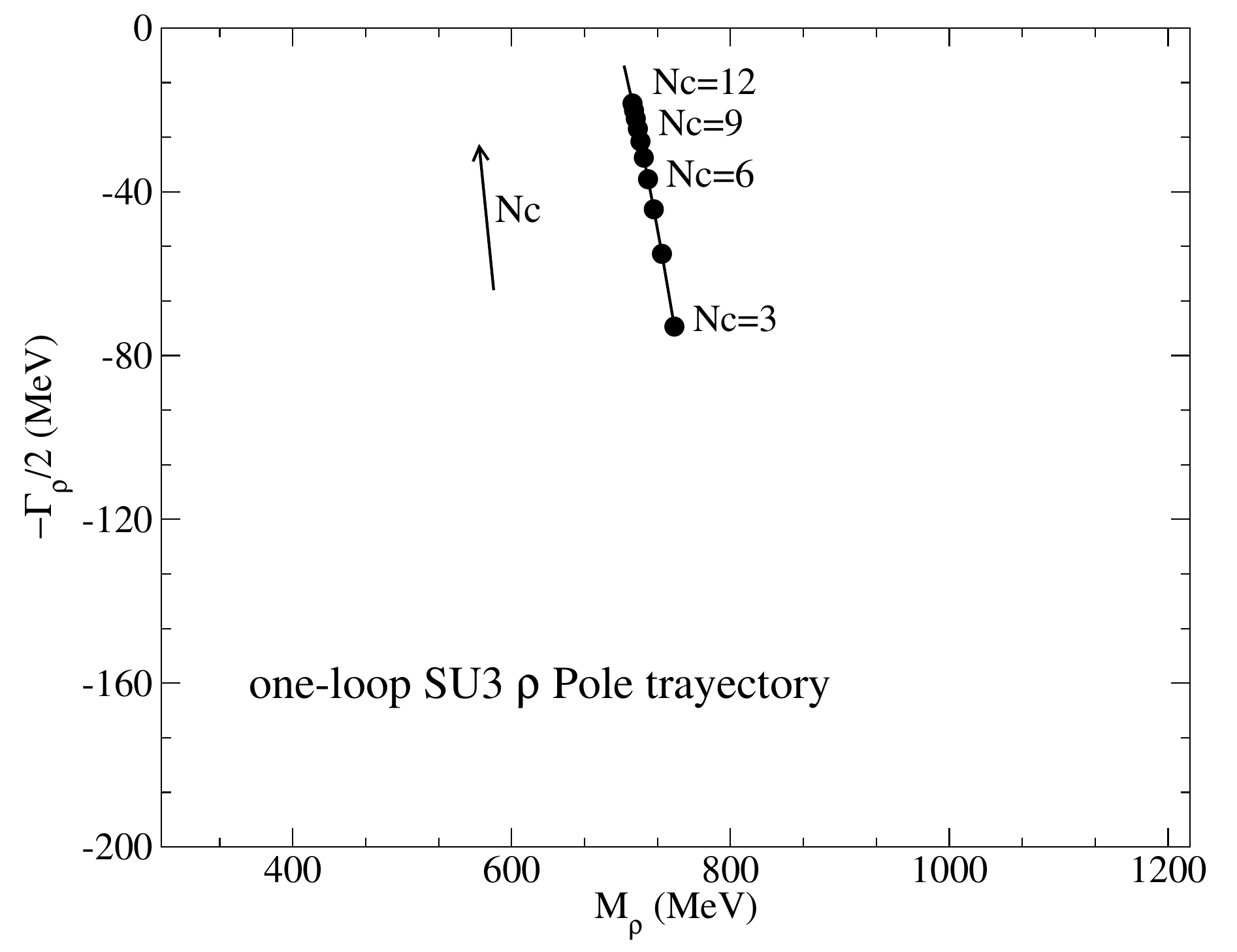}}
  \hspace{-.4cm}
\raisebox{-0.5\height}{%
  \includegraphics[height=3.cm]{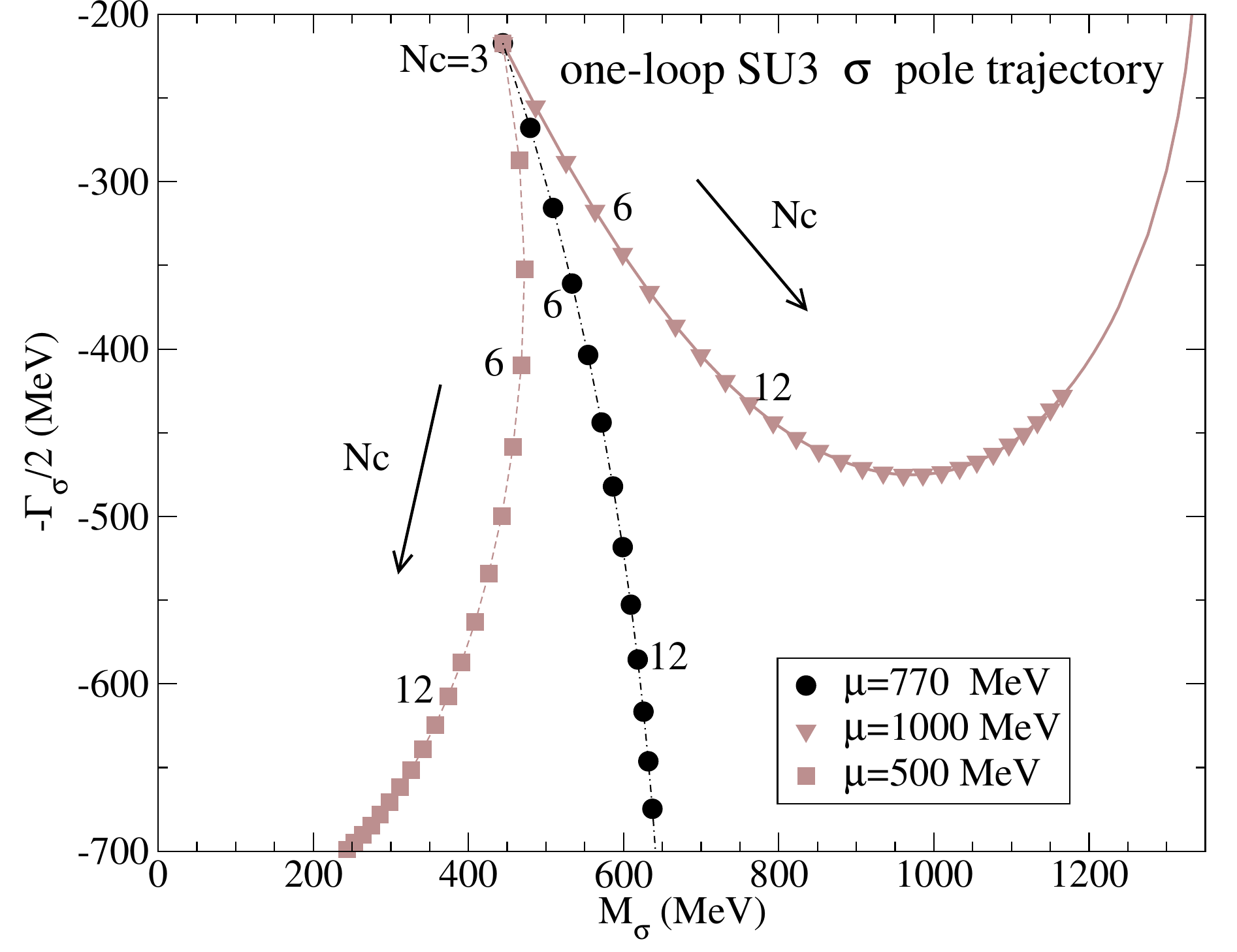} }
  \hspace{-.35cm}
\raisebox{-0.5\height}{%
  \includegraphics[height=3.2cm]{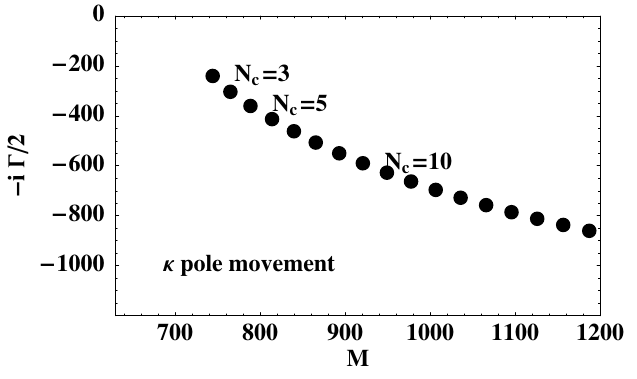} }
\caption{Trajectories of the $\rho(770)$ (left), $\sigma/f_0(500)$ (center) and $\kappa/K^*_0(700)$ poles in the complex plane as $N_c$ is varied away from 3 within NLO ChPT unitarized with the IAM. The lighter curves on the center plot
indicate the uncertainties when varying the regularization
scale $\mu$ in the usual range, as recalculated in \cite{RuizdeElvira:2010cs}.
In the case of the $\rho(770)$ the three lines almost overlap and are not plotted in the left figure. 
Left and center figures taken from \cite{RuizdeElvira:2010cs} and 
right figure taken from \cite{Pelaez:2004xp}.
 \label{fig:Ncuncertainties}}
\end{center}
\end{figure}

That was of course, a very crude approximation. However, the leading $1/N_c$ dependence has been studied to NLO for $\pi\pi$ and $\pi K$ in \cite{Pelaez:2003dy,Pelaez:2004xp}  and to NNLO for $\pi\pi$  in \cite{Pelaez:2006nj}. In this case it is possible to study the $N_c$ behavior of light vectors as well, and it comes completely compatible with the expected $M\sim O(1)$, $\Gamma\sim O(1/N_c)$ of ordinary quark-antiquark mesons. This is illustrated in the left panel of Fig.~\ref{fig:Ncuncertainties} for the $\rho(770)$ meson.
This behavior cannot be found only with LO ChPT, but needs the information on genuine QCD dynamics beyond the spontaneous chiral symmetry scale, which is encoded in the LECs.
In contrast, once again the $\sig$ (Fig.~\ref{fig:Ncuncertainties}, center) and $\kap$ poles (Fig.~\ref{fig:Ncuncertainties}, right) have a behavior at odds with that of ordinary mesons, at least not too far from $N_c=3$, which is the region of interest to understand the physical $\sig$ and $\kap$. This is a rather robust result and has been found in other approaches.
We should nevertheless emphasize that if $N_c$ is made very large, the dominance of meson loops, which are suppressed by $1/N_c$, fades away. In such case, even the tiniest mixture of these states with an ordinary meson will become the dominant component for sufficiently large $N_c$ and in that {\it limit} the pole could behave again as a normal $q\bar q$ state. For instance, the implications of that behavior in that large-$N_c$ {\it limit} for the NN interaction have been studied in \cite{CalleCordon:2009ps}.

Actually, for the $\sig$ case, we see in Fig.\ref{fig:Ncuncertainties} that the sigma pole can turn back \cite{RuizdeElvira:2010cs} to the real axis, but well above 1 GeV. This could be interpreted as the presence, within the physical sigma, of a small mixture with a heavier ``ordinary" state around or above 1 GeV, but that the physical $\sig$ state is only recovered after the unitarized $\pi\pi$ interaction is taken into account, sometimes appearing as an adiitional or companion pole due to unitarization. This behavior seems to be favored in the UChPT NNLO analysis \cite{Pelaez:2006nj} and is also supported by other phenomenological approaches \cite{vanBeveren:1986ea,Achasov:1994iu,vanBeveren:2006ua,Giacosa:2006tf,Nieves:2009ez,Nieves:2011gb,Lukashov:2019dir}.
A relatively similar scenario has also been found for the $\kap$, which generated only after meson-meson interactions are introduced to unitarize the quark model \cite{vanBeveren:1986ea}, whose lightest scalar strange resonance would be above 1 GeV otherwise. There is also further support from another   $N_c$-behavior study \cite{Wolkanowski:2015jtc}  where the $\kap$ appears as a companion pole of the $K^*_0(1430)$, which is predominantly $q\bar q$ whereas the $\kap$ has a predominantly non-ordinary nature. However, for the $\sig$ there is another analysis  \cite{Guo:2011pa,Guo:2012yt} whose behavior is closer to the one of the opposite side of the uncertainty in Fig.\ref{fig:Ncuncertainties}. This behavior reaches the third quadrant at very large $N_c$, which lacks a clear interpretation and is somewhat controversial on the identification of the leading $1/N_c$ terms in $U(3)$ ChPT \cite{Nieves:2011gb}.

Nevertheless, near the physical $N_c=3$ value, it is clear that the observed $\sig$ and $\kap$ predominant composition is not $q \bar q$ or genuine tetraquark, but dominated by a meson-meson interaction or meson cloud (in purity they should not be called molecules since they lie well above threshold). 
In the large $N_c$ {\it limit}, therefore, the $\sig$ and $\kap$ do not survive with similar masses and widths to those observed in nature, although some non-dominant components may do. 
In this sense, the large-$N_c$ {\it limit} is not a good approximation to light meson-meson interactions in the scalar channel. It is important to emphasize this difference between the leading $1/N_c$ behavior close to $N_c=3$, with direct phenomenological implications, and the mathematical large-$N_c$ limit, which for some observables may be detached from the real phenomenology.

It should also be noticed that similar conclusions can be obtained by building observables from the meson-meson scattering phase-shift and the light resonance pole parameters whose sub-dominant $N_c$ corrections are highly suppressed \cite{Nebreda:2011cp}. Using as input the pole parameters either from the model-independent dispersion theory approaches it is seen that the $\sig$ and $\kap$ do not yield the expected values for ordinary mesons or glueballs, for several orders of magnitude.

